\documentclass[aps, twocolumn,superscriptaddress, 
amsmath,amssymb,reprint,numbers,noeprint, floatfix,longbibliography]{revtex4-1}

\usepackage{graphicx}
\graphicspath{
      {figures/}
    }

\begin{document}

\title{Microwave-optical quantum transduction with quantum-well exciton polaritons}
\author{Nicholas S. Yama}
\affiliation{Department of Electrical Engineering and Computer Sciences, University of California, Berkeley, Berkeley, California 94720, USA}
\affiliation{Department of Physics, University of California, Berkeley, Berkeley, California 94720, USA}

\author{Mohammad Khalifa}
\affiliation{NNF Quantum Computing Programme, Niels Bohr Institute, University of Copenhagen, Blegdamsvej 17, 2100 Copenhagen, Denmark.}
\affiliation{Molecular Foundry, Lawrence Berkeley National Laboratory, Berkeley, California 94720, USA}

\author{Yuxuan Wei}
\affiliation{Department of Physics, University of California, Berkeley, Berkeley, California 94720, USA}
\affiliation{Materials Sciences Division, Lawrence Berkeley National Laboratory, Berkeley, California 94720, USA}

\author{Adbhut Gupta}
\affiliation{Department of Electrical and Computer Engineering, Princeton University, Princeton, New Jersey 08544, USA}

\author{Loren N. Pfeiffer}
\affiliation{Department of Electrical and Computer Engineering, Princeton University, Princeton, New Jersey 08544, USA}

\author{Feng Wang}
\affiliation{Department of Physics, University of California, Berkeley, Berkeley, California 94720, USA}
\affiliation{Materials Sciences Division, Lawrence Berkeley National Laboratory, Berkeley, California 94720, USA}
\affiliation{Kavli Energy NanoScience Institute, Berkeley, CA, USA}

\author{Alp Sipahigil}
\affiliation{Department of Electrical Engineering and Computer Sciences, University of California, Berkeley, Berkeley, California 94720, USA}
\affiliation{Department of Physics, University of California, Berkeley, Berkeley, California 94720, USA}
\affiliation{Materials Sciences Division, Lawrence Berkeley National Laboratory, Berkeley, California 94720, USA}

\begin{abstract}
    Linking superconducting quantum processors over optical fiber requires microwave-optical quantum transducers that are efficient, high-bandwidth, and low-noise. 
    This combination has been out of reach because of the weak nonlinearities of conventional electro-optic (EO) materials.
    We propose a transducer with an engineered, strong EO nonlinearity using the quantum-confined Stark effect (QCSE) of quantum-well excitons.
    The QCSE transducer leverages the strong light-matter interactions and extreme polarizability of quantum well excitons.
    Strong coupling to an optical cavity forms exciton-polaritons whose vacuum EO coupling is up to five orders of magnitude larger than in a typical thin-film lithium niobate device. 
    We describe a GaAs/AlGaAs stepped-quantum-well transducer design with predicted transduction efficiencies above 60\% and fewer than 0.01 added noise quanta over a $\sim$60\,MHz bandwidth, using pump powers below 10\,\textmu W and loaded optical and microwave quality factors of only $3\times10^4$ and $10^3$.
    We further show that the cavity protection of the polariton provides robustness against inhomogeneous broadening of the quantum well excitons. 
\end{abstract}

\maketitle

\section{Introduction}
Superconducting (SC) quantum processors are a leading platform for quantum computing with recent experiments demonstrating performance at the limits of classical computing in specific tasks~\cite{larose2026advantage}.
However, fault-tolerant quantum computation will require significant scaling in the number of qubits and their interconnectivity~\cite{google2025belowthreshold,yoder2025tour}.
Such scaling is constrained by the necessity of millikelvin operation temperatures which are difficult to maintain with existing cryogenic technologies as SC processors and their necessary input/output (I/O) architecture increase in size~\cite{lecocq2021control}.
Direct microwave-frequency cryogenic links between processors in separate refrigerators~\cite{magnard2020microwave} may add connectivity but are difficult to scale.

Optical interconnects circumvent these issues by leveraging the high bandwidths and thermal insensitivity of optical ($\sim200$\,THz) photons.
The negligible thermal conductivity of optical fibers combined with frequency-multiplexing would enable the delivery of control signals to many qubits simultaneously, circumventing the I/O bottleneck~\cite{lecocq2021control}.
Coherent transduction between microwave- and optical-frequency photons would enable a high-bandwidth quantum channel for linking quantum processors over arbitrary distances via classical optical networking infrastructure~\cite{kimble2008quantum,lauk2020transduction}.

\begin{figure*}
    \centering
    \includegraphics{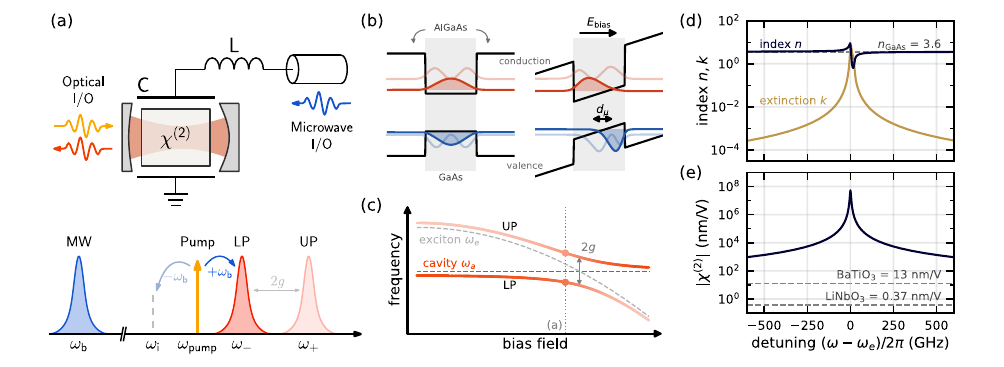}
    \caption{
        \textbf{Quantum transduction with quantum-well exciton polaritons.}
        (a) Schematic of a electro-optic transducer in which a nonlinear $\chi^{(2)}$ medium is placed within a capacitor of an $LC$ circuit and an optical cavity.
        The medium is pumped at a frequency $\omega_\mathrm{pump}$ detuned from an optical resonance (here the lower polariton, LP) at $\omega_-$ by the microwave resonance frequency $\omega_b$.
        An incident microwave photon scatters pump photons primarily into the LP.
        (b) The electron and hole wavefunctions of the QW heterostructure at zero (left) and finite (right) bias.
        Under a bias a large static dipole $d_\mu$ can be formed.
        (c) Energy levels of the QW exciton and cavity as a function of the bias field showing an anti-crossing forming the upper- (UP) and lower-polaritons (LP).
        Effective (d) index $n+ik$ and (e) nonlinear susceptibility $|\chi^{(2)}|$ of the QW medium.
    }
    \label{fig:overview}
\end{figure*}

A myriad of transduction schemes based on electro-optic (EO)~\cite{fan2018superconducting,xu2021bidirectional,sahu2023entangling,warner2025coherent}, electro-optomechanical~\cite{higginbotham2018harnessing,mirhosseini2020transduction,meesala2024quantum}, and magneto-optic~\cite{hisatomi2016bidirectional,zhu2020waveguide,xie2025transduction,khalifa2025robust} interactions have been proposed and experimentally demonstrated.
Recent efforts have shown separately high-efficiency~\cite{higginbotham2018harnessing} and low-noise~\cite{mirhosseini2020transduction,sahu2023entangling,meesala2024quantum,xie2025transduction, zhao2025quantum} conversion.
However, achieving simultaneously high efficiency, large bandwidth, and low noise remains an outstanding challenge.
This is because transduction is mediated by weak optical nonlinearities which require parametric enhancement that can inject noise via parasitic heating and other undesirable nonlinear effects.
Engineered low-loss resonances relax the power requirements~\cite{lauk2020transduction} but introduce additional challenges in fabrication, stability, and bandwidth.

Here we propose a transducer design based on the quantum-confined Stark effect (QCSE)~\cite{miller1984qcse,miller1985electric} of quantum-well (QW) excitons, which form exciton polaritons in an integrated photonic resonator embedded in a superconducting microwave resonator. The lower polariton acts as an EO mode with a strongly enhanced nonlinearity, enabling high-bandwidth and low-noise operation at low pump powers. 
Our approach expands upon earlier work utilizing quantum dots for transduction~\cite{tsuchimoto2017proposal,tsuchimoto2022largebandwidth} by enabling parametric enhancement and strong modal overlap with microwave and optical modes, while also leveraging the highly engineerable QW platform. 
This introduces new applications for electrically-modulated QWs in transduction, beyond sensing~\cite{togan2020sensor} and waveguide modulation~\cite{liran2025femtojoule}.
First we describe a model of the QCSE transduction process and compare it against traditional bulk EO responses.
We then describe a practical transducer design capable of high-efficiency and low-noise transduction at microwatt-level pump powers and optical quality factors of only $30{,}000$.
Finally, we show that cavity protection, previously established for ensemble linewidth narrowing~\cite{diniz2011protecting, zhong2017cavityprotected}, extends to the transduction
efficiency and added noise, making the transducer robust to disorder.

\section{QCSE transduction}
Figure~\ref{fig:overview}a illustrates a typical EO transducer architecture in which a nonlinear $\chi^{(2)}$ medium is embedded within an optical and microwave resonator.
A strong pump field at $\omega_\mathrm{pump}$ is detuned from an optical resonance (here the lower polariton, LP) by the microwave resonator frequency $\omega_b$.
Incident microwave fields are mixed with the pump to form sidebands at the signal $\omega_s = \omega_\mathrm{pump}+\omega_b$ and idler $\omega_i = \omega_\mathrm{pump}-\omega_b$ frequencies.
The signal (anti-Stokes) sideband is the coherent transduction process and is enhanced by the optical resonance, while the idler (Stokes) sideband corresponds to the amplification process which injects noise and is suppressed.
The strength of the nonlinearity, characterized by the nonlinear susceptibility $\chi^{(2)}$, dictates the necessary pump power and resonance quality factor.

The QCSE transducer leverages the engineered confinement of AlGaAs/GaAs QWs as shown in Fig.~\ref{fig:overview}b to achieve a large nonlinear susceptibility.
The optical pump field generates excitons within the QW.
The strong co-localization of the electron and hole increases the binding energy and optical coupling rate $g$ of the QW excitons while also preventing them from disassociating in an applied dc bias field $E_\mathrm{bias}$ normal to the QW.
The applied bias induces a static electric dipole $d_\mu$ and shifts the exciton frequency $\omega_e$.
As the applied bias increases, the exciton hybridizes with the optical cavity resonance at $\omega_a$ to form the upper (UP) and lower (LP) polaritons (Fig.~\ref{fig:overview}c).
The microwave-cavity photons modulate the polariton frequency through the linear quantum-confined Stark effect, giving rise to an effective $\chi^{(2)}$ EO response.

The QCSE Hamiltonian is given by
\begin{align}
    \label{eq:qcse_hamiltonian_base}
    \hat H ={}& \hbar\omega_a \hat a^\dagger \hat a
    + \hbar\omega_b \hat b^\dagger \hat b
    + \hbar\omega_e \hat e^\dagger \hat e \nonumber\\
    & - \hbar g \Big(\hat a^\dagger \hat e + \hat a \hat e^\dagger\Big)
    - \hbar \xi \hat e^\dagger\hat e \Big(\hat b + \hat b^\dagger\Big),
\end{align}
where $\hat a$, $\hat b$, and $\hat e$ are the optical photon, microwave photon, and exciton annihilation operators respectively.
The excitons are treated as bosons, which is valid in the low-density limit~\cite{haugkoch2004quantum}.
The two terms on the second line correspond to the linear exciton-optical interaction and the nonlinear exciton-microwave Stark effect interaction respectively.
The Stark-shift term describes a per-photon shift of the exciton frequency by $\xi = d_\mu E_{b,\mathrm{zpf}}/\hbar$, where $E_{b,\mathrm{zpf}}$ is the microwave zero-point fluctuation amplitude.
Full details are given in the Supplemental Material~\cite{supplement}.

The QCSE process is a resonant nonlinear interaction which distinguishes it from the traditional non-resonant/bulk nonlinearities of typical EO modulators based on, e.g., thin-film lithium niobate (TFLN)~\cite{zhu2021integrated} or barium titanate (BTO)~\cite{abel2019large}.
The large optical dispersion of the QW exciton near its resonance enhances the linear light-matter interaction via an increased group index $n_g$, while increasing its sensitivity to electric-field modulation, simultaneously enhancing the $\chi^{(2)}$ response.
However, the strong dispersion introduces new dissipation channels as required by the Kramers-Kronig relations, which work against increased nonlinearity.
Similar tradeoffs between resonant enhancement and dissipation are well known in rare-earth-ion~\cite{everts2019microwave,rochman2023microwave,xie2025transduction}, Rydberg-atom-based~\cite{kumar2023quantum}, and magnon-exciton~\cite{adak2026magnon} transducer architectures.
These effects are characterized by the internal semi-classical effective index and nonlinear susceptibility of the QW computed in Fig.~\ref{fig:overview}d,e~\cite{supplement}.
While $\chi^{(2)}\sim10^8$\,nm/V on resonance, the polaritons are detuned from the exciton by roughly $g$, suppressing both the nonlinearity and loss by $(\gamma/g)^2\sim10^{-4}$, where $\gamma$ is the intrinsic exciton loss rate which includes emission into non-guided modes and non-radiative decay.
The finite filling factor of the QW and engineered coupling to guided modes further suppresses both, yielding an effective $\chi^{(2)}\sim10^2$\,nm/V (ten times larger than BTO) with losses consistent with quality factors exceeding $10^5$.

\section{Model}
We can convert the QCSE Hamiltonian Eq.~\eqref{eq:qcse_hamiltonian_base} into the polariton frame via the transformation
\begin{align*}
    \hat a_+ &= C^* \hat e + X \hat a, 
        & \hat e &= C \hat a_+ + X \hat a_-,\\
    \hat a_- &= X^* \hat e - C \hat a, 
        & \hat a &= X^* \hat a_+ - C^* \hat a_-,
\end{align*}
where the coefficients $X$ and $C$ are the Hopfield coefficients~\cite{haugkoch2004quantum,hopfield1958theory}
\begin{align*}
    C &= \frac{2g}{\sqrt{(\Delta - \Delta_p)^2 + 4g^2}},
    & 
    X &= \frac{\Delta - \Delta_p}{\sqrt{(\Delta - \Delta_p)^2 + 4g^2}},
\end{align*}
with exciton-optical detuning $\Delta = \omega_e - \omega_a$ and polariton splitting $\Delta_p = \sqrt{\Delta^2 + 4g^2}$.
The polariton frequencies are $\omega_\pm = (\omega_e + \omega_a \pm \Delta_p)/2$.
The Hamiltonian is then
\begin{align}
    \hat H & =  \hbar\omega_+ \hat a_+^\dagger \hat a_+
    + \hbar\omega_- \hat a_-^\dagger \hat a_-
    + \hbar \omega_b \hat b^\dagger\hat b \nonumber\\
    & \quad- \hbar \Big(
        g_\mathrm{eo}^+ \hat a_+^\dagger \hat a_+ 
        + g_\mathrm{eo}^- \hat a_-^\dagger \hat a_-\Big)
    \Big(\hat b + \hat b^\dagger \Big),
\end{align}
where we have eliminated the fast rotating terms $\hat a_\pm^\dagger\hat a_\mp$ under the rotating-wave approximation (RWA) and defined the vacuum electro-optic coupling rate of the upper/lower polariton $g_\mathrm{eo}^+ = \xi|C|^2$ and $g_\mathrm{eo}^-=\xi|X|^2$, respectively.

The QCSE Hamiltonian now has an identical form to a bulk single-mode EO interaction $\hat H_\mathrm{bulk} = -\hbar g_\mathrm{eo}^\mathrm{bulk}\hat a^\dagger\hat a\big(\hat b + \hat b^\dagger\big)$ where $g_\mathrm{eo}^\mathrm{bulk} = (\omega_a/n^2) f \chi^{(2)}_\mathrm{bulk} E_{b,\mathrm{zpf}}$.
Here $f$ is the filling factor of the nonlinear medium and $n$ is the refractive index.
Consequently, the LP can be utilized in the standard sideband modulation scheme (Fig.~\ref{fig:overview}a), in which it acts like a bulk $\chi^{(2)}$ mode with modified nonlinear coupling rate $g_\mathrm{eo}^- = \xi|X|^2$, line width $\kappa_- = |C|^2\kappa + |X|^2\gamma$ and extrinsic coupling $\kappa_{-,\mathrm{ext}} = |C|^2\kappa_\mathrm{ext}$.
In this sense, the exciton loss can be considered a new intrinsic loss channel $|X|^2\gamma$.
Assuming an otherwise identical device, the QCSE nonlinearity is enhanced relative to the bulk by approximately $g_\mathrm{eo}^-/g_\mathrm{eo}^\mathrm{bulk} \sim 80$, which maps to a $6400$-fold reduced pump power (some of which must be used to mitigate the lower quality factors).
The large nonlinearity also permits a compact, low-$Q$ device with capacitance
$C \approx 1$\,fF at $\omega_b/2\pi = 30$\,GHz, which increases
$E_{b,\mathrm{zpf}}$ and hence $g_\mathrm{eo}^-$ by a further $\sim 10^3$
relative to demonstrated TFLN transducers.

Carrying the analogy forward, under an applied pump at $\omega_\mathrm{pump} = \omega_- -\omega_b$, the operators are expanded as $\hat a_- \to \bar a_- + \delta \hat a_-$ where $\bar a_-$ is the classical pump amplitude of the LP mode and $\delta\hat a_-$ describes the quantum fluctuations.
This linearizes the interaction Hamiltonian into the beam splitter form $\hat H_\mathrm{int} = -\hbar g_\mathrm{eo}^- \big(\bar a_- \hat a_-^\dagger \hat b + \bar a_-^* \hat a_- \hat b^\dagger\big)$ (cross terms dropped under the RWA).
Using input-output theory~\cite{gardiner1985input}, the transduction efficiency on resonance ($\omega_\mathrm{pump} = \omega_- -\omega_b$, microwave input at $\omega_b$) is
\begin{equation}
    \label{eq:main:model:efficiency}
    \eta = \eta_-\eta_b \frac{4\mathcal C}{|1+\mathcal C|^2},
\end{equation}
where $\eta_- = \kappa_{-,\mathrm{ext}}/\kappa_-$, $\eta_b = \Gamma_\mathrm{ext}/\Gamma$ are the extrinsic coupling efficiencies for the LP and microwave mode, respectively, and $\mathcal C = 4(g_\mathrm{eo}^-)^2|\bar a_-|^2/\kappa_-\Gamma$ is the pump-enhanced cooperativity.
The efficiency is maximized at impedance matching ($\mathcal C = 1$), where the optically induced damping of the microwave mode equals its loaded line width $\Gamma$; for $\kappa_- \gg \Gamma$, the transduction bandwidth is set by the microwave mode and equals $(1+\mathcal C)\Gamma$.
The added noise at leading order is
\begin{equation}
    \label{eq:main:model:noise}
    n_\mathrm{add} =
    \frac{\Gamma_\mathrm{int}}{\Gamma_\mathrm{ext}}n_\mathrm{th}
    +
    \frac{\mathcal C}{\eta_b} |\chi_-(-\omega_b)|^2,
\end{equation}
where $n_\mathrm{th}$ is the thermal occupation of the microwave mode (negligible at millikelvin temperatures for $\omega_b/2\pi = 30$\,GHz) and $\chi_-(\omega) = [2i(\omega_b-\omega)/\kappa_- + 1]^{-1}$ is the susceptibility of the LP at the idler (in the appropriate rotating frame) that gives rise to the amplification noise and is suppressed by roughly $(\kappa_-/4\omega_b)^2$.

\section{Device design and performance}
Efficient transduction relies on large field overlaps between the microwave mode, the photonic mode, and the active nonlinear QW medium.
Below, we present a design that achieves efficient, low-noise transduction with low pump powers and modest quality factors.

\begin{figure*}
    \centering
    \includegraphics{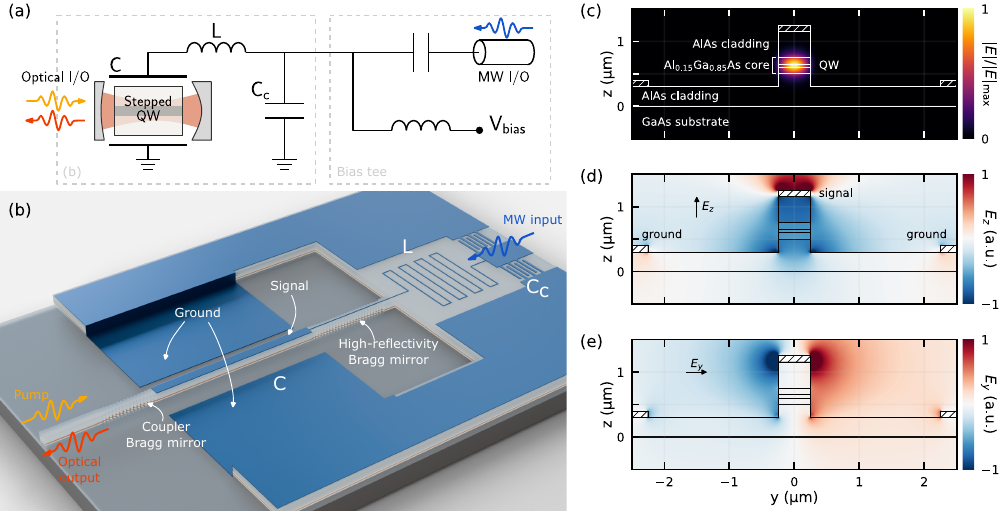}
    \caption{
        \textbf{Transducer design.}
        (a) Circuit diagram of the transducer design including an off-chip bias tee for injecting the microwave signal and bias voltage.
        (b) Illustration of the proposed transducer device.
        The QW is embedded in a 15\% AlGaAs waveguide core with an AlAs cladding.
        The waveguide cavity is etched into the structure with Bragg mirrors on either end and embedded within a microstrip capacitor $C$.
        The series inductor $L$ and shunt coupling capacitor $C_c \gg C$ connect the system to a CPW feedline.
        (c) Simulated fundamental TE mode of the photonic waveguide cavity.
        Simulated (d) out-of-plane $E_z$ and (e) in-plane $E_y$ microwave/bias field profiles in the waveguide cavity.
    }
    \label{fig:device}
\end{figure*}

\subsection{Photonic and microwave design}
Figure~\ref{fig:device} illustrates the QCSE transducer design consisting of a photonic waveguide cavity (formed from two Bragg reflectors) embedded within the microstrip capacitor of a lumped element microwave resonator.
The strip capacitor is in series with a lumped-element inductor made from a high-kinetic-inductance superconductor~\cite{samkharadze2016highkinetic} forming the microwave resonance at $\omega_b = 1/\sqrt{LC}$.
A second large coupling capacitor shunts the transducer circuit at the feedline which, as shown in Fig.~\ref{fig:device}a, carries both the dc bias and microwave signal via an upstream bias tee.
The pump field and transduced photons are injected/emitted through the partially reflective Bragg mirror (Fig.~\ref{fig:device}b).

The ridge waveguide is formed from a 250-nm Al$_{0.15}$Ga$_{0.85}$As core with embedded QW heterostructure and cladded by AlAs.
The 500-nm-wide ridge accommodates a fundamental transverse-electric (TE) mode shown in Fig.~\ref{fig:device}c with intrinsic quality factor $Q^\mathrm{opt}_\mathrm{int} > 10^6$.
Coupling/reflector mirror transmission of $7.5$\% and $0.01$\% respectively yield a loaded quality factor $Q^\mathrm{opt} \approx 30{,}000$ for a 25-\textmu m-long waveguide.
The signal trace is patterned on top of the ridge AlAs cladding yielding the out-/in-plane fields shown in Fig.~\ref{fig:device}d,e and a total capacitance of $C = 1.24$\,fF.
The structure is designed to minimize the in-plane fields which broaden the exciton resonance~\cite{miller1985electric,lederman1976theory,rocke1998exciton}.
Full details of the design are given in the Supplemental Material~\cite{supplement}.

\subsection{Quantum well design}
\begin{figure*}
    \centering
    \includegraphics{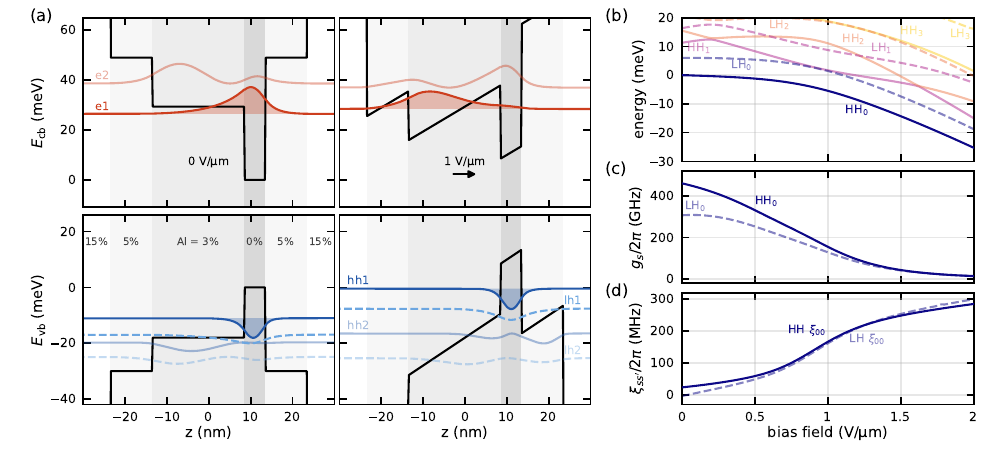}
    \caption{
        \textbf{Stepped quantum well design.}
        (a) The stepped QW wavefunctions at 0 and 1\,V/\textmu m bias.
        The well consists of a 27-nm-wide well at 3\% Al concentration surrounded by a 5\% Al barrier layer and with a 5-nm 0\% Al well on the right edge.
        (b) Computed energies of the $1s$-like exciton states as a function of the applied bias.
        The ground state HH$_0$ is a heavy hole exciton and is separated from the nearest exciton state by more than 5 meV.
        (c) Computed optical coupling rate for the different exciton states as a function of the applied bias.
        (d) Computed diagonal elements $\xi_{00}= g_\mathrm{eo}^-/|X|^2$ of the electro-optic coupling matrix for HH$_0$ and LH$_0$.
    }
    \label{fig:qw}
\end{figure*}

We use a stepped QW design~\cite{morita1990stepped,chen1992shaped} as shown in Fig.~\ref{fig:qw}a which consists of a 27-nm-wide 3\% AlGaAs well with 5\% AlGaAs barriers and a single 5-nm GaAs well located inside on the right edge.
The design leverages the tighter binding of the heavy hole (HH) to localize it in the narrow GaAs well while the light hole (LH) and electron are largely distributed over the full well.
The broken inversion symmetry gives a static dipole moment of about 2\,nm at zero bias which can be further polarized without the need for large bias fields.
Moreover, the selective confinement of the HH produces a large separation in energy of the HH ground state (hh1) from the other states, minimizing their admixture.

Under an applied positive bias the stepped edge of the QW shifts downward creating a well that traps the electron on the opposite end from the hole (Fig.~\ref{fig:qw}a, right).
The resulting $1s$-like exciton ground state HH$_0$ is primarily composed of the electron/HH ground states e1 and hh1, respectively, and takes on the character of an indirect exciton of a double QW~\cite{dignam1991exciton}.
This character is seen in Fig.~\ref{fig:qw}b--d which plots the energy, optical coupling rate $g$, and microwave coupling rate $\xi$ as a function of the applied bias.
Like a typical indirect exciton, increasing the bias further separates the electron and hole causing a reduction in the optical coupling while increasing the microwave coupling.
At a bias of $1$\,V/\textmu m, the optical coupling and single-photon Stark shift are approximately $g/2\pi=154$\,GHz and $\xi/2\pi=170$\,MHz, respectively. 
At zero detuning this corresponds to $g_\mathrm{eo}^-/2\pi \approx 85$\,MHz,
about $10^5$ times larger than demonstrated TFLN transducers
($\approx 1$\,kHz~\cite{warner2025coherent}).
Thus the exciton can sustain strong coupling with the cavity (line width $\kappa/2\pi= 12.5$\,GHz) while maintaining a vacuum EO coupling that exceeds typical microwave cavity line widths.
Full details of the QW are given in the Supplemental Material~\cite{supplement}.

\subsection{Power, efficiency, and noise}
\begin{figure}
    \centering
    \includegraphics{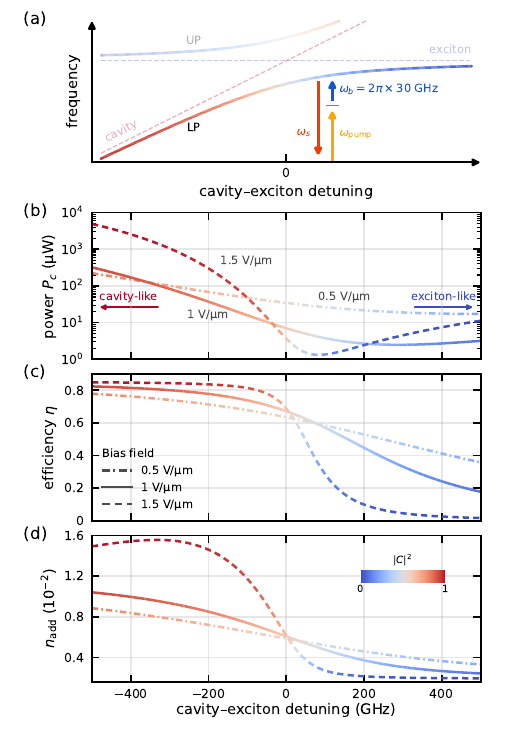}
    \caption{
        \textbf{Transduction performance.}
        (a) Illustration of the polariton levels and transduction process as the cavity is detuned through the HH$_0$ exciton.
        (b) The pump power $P_c$, (c) peak transduction efficiency $\eta$, and (d) added noise $n_\mathrm{add}$ at the impedance matched ($\mathcal C = 1$) point as a function of the cavity-exciton detuning $-\Delta$ for different bias fields.
        The color of the curves corresponds to the character of the polariton with red corresponding to a more cavity-like and blue a more exciton-like polariton.
        The solid line corresponds to 1\,V/\textmu m while dashed/dashed-dotted correspond to 1.5\,V/\textmu m and 0.5\,V/\textmu m, respectively.
        The cavity and exciton line widths are fixed at $\kappa/2\pi=12.5$\,GHz and $\gamma/2\pi=5$\,GHz, respectively, and the calculation includes the higher-order exciton states.
    }
    \label{fig:performance}
\end{figure}

Figure~\ref{fig:performance} shows the power $P_c$, efficiency $\eta$, and added noise $n_\mathrm{add}$ of the QCSE transducer at the impedance-matched ($\mathcal C=1$) point as a function of the cavity-exciton detuning and bias field.
The strong nonlinearity allows high efficiency operation with modest microwave and optical quality factors, and low pump powers. 
We assume a loaded optical quality factor $Q^\mathrm{opt}=30{,}000$ with intrinsic $Q^\mathrm{opt}_\mathrm{int} = 5\times10^5$ ($\eta_a = \kappa_\mathrm{ext}/\kappa = 0.94$).  An even lower $Q^\mathrm{opt}_\mathrm{int} = 10^5$ retains $\eta_a = 0.7$.
The larger optical line width is necessary for efficient transduction by overcoupling the intrinsic exciton homogeneous loss ($\gamma/2\pi=5$\,GHz). 
Low-$Q^\mathrm{opt}$ operation can be afforded due to the large QCSE nonlinearity while also enabling wider tunability of the operation frequency.
However, low $Q^\mathrm{opt}$ necessitates a correspondingly large microwave frequency $\omega_b/2\pi=30$\,GHz to achieve sideband-resolved operation and sufficient suppression of the amplification noise.
The large nonlinearity also relaxes the microwave loss constraints.
We assume a loaded microwave quality factor $Q^\mathrm{mw} = 10^3$ with intrinsic $Q^\mathrm{mw}_\mathrm{int} = 10^4$~\cite{scigliuzzo2020phononic} ($\eta_b = 0.9$) which are well within typical experimentally realizable values.
With these modest quality factors, the transduction efficiency is limited to $\eta < \eta_a \eta_b = 0.846$, and the transduction bandwidth at
impedance matching is $2\Gamma/2\pi \approx 60$\,MHz.
Because $P_c \propto \Gamma$ at fixed $\mathcal C$, bandwidth can be
traded directly against pump power. 
At a bias of 1\,V/\textmu m and zero detuning, the transducer reaches $\eta \approx 0.67$ and $n_\mathrm{add} \approx 0.006$ with $P_c \approx 7$\,\textmu W. 
The photon fraction is $|C|^2 \approx 0.6$ at this point due to the coupling with the higher-energy exciton states.
The region with $\eta > 0.6$, $n_\mathrm{add} < 0.01$, and $P_c < 10$\,\textmu W spans about 100\,GHz in detuning (200\,GHz for $\eta > 0.5$).

The cavity-exciton detuning controls the exciton $|X|^2$ and photon fraction $|C|^2$ of the LP and the effective EO nonlinearity.
In the cavity-like limit $|C|^2\approx 1$, the system behaves like a traditional bulk EO nonlinearity, suppressing the exciton loss channel but reducing the nonlinear interaction strength $\xi|\bar a_-| |X|^2$.
This enables increased efficiency by improving the output coupling efficiency $\eta_-$ but drives the impedance-matching power $P_c$ up considerably (Fig.~\ref{fig:performance}b,c).
In contrast, a larger exciton fraction $|X|^2$ increases the nonlinear interaction, reducing impedance-matching power at the expense of a worse output coupling.
Further increasing $|X|^2$ drives the transduction efficiency $\eta$ down, and eventually starts to increase the required power as the polariton becomes less efficiently pumped.
The noise is largely insensitive, with only a slight preference for more exciton-like operation due to the narrower overall polariton line width $\kappa_- = |C|^2\kappa + |X|^2\gamma$ (where $\gamma<\kappa$).
We find that resonant cavity-exciton operation best balances these competing effects.

Similar to a bulk EO process, excessively large input power is problematic for the QCSE transducer, however via different mechanisms.
The exciton no longer approximates a boson as the density approaches $1/\pi a_B^2\sim 10^3$\,\textmu m$^{-2}$ where $a_B\sim 10$\,nm is the Bohr radius.
This is still considerably larger than the roughly $1$\,\textmu m$^{-2}$ density at our nominal operation point.
However, collisional broadening~\cite{schultheis1986ultrafast,kurdyubov2023nonlinear} and Kerr-like shifts due to exchange interactions~\cite{ciuti1998role} and dipole-dipole interactions~\cite{rosenberg2018dipolar,togan2018enhanced} may still be active at these densities.
While transduction should be more robust to any shifts in the operation frequency, the collisional broadening could reduce the efficiency by increasing the exciton homogeneous line width $\gamma$.
Fortunately, our specific operation mode of pumping the LP at the red-detuned sideband should mitigate the collisional broadening by spectrally isolating the pumped exciton branch from the dark modes that it would otherwise scatter into~\cite{knox1989femtosecond,kuklinski1990real}.
This is analogous to the effect of cavity protection discussed below.

\section{Disorder and cavity protection}
\begin{figure}
    \centering
    \includegraphics{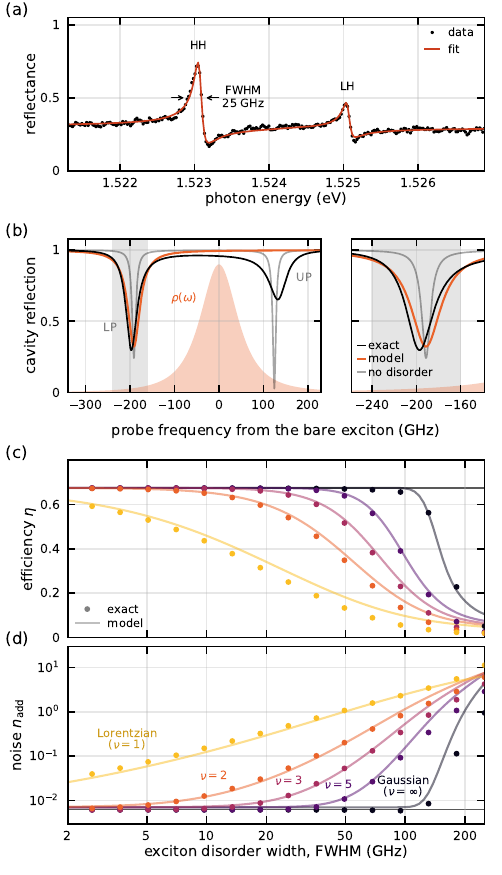}
    \caption{
        \textbf{Cavity protection.}
        (a) Measured reflection spectrum from a 30-nm QW and fit.
        (b) Simulated exact (black), model (orange), and non-disordered (grey) polariton spectra in the presence of an exciton continuum ensemble $\rho(\omega)$.
        The model (presented in the text) linearizes the system response at the LP and thus neglects the UP.
        Simulated impedance-matched (c) efficiency and (d) added noise of the transducer for different $\rho(\omega)$ modeled by Student's $t$-distributions.
        The exact solutions (circles) are generally predicted by the model (solid lines).
        All calculations include only the HH ground state with detuning chosen to achieve the same nominal separation between the LP and bare exciton frequency of $\Delta_-\approx200$\,GHz as in the transducer operation point of Fig.~\ref{fig:performance}.
        Grey lines describe the device without disorder.
    }
    \label{fig:continuum}
\end{figure}

So far, we have only considered homogeneous broadening of the exciton $\gamma$ due to radiative decay into unguided modes and non-radiative decay. 
In practice, the effect of exciton inhomogeneous broadening due to spatial strain or charge disorder deserves particular attention due to the importance of excitonic coherences in the transduction process.
To estimate the experimentally achievable inhomogeneous broadening in QWs, Figure~\ref{fig:continuum}a plots our measured reflectance spectrum of a 30-nm QW (27\% Al barrier).
We fit the spectrum to an oscillator response~\cite{andreani1991radiative,ivchenko2005optical} and determine the total full width at half maximum (FWHM) of the HH resonance to be about 25\,GHz, of which up to about 11\,GHz may be attributed to non-radiative and inhomogeneous broadening.
It is expected that the majority of this broadening is inhomogeneous in nature~\cite{gammon1996homogeneous}, which supports our assumption of an exciton loss rate $\gamma/2\pi=5$\,GHz provided radiative decay is primarily into the waveguide mode.

Inhomogeneity is modeled by the addition of $N$ different exciton branches $\hat e_n$ with random frequency $\omega_n$ distributed about the average bare exciton frequency $\bar\omega_e$.
The optical cavity couples to a specific bright state $\hat B$ superposition of the exciton states
\begin{equation}
    \hat B = \frac{1}{G}\sum_{n=1}^N g_n \hat e_n,
\end{equation}
where $g_n$ are the individual optical coupling rates of each exciton branch $n$ to the optical mode.
The coupling rates are constrained by the oscillator sum rule such that $G^2 = \sum_n |g_n|^2$ where $G$ is the optical coupling rate of the momentum-matched bare exciton state in the absence of disorder ($g$ in the prior sections).
The density of states projected on the bright state follows as $\rho(\omega) = \sum_n p_n \delta(\omega-\omega_n)$ where $p_n = |g_n|^2/G^2$.

If the coupling $G$ sufficiently exceeds the characteristic width of the density of states $\rho(\omega)$, then, to lowest order, the polariton interacts with the ensemble as if it were a single exciton at the ensemble averaged frequency $\bar\omega_e$.
Thus the polaritons occur at frequencies $\omega_p \approx \omega_\pm$ matching the non-disordered system with $\omega_e \to \bar\omega_e$.
Linearizing the cavity response at the LP produces the well-known result that the polariton line width is given by~\cite{diniz2011protecting,houdre1996rabi,whittaker1998inhomogeneous}
\begin{equation}
    \tilde\kappa_- = |C|^2\kappa + |X|^2\gamma + 2\pi|C|^2G^2\rho(\omega_p).
\end{equation}
The first two terms match the line width of the non-disordered polariton, while the new term is an additional broadening due to decay into the dark exciton states orthogonal to $\hat B$.
Because the polariton roughly occurs with detuning $\omega_p - \bar\omega_e \approx G$, it experiences a cavity protection effect wherein the additional loss is suppressed by increasing the coupling $G$ provided the density of states $\rho(\omega)$ falls off faster than $\omega^{-2}$~\cite{diniz2011protecting,zhong2017cavityprotected}.
Figure~\ref{fig:continuum}b plots a representative cavity reflection spectrum with exciton inhomogeneity in the continuum limit.
The linearized model (orange) and exact solution (black) both predict the broadening of the LP in comparison to the homogeneous spectrum without disorder (grey); however, the linearized model misses an additional dispersive shift.

We now consider the effects of cavity protection on the quantum transducer.
We linearize the QCSE Hamiltonian as in the clean, homogeneously broadened system.
The resulting mean-field equations map onto the linear polariton response with the additional broadening $2\pi|C|^2G^2\rho(\omega_p)$, modifying the pump efficiency.
At first order, a similar renormalization of the optical mode occurs with an analogous renormalization of the microwave mode.
It can be shown that the microwave mode is similarly broadened to $\tilde\Gamma = \Gamma+\delta\Gamma$ with
\begin{equation}
    \delta \Gamma = \frac{2\xi^2|C|^2|\bar a_-|^2}{\omega_b^2}G^2\pi\Big[ \rho(\omega_s) - \rho(\omega_i) \Big]
\end{equation}
where $\omega_s = \omega_\mathrm{pump}+\omega_b$ and $\omega_i = \omega_\mathrm{pump} -\omega_b$ are the signal and idler (amplification) frequencies respectively.
The extra broadening on both modes acts like an intrinsic dissipation channel which reduces the output coupling efficiency, e.g.\ as $\eta_- \to \kappa_{-,\mathrm{ext}}/\tilde\kappa_-$.
These additional loss channels also scale as $G^2\rho(\omega)$ and are thus suppressed by cavity protection.

The shared coupling of the optical and microwave modes to the exciton dark state ensemble also introduces a new dissipative coupling channel which can interfere with the dispersive QCSE process.
This modifies the transduction efficiency on resonance as
\begin{equation}
    \eta_\mathrm{ab} = \frac{\kappa_{-,\mathrm{ext}}}{\tilde \kappa_-}
    \frac{\Gamma_{\mathrm{ext}}}{\tilde \Gamma}
    \frac{4\mathcal C}{|1 + \mathcal C (1+i\theta)|^2}
\end{equation}
where $\theta =  2\pi G^2 |C|^2 \rho(\omega_s)/ \omega_b |X|^2$.
Finite $\theta$ shifts the transduction window and introduces a small non-reciprocity through asymmetric interference of the two coupling paths.
In the limit where $G^2\rho(\omega_s) \to 0$ the transduction efficiency reduces to the non-disordered form Eq.~\eqref{eq:main:model:efficiency}, indicating that it is also cavity protected.

The new dissipative channels inject noise via an amplification process with ensemble excitons near the pump frequency.
The resulting noise at lowest order is given by
\begin{align}
    n_\mathrm{add} ={}& 
    \frac{\Gamma_\mathrm{int}}{\Gamma_\mathrm{ext}}n_\mathrm{th}
    +
    \frac{1}{\eta_b} \mathcal C|\chi_-(-\omega_b)|^2 \nonumber\\
    & + \frac{\xi^2|C|^2|\bar a_-|^2}{\omega_b^2\Gamma_\mathrm{ext}}2\pi G^2 \Big(\rho(\omega_i) + \rho(\omega_\mathrm{pump}) \Big) \\
    & - \frac{16\xi^2|C|^2|\bar a_-|^2 |X|^2}{\Gamma_\mathrm{ext}\tilde\kappa_-^2}
    2\pi G^2 \rho(\omega_i) |\chi_-(-\omega_b)|^2. \nonumber
\end{align}
The first row reproduces the non-disordered noise Eq.~\eqref{eq:main:model:noise} while the second corresponds to direct injection of noise from the exciton ensemble at the idler and pump frequencies.
The third row describes the interference of the optical and exciton idler noise paths.
As before, the new terms scale in proportion to $G^2\rho(\omega)$ and are thus suppressed by cavity protection.
Full details of the derivation are given in the Supplemental Material~\cite{supplement}.

Figure~\ref{fig:continuum}c,d plot the efficiency and added noise of the transducer versus the FWHM of various inhomogeneous distributions $\rho(\omega)$ modeled by the Student's $t$-distribution.
The tail parameter $\nu$ determines the decay of the distribution tails as $\omega^{-(\nu+1)}$.
A Lorentzian ($\nu=1$) distribution falls off quadratically and is not cavity protected, injecting significant noise/broadening even at extremely narrow FWHM.
In contrast, a Gaussian ($\nu=\infty$) broadening typical of inhomogeneous distributions behaves almost identically to the non-disordered system for FWHMs less than about half the LP-exciton separation $\Delta_- = |\omega_- -\bar\omega_e|$.
This $\sim$100\,GHz margin also accommodates additional inhomogeneous broadening expected in the stepped, etched device QW from alloy disorder, interface roughness, and sidewalls.
The QCSE system can then leverage both the cavity-exciton detuning and the applied bias (which tunes $G$) to mitigate the effects of the inhomogeneity.

For expected inhomogeneities of $\sim 11$\,GHz, the density of states in the tails near the LP is sufficiently small such that it should be modeled by discrete states~\cite{hegarty1984localized,whittaker1998inhomogeneous,savona2007disorder}.
Thus, in a finite-size device with narrow $\rho(\omega)$, the density of states at the relevant frequencies is often effectively zero and the device behaves as if there were no disorder.
However, the density of states near any single branch becomes significantly larger than the continuum approximation.
Consequently, if an exciton branch lands near the LP, its effect is much more dramatic, causing large dispersive shifts, hybridization, and significant injected noise.
Thus, cavity protection in the discrete limit takes on the form of a suppressed probability for this specific failure mode, rather than a smooth diminishment of efficiency/noise metrics.
A full analysis of the discrete system is given in the Supplemental Material~\cite{supplement}.

\section{Discussion}
We have presented an approach for quantum transduction based on the QCSE of QW exciton polaritons.
The large enhancement of the EO nonlinearity afforded by the QCSE mechanism enables high efficiency ($\eta>0.6$) and low noise ($n_\mathrm{add} < 0.01$) transduction at low pump powers ($<10$\,\textmu W) and over wide bandwidths ($\sim 60$\,MHz).
We show that the device is robust against disorder provided the exciton density of states is small near the polariton which can be tuned directly via the applied bias.
The QCSE transducer deviates from standard EO transducers in both optical and microwave frequencies.
The optical frequency is pinned near 810\,nm ($\sim 370$\,THz) by the AlGaAs/GaAs QW resonance, while the resonance line width must increase to efficiently extract guided photons before the excitons can decay. 
Consequently, the microwave frequency must also increase to around $30$\,GHz to achieve sufficient sideband resolution, and can be adjusted based on experimentally achievable exciton line widths.
Such wavelengths are suitable for linking quantum processors in nearby refrigerators for scaling, and benefit from the highly mature AlGaAs/GaAs material platform.
Quantum wells based on AlGaInAs-on-InP~\cite{zah1994high} could be utilized at telecommunications wavelengths if long-distance connectivity is required.
Similarly, while the $30$-GHz operation point is not directly compatible with typical transmon qubits, high-efficiency Josephson parametric circuits~\cite{abdo2013full} and novel high-frequency qubits/readout~\cite{anferov2024superconducting,kurilovich2025readout} could alleviate this issue.
Altogether this positions the QCSE transducer as a promising approach for optically interconnecting remote quantum processors operating at microwave frequencies.

\section*{Acknowledgments} 
We thank Iacopo Carusotto and Andrei Faraon for fruitful discussions. 
This work was primarily funded by the U.S. Department of Energy, Office of Science, Office of Basic Energy Sciences, Materials Sciences and Engineering Division under Contract No. DE-AC02-05CH11231 within the Quantum Coherent Systems Program (KCAS26). 
The Princeton portion of this work is funded in part by the Gordon and Betty Moore Foundation EPiQS initiative, Grant GBMF14260.
N.S.Y.\ acknowledges support by an appointment to the IC Postdoctoral Research Fellowship Program at the University of California Berkeley administered by ORISE through an interagency agreement between the U.S.\ Dept.\ of Energy and Office of the Director of National Intelligence. 
M.K.\ acknowledges funding support from Novo Nordisk Foundation, Grant number NNF22SA0081175, NNF Quantum Computing Programme.
The authors acknowledge the use of AI (Anthropic \textit{Claude Opus}) for code generation, verifying analytical and numerical calculations, and literature search.

\bibliography{main}

\end{document}


\title{Supplemental Material for ``Microwave-optical quantum transduction with quantum-well exciton polaritons''}

\author{Nicholas S. Yama}
\affiliation{Department of Electrical Engineering and Computer Sciences, University of California, Berkeley, Berkeley, California 94720, USA}
\affiliation{Department of Physics, University of California, Berkeley, Berkeley, California 94720, USA}

\author{Mohammad Khalifa}
\affiliation{NNF Quantum Computing Programme, Niels Bohr Institute, University of Copenhagen, Blegdamsvej 17, 2100 Copenhagen, Denmark.}
\affiliation{Molecular Foundry, Lawrence Berkeley National Laboratory, Berkeley, California 94720, USA}

\author{Yuxuan Wei}
\affiliation{Department of Physics, University of California, Berkeley, Berkeley, California 94720, USA}
\affiliation{Materials Sciences Division, Lawrence Berkeley National Laboratory, Berkeley, California 94720, USA}

\author{Adbhut Gupta}
\affiliation{Department of Electrical and Computer Engineering, Princeton University, Princeton, New Jersey 08544, USA}

\author{Loren N. Pfeiffer}
\affiliation{Department of Electrical and Computer Engineering, Princeton University, Princeton, New Jersey 08544, USA}

\author{Feng Wang}
\affiliation{Department of Physics, University of California, Berkeley, Berkeley, California 94720, USA}
\affiliation{Materials Sciences Division, Lawrence Berkeley National Laboratory, Berkeley, California 94720, USA}
\affiliation{Kavli Energy NanoScience Institute, Berkeley, CA, USA}

\author{Alp Sipahigil}
\affiliation{Department of Electrical Engineering and Computer Sciences, University of California, Berkeley, Berkeley, California 94720, USA}
\affiliation{Department of Physics, University of California, Berkeley, Berkeley, California 94720, USA}
\affiliation{Materials Sciences Division, Lawrence Berkeley National Laboratory, Berkeley, California 94720, USA}

\maketitle            
\tableofcontents

\clearpage
\section{System Hamiltonian}

\subsection{Excitons}
The electronic states of the system are modeled by the Bloch wavefunctions $\psi_{\lambda\vb k}(\vb r) = \exp(i\vb k\cdot\vb r) u_{\lambda\vb k}(\vb r)$ where $\lambda$ is the band index, $\vb k$ is the wavevector, and $u_{\lambda\vb k}(\vb r)$ is the lattice-periodic Bloch wavefunction.
We will assume a two-band model consisting of a conduction and valence band $\lambda \in\{c,v\}$ with parabolic energy dispersion $E_{\lambda\vb k} = V_\lambda + \hbar^2k^2/2m_\lambda$ about the $\Gamma$ ($|\vb k| =0$) point.
We take the effective-mass approximation and assume $u_{\lambda\vb k}(\vb r) \approx u_{\lambda\vb 0}(\vb r)$ for all states of interest.
In a generic heterostructure (defined by spatially varying effective mass $m_\lambda(\vb r)$ and bandgap $V_\lambda(\vb r)$), the effective Hamiltonian is given in the position-space representation
\begin{equation}
    \hat H_\lambda = -\frac{\hbar^2}{2}\nabla\cdot\left(\frac{1}{m_\lambda(\vb r)} \nabla\right) + V_\lambda(\vb r) + V_\mathrm{ext}(\vb r),
\end{equation}
where $V_\mathrm{ext}(\vb r)$ describes an externally applied potential.

We assume that the structure is grown along the $z$ axis and is uniform in the transverse $(x,y)$ directions with large spatial extent and infinite-potential walls on its boundaries.
This allows the system to be factored into a transverse component $\phi_{\vb k}(x,y)$ and longitudinal component $\zeta_{\lambda n}(z)$ each satisfying
\begin{subequations}
    \begin{align}
        \left(\nabla_\perp^2 + k^2\right) \phi_{\vb k}(x,y) &= 0, \\
        \left[-\frac{\hbar^2}{2}\pdv{z}\left(\frac{1}{m_\lambda(z)} \pdv{z}\right) + V_\lambda(z) + V_\mathrm{ext}(z) \right]\zeta_{\lambda n}(z) &= E_{\lambda n}\zeta_{\lambda n}(z).
    \end{align}
\end{subequations}
respectively, where $\vb k$ is the in-plane wavevector, and $n$ indexes the subbands of $\lambda$.
Each of the wavefunction components are normalized $\iint\dd{x}\dd{y} \phi_{\vb k}^*\phi_{\vb k^\prime} = \delta_{\vb k\vb k^\prime}$ and $\int\dd{z} \zeta_{\lambda n}^* \zeta_{\lambda n^\prime} = \delta_{nn^\prime}$.
The final states $\ket{\lambda n \vb k}$ have wavefunctions $\braket{\vb r}{\lambda n\vb k} = \phi_{\vb k}(x,y)\zeta_{\lambda n}(z) u_{\lambda 0}(\vb r)$.
Orthogonality $\braket{\lambda n\vb k}{\lambda^\prime n^\prime\vb k^\prime}=\delta_{\lambda\lambda^\prime}\delta_{nn^\prime}\delta_{\vb k\vb k^\prime}$ is ensured by the Bloch function normalization over the unit-cell volume $\Omega$ as $\int_\Omega\dd[3]{r} u_{\lambda0}^*u_{\lambda^\prime 0} = \Omega\delta_{\lambda\lambda^\prime}$ which is factored out of the volume integrals under the slowly varying envelope approximation.

Excitons are a collective excitation of the many-body crystal states comprised of a conduction-band electron and a valence-band hole.
The electron and hole interact via the Coulomb interaction with total Hamiltonian in the $(r_e,r_h)$ position space
\begin{align}
    \hat H = {}
    &-\frac{\hbar^2}{2m_c^*}\nabla_{\parallel e}^2 - \frac{\hbar^2}{2}\pdv{z_e}\left(\frac{1}{m_c(z_e)}\pdv{z_e}\right) + V_c(z_e) + eE_\mathrm{ext}z_e \nonumber\\
    &-\frac{\hbar^2}{2m_v^*}\nabla_{\parallel h}^2 - \frac{\hbar^2}{2}\pdv{z_h}\left(\frac{1}{m_v(z_h)}\pdv{z_h}\right) + V_v(z_h) - eE_\mathrm{ext}z_h \\
    &-\frac{e^2}{4\pi\epsilon_0 \epsilon_r |\vb r_e - \vb r_h|}, \nonumber
\end{align}
where $E_\mathrm{ext}$ is the applied external field and the in-plane effective masses $m_\lambda^*$ are taken as constant across the structure.
In general, the Coulomb interaction couples the longitudinal and transverse motion such that the exciton has finite projection onto all subbands.
For the quantum-well (QW) system, we will assume that the longitudinal confinement is sufficiently narrow such that the system is quasi-two-dimensional and the eigenstates can be expanded
\begin{equation}
    \braket{\vb r_e \vb r_h}{\vb K \nu} = \Phi_{\vb K}(\vb R) \psi_\nu(\bm\rho) Z_\nu(z_e, z_h) u_{c0}(\vb r_e) u_{v0}^*(\vb r_h)
\end{equation}
where we have moved into the center-of-mass (CM) frame via
\begin{align*}
    \vb R = \frac{m_c^*\vb r_{\parallel e} + m_v^*\vb r_{\parallel h}}{M}, \qquad\qquad
    \bm \rho = \vb r_{\parallel e} - \vb r_{\parallel h}, \qquad\qquad
    M = m_c^* + m_v^*, \qquad\qquad
    \mu = m_c^* m_v^* / M.
\end{align*}
Here $\Phi_{\vb K}(\vb R)$ is the CM motion wavefunction with in-plane wavevector $\vb K$ (analogous to the in-plane wavevector $\vb k$ of the single-electron states), $\psi_\nu(\bm\rho)$ is the in-plane relative motion with quantum numbers $\nu$ and $Z_\nu(z_e,z_h)$ is the joint longitudinal wavefunction of the electron and hole for state $\nu$.
We second quantize the exciton states with operators $\hat e_{\vb K\nu}^\dagger$($\hat e_{\vb K\nu}$) which create/annihilate excitons.
More precisely, they raise/lower the many-body crystal system from the crystal ground state to the first excited-state manifold.
It can be shown that they are approximately bosonic provided the number density is small~\cite{haugkoch2004quantum}
\begin{equation}
    [\hat e_{\vb K \nu}, \hat e_{\vb K\nu}^\dagger] = 1 - \mathcal O(na_0^2)
\end{equation}
where $n$ is the number density per unit area and $a_0$ is a characteristic wavefunction scale (e.g.\ the 2-d Bohr radius).
We assume this limit and verify it later. 
The exciton Hamiltonian is thus
\begin{equation}
    \hat H_x = \sum_{\vb K,\nu} \hbar\omega_{\vb K\nu} \hat e_{\vb K\nu}^\dagger \hat e_{\vb K\nu},
\end{equation}
with energies $\hbar \omega_{\vb K\nu} = \hbar^2 K^2/2M + E_\nu$ and we assume $[\hat e_{\vb K \nu}, \hat e_{\vb K^\prime\nu^\prime}^\dagger] = \delta_{\vb K\vb K^\prime} \delta_{\nu\nu^\prime}$.

\subsection{Electromagnetic fields}

The optical modes $\bm\psi_m(\vb r)$ are the eigenfunctions of the electromagnetic wave equation with corresponding frequencies $\omega_m$.
We second quantize these modes as $\hat a_m$ with the total field operator
\begin{equation}
    \vu E_o(\vb r) = \sum_m E_{m,\mathrm{zpf}} \Big(\hat a_m \bm\psi_m(\vb r) + \hat a_m^\dagger \bm\psi_m^*(\vb r)\Big)
\end{equation}
where $E_{m,\mathrm{zpf}} = \sqrt{\hbar \omega_m/ 2\epsilon_0 n^2 V_m}$ is the electric field zero-point fluctuation (ZPF) amplitude and 
\begin{equation}
    n^2 V_m = \int\dd[3]{r} \bm\psi_m^*(\vb r) \cdot n^2(\vb r) \bm\psi_m(\vb r)
\end{equation}
is the mode volume for index distribution scaled by the index value $n^2$ at the mode maximum.
This implicitly assumes that the modes are unit-normalized $\bm\psi_m(\vb r)/\max|\bm\psi_m|$.
It is also necessary to define the vector potential operator
\begin{equation}
    \label{eq:si:optical-vector-potential}
    \vu A_o(\vb r) = \sum_m \frac{E_{m,\mathrm{zpf}}}{i\omega_m} \Big(\hat a_m \bm\psi_m(\vb r) - \hat a_m^\dagger \bm\psi_m^*(\vb r)\Big)
\end{equation}
which arises from the classical requirement that $\vb E = -\partial_t \vb A$.

The microwave fields are modeled as originating from a lumped-element $LC$ circuit.
Directly quantizing the circuit in terms of flux $\hat \Phi$ and charge $\hat Q$ enables determination of the voltage operator which we choose as
\begin{equation}
    \hat V = \frac{\hat Q}{C} = V_\mathrm{zpf} \Big(\hat b + \hat b^\dagger\Big),
\end{equation}
where $V_\mathrm{zpf} = \sqrt{\hbar \omega_\mu/2C}$ is the voltage ZPF amplitude for a mode with frequency $\omega_\mu = 1/\sqrt{LC}$.
In the lumped-element, quasi-static limit, the scalar potential $\hat \phi(\vb r) = \hat V \phi(\vb r)$ where $\phi(\vb r)$ solves the corresponding electrostatics problem with unit voltage difference between the signal and ground traces.
The microwave electric field operator follows directly as
\begin{equation}
    \vu E_\mu = -\nabla\hat\phi(\vb r) = E_{\mu,\mathrm{zpf}}\Big(\hat b + \hat b^\dagger\Big) \bm\psi_\mu(\vb r)
\end{equation}
where $E_{\mu,\mathrm{zpf}} = V_\mathrm{zpf} |\nabla\phi(\vb r_0)|$ is the ZPF amplitude and $\bm\psi_\mu(\vb r) = -\nabla\phi(\vb r) / |\nabla\phi(\vb r_0)|$ is the field profile normalized at a specific point $\vb r_0$.
This normalization is chosen since $\nabla\phi(\vb r)$ is generally divergent.

The total electric field operator is given directly by $\vu E = \vu E_o + \vu E_\mu$.
The Hamiltonian is
\begin{equation}
    \hat H_{f} = \hat H_o + \hat H_\mu = \sum_m \hbar\omega_m \hat a_m^\dagger \hat a_m + \hbar\omega_\mu \hat b^\dagger \hat b.
\end{equation}

\subsection{Exciton-optical interaction} \label{sec:si:hamiltonian:optical-interaction}
The exciton-optical interaction is derived from the minimal-coupling Hamiltonian $\hat H_{eo} = (e/m_0)\sum_j \vu A(\vu r_j) \cdot \vu p_j$ where $m_0$ is the free electron mass, $-e$ is the electron charge, and the sum is taken over all electrons in the system $j$.
This is a one-body interaction for the electrons and can be expanded in terms of second-quantized electron operators $\alpha_{\lambda n\vb k}$ as
\begin{equation}
    \hat H_{eo} = \frac{e}{m_0}\sum_{\lambda n\vb k, \lambda^\prime n^\prime \vb k^\prime} \matrixel{\lambda n \vb k}{\vu A(\vu r) \cdot \vu p}{\lambda^\prime n^\prime \vb k^\prime} \hat \alpha_{\lambda n \vb k}^\dagger \hat \alpha_{\lambda^\prime n^\prime \vb k^\prime}.
\end{equation}
In the low-density (linear) regime we can approximate the crystal Hilbert space to be spanned by the crystal ground state $\ket{0}$ (valence band filled with electrons) and the single-exciton (single excitation) states $\ket{\vb K\nu}$.
Projecting onto this subspace and defining $\hat e_{\vb K\nu} = \dyad{0}{\vb K\nu}$ we find
\begin{equation}
    \hat H_{eo} = \frac{e}{m_0} \sum_{\lambda n\vb k, \lambda^\prime n^\prime \vb k^\prime} \sum_{\vb K\nu}
    \matrixel{\lambda n \vb k}{\vu A(\vu r) \cdot \vu p}{\lambda^\prime n^\prime \vb k^\prime}
    \Big(
        \matrixel{\vb K\nu}{\hat \alpha_{\lambda n \vb k}^\dagger \hat \alpha_{\lambda^\prime n^\prime \vb k^\prime}}{0} \hat e_{\vb K\nu}^\dagger
        +
        \matrixel{0}{\hat \alpha_{\lambda n \vb k}^\dagger \hat \alpha_{\lambda^\prime n^\prime \vb k^\prime}}{\vb K\nu} \hat e_{\vb K\nu}
    \Big).
\end{equation}
The electron operators $\alpha_{\lambda n\vb k}^\dagger$($\alpha_{\lambda n\vb k}$) create/annihilate electrons and so only $\hat \alpha_{v n\vb k}\ket{0}$ is non-vanishing.
Furthermore, since they must bridge the crystal ground state $\ket{0}$ to an exciton state $\ket{\vb K\nu}$, this forces $\lambda\neq \lambda^\prime$.
Expanding the vector potential as in Eq.~\eqref{eq:si:optical-vector-potential} we can write the optical interaction as
\begin{equation}
    \label{eq:si:exciton-optical-hamiltonian}
    \hat H_{eo} = -\sum_m \sum_{\vb K\nu} \Big(\hbar g_{m,\vb K\nu} \hat a_m^\dagger \hat e_{\vb K\nu} + \mathrm{H.c.}\Big),
\end{equation}
where we have taken the RWA to eliminate the off-resonant terms and define the coupling rate
\begin{equation}
    \hbar g_{m,\vb K\nu} = \frac{eE_{m,\mathrm{zpf}}}{i\omega_m m_0} \sum_{n\vb k, n^\prime \vb k^\prime}
    \matrixel{v n \vb k}{\bm\psi_m^*(\vu r) \cdot \vu p}{c n^\prime\vb k^\prime} 
    \braket{c n^\prime \vb k^\prime, v n \vb k}{\vb K\nu}.
\end{equation}
Note that $\ket{c n^\prime \vb k^\prime, v n \vb k}$ is an electron-hole pair with wavefunction $\braket{\vb r_e \vb r_h}{c n^\prime \vb k^\prime, v n \vb k} = \braket{\vb r_e}{c n^\prime \vb k^\prime} (\braket{\vb r_h}{v n \vb k})^*$.
The hole part is conjugated because it is the time-reversed electron wavefunction.

The first factor is simplified in position space upon resolution of the identity $\hat I = \iint\dd[3]{r}\dd[3]{r^\prime} \dyad{\vb r\vb r^\prime}$ as
\begin{equation}
    \matrixel{v n \vb k}{\bm\psi_m^*(\vu r) \cdot \vu p}{c n^\prime\vb k^\prime} 
    =
    \int \dd[2]{r_\parallel} \phi_{\vb k}^*(\vb r_\parallel) \phi_{\vb k^\prime}(\vb r_\parallel) \int \dd{z} 
    \zeta_{vn}^*(z) \zeta_{cn^\prime}(z) 
    \Big(\bm\psi_m^*(\vb r_\parallel, z) \cdot \vb p_{vc} \Big),
\end{equation}
where
\begin{equation}
    \vb p_{vc} = \int_\Omega \frac{\dd[3]{r}}{\Omega} u_{v0}^*(\vb r) \matrixel{\vb r}{\vu p}{\vb r} u_{c0}(\vb r)
\end{equation}
is the inter-band momentum matrix element.
This term describes the geometric overlap of the optical field and the exciton wavefunction.
The dot product selects the polarization while the transverse integral over $\vb r_\parallel$ ensures that the in-plane momentum is conserved upon absorption.
Importantly, if the optical field varies slowly over the width of the QW, then the $z$ integral reduces to a simple integral overlap between the electron and hole wavefunctions.
The second factor is likewise
\begin{equation}
    \braket{c n^\prime \vb k^\prime, v n \vb k}{\vb K\nu} =
    \int \dd[2]{R}\int\dd[2]{\rho} \phi_{\vb k^\prime}^*(\vb r_{\parallel e}) \phi_{\vb k}(\vb r_{\parallel h}) \Phi_{\vb K}(\vb R) \psi_{\nu}(\bm\rho)\iint\dd{z_e}\dd{z_h} \zeta_{cn^\prime}^*(z_e) \zeta_{vn}(z_h) Z_\nu(z_e,z_h),
\end{equation}
and describes the projection of the exciton state onto the specific electron-hole pair.

Combining these factors we use the fact that the in-plane wavefunction $\phi_{\vb k}$ form a basis and thus $\sum_{\vb k} \phi_{\vb k}^*(\vb r) \phi_{\vb k} (\vb r^\prime) = \delta(\vb r - \vb r^\prime)$ and thus we have $\vb r_\parallel = \vb r_{\parallel e} = \vb r_{\parallel h}$ via the two summations over $\vb k$ and $\vb k^\prime$.
This eliminates two of the transverse integrals setting $\vb R = \vb r_\parallel$ and $\bm\rho = 0$ giving
\begin{equation}
    \label{eq:si:exciton-optical-rate}
    \hbar g_{m,\vb K\nu} = E_{m,\mathrm{zpf}}\vb d_{vc} \psi_\nu(\bm\rho=0) \cdot \sum_{n, n^\prime} Z_{\nu,nn^\prime}
    \int \dd[2]{r_\parallel}  \int \dd{z} 
    \zeta_{vn}^*(z) \zeta_{cn^\prime}(z) 
    \bm\psi_m^*(\vb r_\parallel, z)
    \Phi_{\vb K}(\vb r_\parallel)
\end{equation}
where $Z_{\nu,nn^\prime} =\iint\dd{z_e}\dd{z_h} \zeta_{cn^\prime}^*(z_e) \zeta_{vn}(z_h) Z_\nu(z_e,z_h)$ is the overlap of the exciton and electron-hole pair longitudinal wavefunctions.
And finally, we have defined the inter-band dipole matrix element $\vb d_{vc} = e\vb p_{vc}/i\omega_m m_0$ for a near-resonant optical field.
This yields a form similar to what one would expect in the dipole approximation but avoids the pitfalls of computing the inter-band position matrix element.
There are several points of interest.
First is that it scales with the relative electron-hole wavefunction when they coincide $\bm\rho = 0$.
For a $1s$-like 2-d exciton state this scales as $1/a_0$, where $a_0$ is the Bohr radius --- the QW is enhanced due to the $z$ confinement increasing the binding energy.
Similarly, the coupling rate depends on the electron-hole longitudinal wavefunction overlap (weighted by the optical mode profile).
Lastly, the in-plane overlap integral of the optical mode against the CM wavefunction $\Phi_{\vb K}$ implies conservation of the in-plane momentum.
For structures with sufficiently large transverse extent, the boundary conditions of the optical and exciton fields can become quite similar, allowing for each optical mode to selectively couple to a specific exciton mode.

\subsection{Exciton-microwave interaction} \label{sec:si:hamiltonian:microwave-interaction}
The microwave interaction is modeled by a dipole interaction~\cite{miller1984qcse,miller1985electric}
\begin{equation}
    \hat H_{e\mu} = -\sum_j e \hat E_{\mu,z}(\vu R_j, z_0) \big(\hat z_{h,j} - \hat z_{e,j}\big),
\end{equation}
where we sum over all excitons $j$.
We also assume that the microwave field varies sufficiently slowly compared to the electron-hole relative motion such that it is well approximated by a constant field evaluated at the CM position $\vb R_j$ in the center of the well $z=z_0$.
Note that the in-plane components do not contribute anything to lowest order since the exciton does not have an in-plane dipole moment.
In second quantization the operator can be expressed
\begin{equation}
    \hat H_{e\mu} = -\sum_{\vb K\nu, \vb K^\prime\nu^\prime} \hbar\xi_{\vb K\nu, \vb K^\prime\nu^\prime}
    \hat e_{\vb K\nu}^\dagger \hat e_{\vb K^\prime\nu^\prime}
    \Big(\hat b + \hat b^\dagger\Big),
\end{equation}
where the coupling rate is given by
\begin{equation}
    \hbar\xi_{\vb K\nu, \vb K^\prime\nu^\prime}
    =
    e E_{\mu,\mathrm{zpf}} \matrixel{\vb K\nu}{\bm\psi_{\mu,z}(\vu R,z_0)\big(\hat z_h - \hat z_e\big)}{\vb K^\prime\nu^\prime}.
\end{equation}
Expanding this matrix element in position space we obtain
\begin{align}
    \hbar\xi_{\vb K\nu, \vb K^\prime\nu^\prime}
    = {}
    & e E_{\mu,\mathrm{zpf}} \iint\dd{z_e}\dd{z_h} \big(z_h - z_e\big) Z_{\nu}^*(z_e,z_h) Z_{\nu^\prime}(z_e,z_h) 
    \int\dd[2]{\rho} \psi_{\nu}^*(\bm\rho) \psi_{\nu^\prime}(\bm\rho) \nonumber\\
    &\times \int\dd[2]{R} \Phi_{\vb K}^*(\vb R) \Phi_{\vb K^\prime}(\vb R) \psi_{\mu,z}(\vb R,z_0).
\end{align}
The first row integrals over $z$ and $\bm\rho$ describe the (inter-)exciton dipole moment.
The second row integral over the CM coordinate $\vb R$ describes the in-plane momentum conservation requirement.

\subsection{QCSE Hamiltonian}
The total quantum-confined Stark effect (QCSE) Hamiltonian is thus
\begin{align}
    \hat H = {}
        & \sum_m \hbar\omega_m \hat a_m^\dagger \hat a_m + \hbar\omega_\mu \hat b^\dagger \hat b & & \mathrm{(photons)} \nonumber\\
        & + \sum_{\vb K \nu} \hbar\omega_{\vb K\nu} \hat e_{\vb K\nu}^\dagger \hat e_{\vb K\nu} & & \mathrm{(excitons)} \nonumber\\
        & -\sum_m \sum_{\vb K\nu} \Big(\hbar g_{m,\vb K\nu} a_m^\dagger e_{\vb K\nu} + \mathrm{H.c.}\Big) & & \mathrm{(exciton{-}optical)} \nonumber\\
        & -\sum_{\vb K\nu, \vb K^\prime\nu^\prime} \hbar\xi_{\vb K\nu, \vb K^\prime\nu^\prime} \hat e_{\vb K\nu}^\dagger \hat e_{\vb K^\prime\nu^\prime} \Big(\hat b + \hat b^\dagger\Big). & & \mathrm{(exciton{-}microwave)}
\end{align}

\clearpage
\section{Semiclassical susceptibility}
Here we derive an effective semiclassical optical nonlinear susceptibility by treating the QW structure as an effective resonant optical medium which is used in the main text Fig.~1d,e.
The strategy is to evaluate the macroscopic optical dipole $\ev*{\vu d} = -\sum_j e\ev*{\vu r_j}$ in response to a classical applied field $\mathcal E(\omega)$ from which the polarization density is related $\mathcal P(\omega) = \ev*{\vu d}/V$.
The linear susceptibility is recovered by relating $\mathcal P(\omega) = \epsilon_0 \chi^{(1)}(\omega) \mathcal E(\omega)$.
The nonlinear susceptibility is obtained perturbatively by considering the modification of the linear response in the presence of a second low-frequency field.

Directly evaluating $\ev*{\vu d}$ via second quantization would effectively re-trace the steps used to derive the fully-quantum exciton-optical interaction Eq.~\eqref{eq:si:exciton-optical-hamiltonian}.
A simpler heuristic argument is to impose a semiclassical approximation to the interaction Hamiltonian by linearizing the field operators $\hat a_m \to \alpha_m e^{-i\omega t}$ and defining the classical applied field $\mathcal E_m(\vb r,t) = E_{m,\mathrm{zpf}} \psi_m(\vb r) \alpha_m e^{-i\omega t}$.
For further simplification we assume that the optical field is uniform (so only $\vb K=0$ is relevant) and that the exciton state of interest is limited to a single ground state $\nu = 0$ whose $z$ motion is separable into the ground state $Z_{\nu}(z_e,z_h) = \zeta_{c1}(z_e) \zeta_{v1}^*(z_h)$.
Then, in the dipole approximation, we define the semiclassical interaction Hamiltonian for the single exciton state as
\begin{equation}
    \hat H_{eo} = -\Big(\mathcal E^*(t) d_{vc,\mathrm{eff}} \hat e + \mathcal E(t) d_{cv,\mathrm{eff}} \hat e^\dagger\Big),
\end{equation}
and dipole moment
\begin{equation}
    \label{eq:si:semiclassical-dipole-moment}
    \ev*{\vu d} = \Big( d_{vc,\mathrm{eff}} \ev{\hat e} + d_{cv,\mathrm{eff}} \ev*{\hat e^\dagger}\Big).
\end{equation}
The effective dipole moment $d_{vc,\mathrm{eff}}$ is simplified from Eq.~\eqref{eq:si:exciton-optical-rate} as
\begin{equation}
    d_{vc,\mathrm{eff}} = d_{vc} L \psi_\nu(\bm\rho=0) \int\dd{z} \zeta^*_{v1}(z) \zeta_{c1}(z).
\end{equation}
The factor of $L$ (the transverse extent) originates from the in-plane integral over $\vb r_\parallel$ in Eq.~\eqref{eq:si:exciton-optical-rate}: the integral itself yields a factor $L^2$ but the in-plane exciton wavefunction carries a normalization $1/\sqrt{L^2}$ --- the optical mode is unit normalized and does not have any $L$ dependence.
Note that $d_{vc} = ep_{vc}/im_0\omega$ is the inter-band dipole and we are neglecting the polarization dependence.

The semiclassical Hamiltonian for this exciton state is
\begin{equation}
    \hat H = \hbar \Big(\omega_0 - \frac{d_\mu \mathcal E_\mu(t)}{\hbar}\Big)\hat e^\dagger \hat e
    -
    \Big(\mathcal E^*(t) d_{vc,\mathrm{eff}} \hat e + \mathcal E(t) d_{cv,\mathrm{eff}} \hat e^\dagger\Big)
\end{equation}
where $d_\mu$ and $\mathcal E_\mu(t)$ are the microwave dipole and field respectively.
We can work in a quasi-static approximation and treat the Stark shift perturbatively $\hat e = \hat e^{(0)} + \epsilon \hat e^{(1)}$.
The Heisenberg equations of motion follow as
\begin{subequations}
    \begin{align}
        \dv{\hat e^{(0)}}{t} + i\Big(\omega_0 - i\frac{\gamma}{2}\Big)\hat e^{(0)} &= \frac{i}{\hbar} \mathcal E(t) d_{cv,\mathrm{eff}}, \\
        \dv{\hat e^{(1)}}{t} + i\Big(\omega_0 - i\frac{\gamma}{2}\Big)\hat e^{(1)} &= \frac{i}{\hbar} \mathcal E_\mu(t) d_{\mu} \hat e^{(0)}(t),
    \end{align}
\end{subequations}
where we introduced the phenomenological decay rate $\gamma$.
Analogous equations exist for $\hat e^\dagger$, driven by the negative-frequency part of the real field.
Evaluating the zeroth-order equation via Fourier transform
\begin{equation}
    \hat e^{(0)}(\omega) = \frac{d_{cv,\mathrm{eff}} \mathcal E(\omega)}{\hbar(\omega_0 - i\gamma/2 - \omega)}
\end{equation}
which can be inserted into Eq.~\eqref{eq:si:semiclassical-dipole-moment} to obtain the linear susceptibility
\begin{equation}
    \chi^{(1)}(\omega) = \frac{|\vb d_{cv}|^2|\psi_\nu(\bm\rho=0)|^2}{ \epsilon_0 \hbar L_z }
    \abs{\int\dd{z} \zeta^*_{v1}(z) \zeta_{c1}(z)}^2 
    \bigg(\frac{1}{\omega_0 - i\gamma/2 - \omega} + \frac{1}{\omega_0 + i\gamma/2 + \omega}\bigg)
\end{equation}
where we normalized the polarization to the volume $V = L^2 L_z$ with confinement length $L_z$ in $z$.
For an exactly 2-d exciton in a $1s$-like state $|\psi_{1s}(\bm\rho=0)|^2 = 8/\pi a_0^2$ where $a_0$ is the 3-d Bohr radius.
This yields the standard form consistent with Ref.~\cite{haugkoch2004quantum}.

To handle the first-order equation we assume the applied microwave field has slow frequency $\mathcal E_\mu(t) = \mathcal E_\mu(\Omega) \exp(-i\Omega t)$ and that the zeroth-order field is driven optically at a fixed frequency $\omega$, $\hat e^{(0)}(t) = \hat e^{(0)}(\omega)\exp(-i\omega t)$.
Then solving for the steady-state solution (via Fourier transform)
\begin{equation}
    \hat e^{(1)}(\omega,\Omega) =  \frac{d_{cv,\mathrm{eff}} d_\mu \mathcal E(\omega) \mathcal E_\mu(\Omega)}{\hbar^2(\omega_0 - i\gamma/2 - \omega)(\omega_0 - i\gamma/2 - \omega - \Omega)}.
\end{equation}
Plugging in the total exciton operator $\hat e = \hat e^{(0)} + \epsilon \hat e^{(1)}$ into Eq.~\eqref{eq:si:semiclassical-dipole-moment} and comparing to the nonlinear polarization $\mathcal P(\omega+\Omega) = 2\epsilon_0\chi^{(2)}(\omega,\Omega)\mathcal E(\omega)\mathcal E_\mu(\Omega)$ we find
\begin{align}
    \chi^{(2)}(\omega,\Omega) = \frac{d_\mu |\vb d_{cv}|^2 |\psi_\nu(\bm\rho=0)|^2}{2 \epsilon_0 \hbar^2 L_z }
    \abs{\int\dd{z} \zeta^*_{v1}(z) \zeta_{c1}(z)}^2 
    \bigg(&\frac{1}{(\omega_0 - i\gamma/2 - \omega)(\omega_0 - i\gamma/2 - \omega - \Omega)} \nonumber\\
    &+ \frac{1}{(\omega_0 + i\gamma/2 + \omega)(\omega_0 + i\gamma/2 + \omega + \Omega)}\bigg)
\end{align}
Note that this is essentially equivalent to writing $\chi^{(2)} = (1/2)\dv*{\chi^{(1)}}{\mathcal E_\mu}$ in the limit $\Omega\to0$.

\clearpage
\section{Transduction efficiency and added noise}
Here we provide an in-depth analysis of the proposed QCSE transduction model.
We will utilize the standard input-output theory via the Heisenberg-Langevin (HL) equations~\cite{gardiner1985input}
\begin{subequations}
    \begin{align}
        \dv{\hat X}{t} &= \frac{i}{\hbar} [\hat H, \hat X]
        - \sum_j\bigg[\Big(\sqrt{\Gamma_j}\hat s_{j+}^\dagger + \frac{\Gamma_j}{2} \hat L_j^\dagger\Big) [\hat L_j, \hat X]
        -
        [\hat L_j^\dagger, \hat X]
        \Big(\sqrt{\Gamma_j} \hat s_{j+} + \frac{\Gamma_j}{2} \hat L_j\Big) \bigg], \\
        \hat s_{j-} &= \hat s_{j+} + \sqrt{\Gamma_j}\hat L_j,
    \end{align}
\end{subequations}
where $\hat X$ is a generic system operator and $j$ indexes the reservoirs with associated jump operator $\hat L_j$ and rate $\Gamma_j$. 
The operators $\hat s_{j\pm}$ are the input/output operators for the given bath.

We consider the simplest form of the system consisting of a single exciton mode, single optical mode, and single microwave mode with Hamiltonian
\begin{align}
    \hat H = \underbrace{\hbar\omega_a \hat a^\dagger \hat a 
    + \hbar\omega_b \hat b^\dagger \hat b}_{\mathrm{fields}}
    + \underbrace{\hbar\omega_e \hat e^\dagger_{} \hat e}_{\mathrm{exciton}}
    - \underbrace{\hbar \Big(g \hat a^\dagger \hat e + g^* \hat a \hat e^\dagger\Big)}_{\mathrm{exciton{-}optical}}
    -\underbrace{\hbar 
        \xi\hat e^\dagger \hat e
    \Big(\hat b + \hat b^\dagger\Big)}_{\mathrm{exciton{-}microwave}}.
\end{align}
The following reservoirs with corresponding rates and jump operators are listed:
\begin{itemize}
    \item Optical cavity loss: $\kappa$, $\hat a$,
    \item Microwave cavity loss: $\Gamma$, $\hat b$,
    \item Exciton loss: $\gamma$, $\hat e$.
\end{itemize}
In all cases the total rate can be split into extrinsic and intrinsic parts (e.g.\ $\kappa = \kappa_\mathrm{ext} + \kappa_\mathrm{int}$) with the extrinsic part describing the coupling loss that can be collected.

\subsection{Scattering matrix}
This section describes the analysis of the intra-exciton (sideband) scheme of a single exciton mode which gives the quoted analytical results in the main text.
This is analogous to the typical sideband-resolved regime of cavity optomechanics.
Dropping the subscripts we have HL equations
\begin{subequations}
    \label{eq:si:efficiency:heisenberg-langevin}
    \begin{align}
        \dv{\hat a}{t} &= -\Big(i\omega_a + \frac{\kappa}{2}\Big)\hat a + ig \hat e - \sqrt{\kappa_\mathrm{ext}}\hat a_\mathrm{in,ext} - \sqrt{\kappa_\mathrm{int}}\hat a_\mathrm{in,int}, \\
        \dv{\hat b}{t} &= -\Big(i\omega_b + \frac{\Gamma}{2}\Big)\hat b + i \xi \hat e^\dagger \hat e - \sqrt{\Gamma_\mathrm{ext}}\hat b_\mathrm{in,ext} - \sqrt{\Gamma_\mathrm{int}}\hat b_\mathrm{in,int}, \\
        \dv{\hat e}{t} &= -\Big(i\omega_e + \frac{\gamma}{2}\Big)\hat e + ig^* \hat a + i\xi \hat e\Big(\hat b + \hat b^\dagger\Big) - \sqrt{\gamma}\hat e_\mathrm{in}, 
    \end{align}
    for the co-rotating terms.
    We also have
    \begin{align}
        \dv{\hat a^\dagger}{t} &= -\Big(-i\omega_a + \frac{\kappa}{2}\Big)\hat a^\dagger - ig^* \hat e^\dagger - \sqrt{\kappa_\mathrm{ext}}\hat a_\mathrm{in,ext}^\dagger - \sqrt{\kappa_\mathrm{int}}\hat a_\mathrm{in,int}^\dagger, \\
        \dv{\hat b^\dagger}{t} &= -\Big(-i\omega_b + \frac{\Gamma}{2}\Big)\hat b^\dagger - i \xi \hat e^\dagger \hat e - \sqrt{\Gamma_\mathrm{ext}}\hat b_\mathrm{in,ext}^\dagger - \sqrt{\Gamma_\mathrm{int}}\hat b_\mathrm{in,int}^\dagger, \\
        \dv{\hat e^\dagger}{t} &= -\Big(-i\omega_e + \frac{\gamma}{2}\Big)\hat e^\dagger - ig \hat a^\dagger - i\xi \hat e^\dagger\Big(\hat b + \hat b^\dagger\Big) - \sqrt{\gamma}\hat e_\mathrm{in}^\dagger, 
    \end{align}
\end{subequations}
for the counter-rotating terms, which are coupled via the nonlinear interaction.

To proceed we linearize the system in the frame rotating at the pump frequency, expanding $\hat a = (\bar a + \delta \hat a)\exp(-i\omega_\mathrm{pump} t)$ and $\hat e = (\bar e + \delta \hat e)\exp(-i\omega_\mathrm{pump} t)$ where $\bar a$, $\bar e$ are classical coherent field amplitudes and $\delta \hat a$, $\delta\hat e$ describe the fluctuations.
Likewise the optical drive is separated $\hat a_\mathrm{in,ext} = (\bar a_\mathrm{in,ext} + \delta\hat a_\mathrm{in,ext})\exp(-i\omega_\mathrm{pump} t)$, and all other inputs have zero mean.
The microwave field is not rotated and only contains the quantum field without a classical pump, i.e.\ $\bar b = 0$.
Collecting terms at each order in the fluctuations gives the coherent amplitude evolution as
\begin{subequations}
    \label{eq:si:model:linearized-order-0}
    \begin{align}
        \dv{\bar a}{t} &= -\Big(i(\omega_a - \omega_\mathrm{pump}) + \frac{\kappa}{2}\Big)\bar a + ig \bar e - \sqrt{\kappa_\mathrm{ext}}\bar a_\mathrm{in,ext}, \\
        \dv{\bar e}{t} &= -\Big(i(\omega_e - \omega_\mathrm{pump}) + \frac{\gamma}{2}\Big)\bar e + ig^* \bar a, 
    \end{align}
\end{subequations}
For a constant classical drive $\bar a_\mathrm{in,ext}$, the steady state $\dv*{\bar a}{t} = \dv*{\bar e}{t} = 0$ gives
\begin{subequations}
    \begin{align}
        \bar a &= \frac{i\sqrt{\kappa_\mathrm{ext}}\,\big(\omega_e - \omega_\mathrm{pump} - i\gamma/2\big)}{\big(\omega_a - \omega_\mathrm{pump} - i\kappa/2\big)\big(\omega_e - \omega_\mathrm{pump} - i\gamma/2\big) - |g|^2} \bar a_\mathrm{in,ext}, \\
        \bar e &= \frac{ig^*\sqrt{\kappa_\mathrm{ext}}}{\big(\omega_a - \omega_\mathrm{pump} - i\kappa/2\big)\big(\omega_e - \omega_\mathrm{pump} - i\gamma/2\big) - |g|^2} \bar a_\mathrm{in,ext}.
    \end{align}
\end{subequations}
The conjugate amplitudes are simply $\bar a^*$ and $\bar e^*$.

The first-order equations follow
\begin{subequations}
    \label{eq:si:model:linearized-order-1}
    \begin{align}
        \dv{\delta\hat a}{t} &= -\Big(i\Delta_a + \frac{\kappa}{2}\Big)\delta\hat a + ig \delta\hat e - \sqrt{\kappa_\mathrm{ext}}\delta\hat a_\mathrm{in,ext} - \sqrt{\kappa_\mathrm{int}}\delta\hat a_\mathrm{in,int}, \\
        \dv{\delta\hat b}{t} &= -\Big(i\omega_b + \frac{\Gamma}{2}\Big)\delta\hat b + i \xi \Big( \bar e^* \delta \hat e + \delta \hat e^\dagger \bar e \Big) - \sqrt{\Gamma_\mathrm{ext}}\delta\hat b_\mathrm{in,ext} - \sqrt{\Gamma_\mathrm{int}} \delta \hat b_\mathrm{in,int}, \\
        \dv{\delta \hat e}{t} &= -\Big(i\Delta_e + \frac{\gamma}{2}\Big)\delta \hat e + ig^* \delta \hat a + i\xi \bar e\Big(\delta\hat b + \delta\hat b^\dagger\Big) - \sqrt{\gamma}\delta \hat e_\mathrm{in}, 
    \end{align}
\end{subequations}
where $\Delta_a = \omega_a - \omega_\mathrm{pump}$, $\Delta_e = \omega_e - \omega_\mathrm{pump}$ are the optical and exciton detunings from the pump field.
The counter-rotating equations can similarly be formed as before.
These equations can be written compactly as a linear system
\begin{subequations}
    \label{eq:si:model:scattering-problem}
    \begin{align}
        \dv{\vb x}{t} &= -H \vb x - K \vb s_\mathrm{in}, \\
        \vb s_\mathrm{out} &= \vb s_\mathrm{in} + K^\dagger \vb x,
    \end{align}
\end{subequations}
where
\begin{align*}
    \vb x = 
    \begin{bmatrix}
        \delta\hat e(t) \\ \delta\hat a(t) \\ \delta\hat b(t) \\
        \delta\hat e^\dagger(t) \\ \delta\hat a^\dagger(t) \\ \delta\hat b^\dagger(t) 
    \end{bmatrix},
    \qquad\qquad
    \vb s_\mathrm{in} = 
    \begin{bmatrix}
        \delta\hat e_\mathrm{in}(t) \\ \delta\hat a_\mathrm{in,int}(t) \\ \delta\hat a_\mathrm{in,ext}(t) \\ \delta\hat b_\mathrm{in,int}(t) \\ \delta\hat b_\mathrm{in,ext}(t) \\
        \delta\hat e_\mathrm{in}^\dagger(t) \\ \delta\hat a_\mathrm{in,int}^\dagger(t) \\ \delta\hat a_\mathrm{in,ext}^\dagger(t) \\ \delta\hat b_\mathrm{in,int}^\dagger(t) \\ \delta\hat b_\mathrm{in,ext}^\dagger(t)
    \end{bmatrix},
\end{align*}
with $\vb s_\mathrm{out}$ defined likewise from the output fields, and
\begin{equation*}
    H = 
    \begin{bmatrix}
        i\Delta_e + \gamma/2 & -ig^* & -i\xi\bar e & 0 & 0 & -i\xi\bar e \\
        -ig & i\Delta_a + \kappa/2 & 0 & 0 & 0 & 0 \\
        -i\xi\bar e^* & 0 & i\omega_b + \Gamma/2 & -i\xi\bar e & 0 & 0 \\
        0 & 0 & i\xi\bar e^* & -i\Delta_e + \gamma/2 & ig & i\xi\bar e^* \\
        0 & 0 & 0& ig^* & -i\Delta_a + \kappa/2 & 0 \\
        i\xi\bar e^* & 0 & 0 & i\xi\bar e & 0 & -i\omega_b + \Gamma/2 \\
    \end{bmatrix},
\end{equation*}
\begin{equation*}
    K = 
    \begin{bmatrix}
        \sqrt{\gamma} & 0 & 0 & 0 & 0 & 0 & 0 & 0 & 0 & 0 \\
        0 & \sqrt{\kappa_\mathrm{int}} & \sqrt{\kappa_\mathrm{ext}} & 0 & 0 & 0 & 0 & 0 & 0 & 0 \\
        0 & 0 & 0 & \sqrt{\Gamma_\mathrm{int}} & \sqrt{\Gamma_\mathrm{ext}} & 0 & 0 & 0 & 0 & 0 \\
        0 & 0 & 0 & 0 & 0 & \sqrt{\gamma} & 0 & 0 & 0 & 0 \\
        0 & 0 & 0 & 0 & 0 & 0 &\sqrt{\kappa_\mathrm{int}} & \sqrt{\kappa_\mathrm{ext}} & 0 & 0 \\
        0 & 0 & 0 & 0 & 0 & 0 & 0 & 0 & \sqrt{\Gamma_\mathrm{int}} & \sqrt{\Gamma_\mathrm{ext}} \\
    \end{bmatrix}.
\end{equation*}
This is solved by Fourier transform $\hat x(\omega) = \int\dd{t}\exp(i\omega t)\hat x(t)$ (with inverse $\hat x(t) = (1/2\pi)\int\dd{\omega}\exp(-i\omega t)\hat x(\omega)$) from which we obtain $\vb s_\mathrm{out}(\omega) = S(\omega) \vb s_\mathrm{in}(\omega)$ where
\begin{equation}
    S(\omega) = 1 + K^\dagger\big( i\omega - H \big)^{-1} K,
\end{equation}
is the scattering matrix whose elements relate the complex scattering amplitudes mapping fluctuations between incoming/outgoing waves in different reservoirs.
An important notational point is that the Fourier transforms are taken on the conjugated operators such that the frequency domain expressions are written in terms of $\hat x^\dagger(\omega) \equiv \mathcal F[\hat x^\dagger(t)]$.
This is explicitly \textit{not} the conjugate of the transform $\mathcal F[\hat x(t)]^\dagger = \big(\hat x(\omega)\big)^\dagger$ which is related as $\hat x^\dagger(\omega) = \big(\hat x(-\omega)\big)^\dagger$.

\vspace{1em}\noindent\textbf{Transduction efficiency:} The total external microwave-to-optical transduction efficiency $\eta_{ab}(\omega)$ corresponds directly to the $S$-matrix element $S_{ab}$ mapping the extrinsic microwave input $\delta\hat b_\mathrm{in,ext}$ to the extrinsic optical output $\delta\hat a_\mathrm{out,ext}$,
\begin{equation}
    \eta_{ab}(\omega) \equiv |S_{ab}(\omega)|^2
\end{equation}
In the absence of strong dissipative coupling the system is typically reciprocal such that the optical-to-microwave transduction efficiency is identical, i.e.\ $|S_{ab}(\omega)|^2 = |S_{ba}(\omega)|^2$.
In such case a single transduction efficiency $\eta(\omega) = \eta_{ab}(\omega) = \eta_{ba}(\omega)$ can be defined.
A relevant case where the reciprocity can be broken is discussed in Sec.~\ref{sec:si:disorder:transduction}.

\vspace{1em}\noindent\textbf{Added noise:} More generally, the output field of a given port is a sum over all the input port fluctuations, for example
\begin{equation}
    \delta\hat a_\mathrm{out,ext}(\omega)
    =
    \sum_j S_{a,j}(\omega) \delta\hat s_{\mathrm{in},j}(\omega),
\end{equation}
where $j$ runs over all entries of $\vb s_\mathrm{in}$, including the conjugate inputs.
The inputs obey the commutator $\big[\delta\hat s_{\mathrm{in},j}(t), \delta\hat s_{\mathrm{in},j}^\dagger(t^\prime)\big] = \pm\delta(t-t^\prime)$, with the sign negative for the conjugate entries, and in the standard Markovian approximation
\begin{equation}
    \big\langle\delta\hat s_{\mathrm{in},j}^\dagger(t)\delta\hat s_{\mathrm{in},k}(t^\prime)\big\rangle = N_j\delta_{jk}\delta(t-t^\prime),
\end{equation}
where $N_j = n_j$ for a field entry and $N_j = n_j + 1$ for a conjugate entry, which is anti-normally ordered so the commutator adds one quantum, and $n_j$ is the thermal occupation of the corresponding reservoir.
The expected output flux is thus
\begin{equation}
    \big\langle\delta\hat a_\mathrm{out,ext}^\dagger\delta\hat a_\mathrm{out,ext}(\omega)\big\rangle
    =
    \sum_j \big|S_{a,j}(\omega)\big|^2 N_j.
\end{equation}
This can be re-expressed in terms of $\big\langle\delta\hat a_\mathrm{out,ext}^\dagger\delta\hat a_\mathrm{out,ext}(\omega)\big\rangle = \eta(n_\mathrm{signal} + n_\mathrm{add})$ where $n_\mathrm{signal}$ is the thermal occupation of the desired input mode to be transduced and $n_\mathrm{add}$ is added noise.
The added noise is thus
\begin{equation}
    \label{eq:si:model:added-noise-general}
    n_\mathrm{add} = \frac{1}{\eta} \sum_{j\neq\mathrm{signal}} \big|S_{a,j}(\omega)\big|^2 N_j.
\end{equation}
Recall that for thermal occupation at temperature $T$ the modes follow Bose-Einstein statistics $n_j = [\exp(\hbar\omega_j/k_BT) - 1]^{-1}$, evaluated at the lab-frame frequency $\omega_j$ of that reservoir ($\omega$ for the microwave reservoirs and $\omega_\mathrm{pump} + \omega$ for the optical and exciton reservoirs, for which $n_j \approx 0$).
This effectively performs a weighted sum along the $\delta\hat a_\mathrm{out,ext}$ row of the $S$-matrix, skipping the signal column.

\subsection{Strong-coupling limit}
Alternatively one can consider the case of strong coupling in which the optical coupling $|g| > \kappa,\gamma$.
In this limit the system is conveniently re-expressed in the polariton basis via the transformation
\begin{align*}
    \hat a_+ &= C^* \hat e + X \hat a, 
        & \hat e &= C \hat a_+ + X \hat a_-,\\
    \hat a_- &= X^* \hat e - C \hat a, 
        & \hat a &= X^* \hat a_+ - C^* \hat a_-,
\end{align*}
where $\hat a_+$ and $\hat a_-$ are the upper (UP) and lower (LP) polaritons and the coefficients $X$ and $C$ are the Hopfield coefficients~\cite{hopfield1958theory,haugkoch2004quantum}
\begin{align*}
    C &= \frac{2g^*}{\sqrt{(\Delta - \Delta_p)^2 + 4|g|^2}},
    & 
    X &= \frac{\Delta - \Delta_p}{\sqrt{(\Delta - \Delta_p)^2 + 4|g|^2}},
\end{align*}
with exciton-optical detuning $\Delta = \omega_e - \omega_a$ and polariton splitting $\Delta_p = \sqrt{\Delta^2 + 4|g|^2}$.
The polariton frequencies are $\omega_\pm = (\omega_e + \omega_a \pm \Delta_p)/2$.
The Hamiltonian is then
\begin{align}
    \hat H & =  \hbar\omega_+ \hat a_+^\dagger \hat a_+
    + \hbar\omega_- \hat a_-^\dagger \hat a_-
    + \hbar \omega_b \hat b^\dagger\hat b \nonumber\\
    & \quad- \hbar \Big(
        g_\mathrm{eo}^+ \hat a_+^\dagger \hat a_+ 
        + g_\mathrm{eo}^{+-} \hat a_+^\dagger \hat a_- 
        + g_\mathrm{eo}^{+-*} \hat a_-^\dagger \hat a_+ 
        + g_\mathrm{eo}^- \hat a_-^\dagger \hat a_-\Big)
    \Big(\hat b + \hat b^\dagger \Big).
\end{align}
The H-L equations are
\begin{subequations}
    \begin{align}
        \dv{\hat a_+}{t} &= - \Big(i\omega_+ + \frac{\kappa_+}{2}\Big) \hat a_+
            + i\Big(g_\mathrm{eo}^+ \hat a_+ + g_\mathrm{eo}^{+-} \hat a_-\Big)\Big(\hat b + \hat b^\dagger\Big)
            + g_\mathrm{bath} \hat a_-
            - \sqrt{\kappa_{+,\mathrm{ext}}} \hat a_\mathrm{in,ext} 
            - \sqrt{\kappa_{+,\mathrm{int}}} \hat a_\mathrm{in,int} 
            - \sqrt{\gamma_+} \hat e_\mathrm{in}, \\
        \dv{\hat a_-}{t} &= - \Big(i\omega_- + \frac{\kappa_-}{2}\Big) \hat a_-
            + i\Big(g_\mathrm{eo}^{+-*} \hat a_+ + g_\mathrm{eo}^- \hat a_-\Big)\Big(\hat b + \hat b^\dagger\Big)
            + g_\mathrm{bath}^* \hat a_+
            + \sqrt{\kappa_{-,\mathrm{ext}}} \hat a_\mathrm{in,ext}
            + \sqrt{\kappa_{-,\mathrm{int}}} \hat a_\mathrm{in,int}
            - \sqrt{\gamma_-} \hat e_\mathrm{in}, \\
        \dv{\hat b}{t} &= - \Big(i\omega_b + \frac{\Gamma}{2}\Big) \hat b
            + i\Big(
                g_\mathrm{eo}^+ \hat a_+^\dagger \hat a_+ 
                + g_\mathrm{eo}^{+-} \hat a_+^\dagger \hat a_- 
                + g_\mathrm{eo}^{+-*} \hat a_-^\dagger \hat a_+ 
                + g_\mathrm{eo}^- \hat a_-^\dagger \hat a_-\Big)
            - \sqrt{\Gamma_\mathrm{ext}} \hat b_\mathrm{in,ext}
            - \sqrt{\Gamma_\mathrm{int}} \hat b_\mathrm{in,int}
    \end{align}
\end{subequations}
where
\begin{align*}
    g_\mathrm{eo}^+ &= \xi |C|^2 ,
        & g_\mathrm{eo}^- &= \xi |X|^2,
        & g_\mathrm{eo}^{+-} &= \xi C^*X,
\end{align*}
are the coupling rates for intra-polariton ($g_\mathrm{eo}^\pm$) and inter-polariton ($g_\mathrm{eo}^{+-}$) modulation and
\begin{align*}
    \kappa_+ &= |X|^2\kappa + |C|^2\gamma,
        & \kappa_- &= |C|^2\kappa + |X|^2\gamma, \\
    \sqrt{\kappa_{+,j}} &= X\sqrt{\kappa_j},
        & \sqrt{\kappa_{-,j}} &= C\sqrt{\kappa_j}, \\
    \sqrt{\gamma_+} &= C^*\sqrt{\gamma},
        & \sqrt{\gamma_-} &= X^*\sqrt{\gamma}, 
\end{align*}
are the loss rates.
Note that the input-output relations are also modified
\begin{subequations}
    \begin{align}
        \hat a_{\mathrm{out},j} &= \hat a_{\mathrm{in},j} + \sqrt{\kappa_{+,j}}^* \hat a_+ - \sqrt{\kappa_{-,j}}^* \hat a_-, \\
        \hat e_\mathrm{out} &= \hat e_\mathrm{in} + \sqrt{\gamma_+}^* \hat a_+ + \sqrt{\gamma_-}^* \hat a_-, \\
        \hat b_{\mathrm{out},j} &= \hat b_{\mathrm{in},j} + \sqrt{\Gamma_j} \hat b.
    \end{align}
\end{subequations}
A notable additional term is the bath-mediated coupling with rate $g_\mathrm{bath} = C^*X(\kappa-\gamma)/2$ that arises from the coupling of energy between the orthogonal modes via the shared bath.
This term is particularly important in the weak-coupling regime when the polariton splitting $\Delta_p \sim 2|g|$ is on the order of the line widths as it ensures unitarity of the scattering process.
On the other hand, it is generally weak ($|g_\mathrm{bath}| \leq |\kappa - \gamma|/4$ since $|CX| \leq 1/2$) and may thus be neglected in the strong coupling regime where $|g| > \kappa, \gamma$.

We now make several simplifying approximations.
First, under the assumption $|g|\gg\kappa,\gamma,\omega_b$, the polariton splitting $\Delta_p \geq 2|g| \gg \omega_b$ and so can take the rotating-wave approximation (RWA) to eliminate the cross coupling terms $\propto g_\mathrm{eo}^{+-}$.
Likewise, the bath coupling terms $\propto g_\mathrm{bath} \ll \Delta_p$ can also be dropped under the RWA for the same reason.
Finally, we also assume that our optical inputs occur at frequencies close to or below the LP such that population of the UP is negligible.
Under these assumptions we eliminate the UP from the dynamics and drop the excess sub/superscripts to get the effective Hamiltonian
\begin{equation}
    \hat H = \hbar\omega_- \hat a_-^\dagger \hat a_-
    + \hbar\omega_b \hat b^\dagger \hat b
    - \hbar g_\mathrm{eo} \hat a_-^\dagger \hat a_- \Big(\hat b + \hat b^\dagger \Big),
\end{equation}
where $g_\mathrm{eo}=g_\mathrm{eo}^{-} = \xi|X|^2$. 
The corresponding system of equations is
\begin{subequations}
    \begin{align}
        \dv{\hat a_-}{t} &= -\Big(i\omega_- + \frac{\kappa_-}{2}\Big)\hat a_-
            + i g_\mathrm{eo} \hat a_- \Big(\hat b + \hat b^\dagger \Big)
            + \sqrt{\kappa_{-,\mathrm{ext}}} \hat a_\mathrm{in,ext}
            + \sqrt{\kappa_{-,\mathrm{int}}} \hat a_\mathrm{in,int}
            - \sqrt{\gamma_{-}} \hat e_\mathrm{in}, \\
        \dv{\hat b}{t} &= -\Big(i\omega_b + \frac{\Gamma}{2}\Big)\hat b
            + i g_\mathrm{eo} \hat a_-^\dagger \hat a_-
            - \sqrt{\Gamma_{\mathrm{ext}}} \hat b_\mathrm{in,ext}
            - \sqrt{\Gamma_{\mathrm{int}}} \hat b_\mathrm{in,int},
    \end{align}
    with input-output relations
    \begin{align}
        \hat a_{\mathrm{out},j} &= \hat a_{\mathrm{in},j} - \sqrt{\kappa_{-,j}}^* \hat a_-, \\
        \hat e_{\mathrm{out}} &= \hat e_{\mathrm{in}} + \sqrt{\gamma_{-}}^* \hat a_-, \\
        \hat b_{\mathrm{out},j} &= \hat b_{\mathrm{in},j} + \sqrt{\Gamma_{j}} \hat b,
    \end{align}
\end{subequations}
where the rates are
\begin{align*}
    \kappa_{-,j} &= |C|^2 \kappa_j &
    \gamma_{-} &= |X|^2 \gamma &
    \kappa_{-} &= |C|^2 \kappa + |X|^2 \gamma = \kappa_{-,\mathrm{int}} + \kappa_{-,\mathrm{ext}} +\gamma_- .
\end{align*}
Thus the LP acts like an optical mode with extrinsic coupling $\kappa_{-,\mathrm{ext}}$ and intrinsic loss rate $\kappa_{-,\mathrm{int}} + \gamma_-$.

From here we can proceed as before, first linearizing the system via $\hat a_- = (\bar a_- + \delta\hat a_-)\exp(-i\omega_\mathrm{pump}t)$, $\hat b = \bar b + \delta\hat b$, and so on.
We then absorb the dc offset of the microwave field $\bar b=0$ as a renormalization of the polariton frequency and get
\begin{equation}
    \bar a_- = \frac{\sqrt{\kappa_{-,\mathrm{ext}}} \bar a_\mathrm{in,ext}}{i(\omega_- - \omega_\mathrm{pump}) + \kappa_-/2}.
\end{equation}
The first-order equations become
\begin{subequations}
    \begin{align}
        \dv{\delta\hat a_-}{t} &= -\Big(i\Delta_- + \frac{\kappa_-}{2}\Big)\delta\hat a_-
            + i g_\mathrm{eo} \bar a_- \Big(\delta\hat b + \delta\hat b^\dagger \Big)
            - \hat A_\mathrm{in}, \\
        \dv{\delta\hat b}{t} &= -\Big(i\omega_b + \frac{\Gamma}{2}\Big)\delta\hat b
            + i g_\mathrm{eo} \Big(\bar a_-^* \delta\hat a_- + \bar a_- \delta\hat a_-^\dagger \Big)
            - \hat B_\mathrm{in}, \\
        \dv{\delta\hat a_-^\dagger}{t} &= -\Big(-i\Delta_- + \frac{\kappa_-}{2}\Big)\delta\hat a_-^\dagger
            - i g_\mathrm{eo} \bar a_-^* \Big(\delta\hat b + \delta\hat b^\dagger \Big)
            - \hat A_\mathrm{in}^\dagger, \\
        \dv{\delta\hat b^\dagger}{t} &= -\Big(-i\omega_b + \frac{\Gamma}{2}\Big)\delta\hat b^\dagger
            - i g_\mathrm{eo} \Big(\bar a_-^* \delta\hat a_- + \bar a_- \delta\hat a_-^\dagger \Big)
            - \hat B_\mathrm{in}^\dagger,
    \end{align}
\end{subequations}
where $\Delta_- = \omega_- - \omega_\mathrm{pump}$ and
\begin{align*}
    \hat A_\mathrm{in} &= 
        - \sqrt{\kappa_{-,\mathrm{ext}}} \delta\hat a_\mathrm{in,ext}
        - \sqrt{\kappa_{-,\mathrm{int}}} \delta\hat a_\mathrm{in,int}
        + \sqrt{\gamma_{-}} \delta\hat e_\mathrm{in}, \\
    \hat B_\mathrm{in} &= 
        \sqrt{\Gamma_{\mathrm{ext}}} \delta\hat b_\mathrm{in,ext}
        + \sqrt{\Gamma_{\mathrm{int}}} \delta\hat b_\mathrm{in,int}.
\end{align*}
These equations can be written in the same form as the general scattering matrix problem Eq.~\eqref{eq:si:model:scattering-problem}.

To facilitate a tractable analytical solution we take the RWA again, this time dropping the counter-rotating microwave field $\delta\hat b^\dagger$ which is suppressed by roughly $\Gamma/4\omega_b$.
The optical idler $\delta\hat a_-^\dagger$ is kept as it is only suppressed by order $\kappa_-/4\omega_b$.
Taking a Fourier transform with the notation $\delta\hat a^\dagger = \delta a^\dagger(\omega) = [\delta a(-\omega)]^\dagger$ we can write the idler field $\delta\hat a_-^\dagger$ as
\begin{equation}
    \delta\hat a_-^\dagger = \frac{2}{\kappa_-}\chi_-^*(-\omega) \Big( - ig_\mathrm{eo} \bar a_-^* \delta\hat b - \hat A_\mathrm{in}^\dagger \Big)
\end{equation}
where $\chi_-(\omega) = [1 - 2i(\omega - \Delta_-)/\kappa_-]^{-1}$ is the normalized susceptibility of the LP.
This drives a self-energy and noise term on the microwave mode which can be absorbed as a renormalization of the microwave frequency and loss
\begin{equation}
    i\tilde\omega_b + \frac{\tilde\Gamma}{2} \equiv
    i\omega_b + \frac{\Gamma}{2} - \frac{2 g_\mathrm{eo}^2 |\bar a_-|^2}{\kappa_-} \chi_-^*(-\omega_b),
\end{equation}
which is valid for $|\omega - \omega_b|\ll2\omega_b$.
The microwave field is given by
\begin{equation}
     \delta\hat b = 
        \frac{2}{\tilde\Gamma} \chi_b(\omega)
        \bigg( i g_\mathrm{eo} \bar a_-^* \delta\hat a_-
        - \hat B_\mathrm{in}
        - \frac{2i g_\mathrm{eo} \bar a_-}{\kappa_-} \chi_-^*(-\omega) \hat A_\mathrm{in}^\dagger \bigg),
\end{equation}
where $\chi_b(\omega) = [1 - 2i(\omega - \tilde\omega_b)/\tilde\Gamma]^{-1}$ is the normalized susceptibility of the microwave field.
Solving for the LP field then gives
\begin{equation}
    \delta\hat a_- = \frac{2}{\kappa_-} \frac{\chi_-(\omega)}{1 + \mathcal C \chi_b(\omega) \chi_-(\omega)} \bigg[
        - \hat A_\mathrm{in}
        - \frac{2i g_\mathrm{eo} \bar a_-}{\tilde\Gamma} \chi_b(\omega) \hat B_\mathrm{in}
        + \mathcal C \frac{\bar a_-}{\bar a_-^*} \chi_b(\omega) \chi_-^*(-\omega) \hat A_\mathrm{in}^\dagger
    \bigg]
\end{equation}
where $\mathcal C = 4 g_\mathrm{eo}^2 |\bar a_-|^2 / \kappa_- \tilde\Gamma$ is the pump enhanced cooperativity.
This gives the output optical field
\begin{align}
    \delta\hat a_\mathrm{out,ext} ={}& 
        \delta\hat a_\mathrm{in,ext}
        + \frac{2\sqrt{\kappa_{-,\mathrm{ext}}}^*}{\kappa_-} \frac{\chi_-(\omega)}{1 + \mathcal C \chi_b(\omega) \chi_-(\omega)} \hat A_\mathrm{in}
        \nonumber\\
        &
        + \frac{4ig_\mathrm{eo} \bar a_- \sqrt{\kappa_{-,\mathrm{ext}}}^*}{\kappa_-\tilde\Gamma} \frac{\chi_b(\omega)\chi_-(\omega)}{1 + \mathcal C \chi_b(\omega) \chi_-(\omega)} \hat B_\mathrm{in}
        - \frac{2\mathcal C \bar a_-\sqrt{\kappa_{-,\mathrm{ext}}}^*}{\bar a_-^*\kappa_-} \frac{\chi_b(\omega)\chi_-(\omega) \chi_-^*(-\omega)}{1 + \mathcal C \chi_b(\omega) \chi_-(\omega)} 
        \hat A_\mathrm{in}^\dagger.
\end{align}

\vspace{1em}
\noindent\textbf{Transduction efficiency:}
The transduction efficiency can be directly read off from the coefficient mapping $\delta\hat b_\mathrm{in,ext}$ to $\delta\hat a_\mathrm{out,ext}$.
It is given by
\begin{equation}
    \eta_{ab}(\omega) = \eta_- \eta_b \frac{4\mathcal C \big|\chi_b(\omega) \chi_-(\omega)\big|^2}{\big|1 + \mathcal C\chi_b(\omega) \chi_-(\omega)\big|^2},
\end{equation}
where $\eta_- = \kappa_{-,\mathrm{ext}}/\kappa_-$ and $\eta_b = \Gamma_\mathrm{ext}/\tilde\Gamma$ are the output coupling efficiencies.
On resonance $\omega = \Delta_- = \tilde\omega_b$ the susceptibilities $\chi_b = \chi_- = 1$ and we recover Eq.~(3) from the main text
\begin{equation}
    \eta = \eta_- \eta_b \frac{4\mathcal C}{(1+\mathcal C)^2}
\end{equation}
with renormalization of $\tilde\Gamma$ explicit.

\vspace{1em}
\noindent\textbf{Added noise:}
The added noise is straightforwardly calculated as described in Eq.~\eqref{eq:si:model:added-noise-general}.
Assuming the optical and exciton modes have zero thermal population and that the microwave reservoirs have occupation $n_\mathrm{th}$ we find
\begin{equation}
    n_\mathrm{add}(\omega) = \frac{\Gamma_\mathrm{int}}{\Gamma_\mathrm{ext}} n_\mathrm{th} + \frac{\mathcal C}{\eta_b} \big|\chi_-(-\omega)\big|^2
\end{equation}
where the first term corresponds to direct thermal noise injection from the microwave reservoirs and the second term corresponds to noise generated by the parametric amplification process.
For $\omega = \tilde\omega_b$ this corresponds to Eq.~(4) of the main text.
Note that in dropping the UP and counter-rotating microwave mode $\hat b^\dagger$ we have dropped noise terms of scale $|\kappa_+/4\Delta_p|^2$ and $|\Gamma/4\omega_b|^2$, respectively.

\subsection{Generalization to more exciton states}
In general, the QW structure contains several exciton modes including the light-hole and heavy-hole excited states (e.g.\ main text Fig.~3). 
An important consideration in the design of the stepped QW was to create sufficient separation of these extra states such that the single-exciton description remains valid.
Nevertheless, even with $>5$\,meV splittings of the heavy-hole ground state from the other states, the strong optical couplings $\hbar g\sim 1$\,meV can still enable substantial hybridization with multiple excitons.
The analytical treatment above is readily extended to such situations (a more thorough model is presented later in Sec.~\ref{sec:si:disorder}).
Main text Fig.~4 includes the higher order stepped QW states in the performance calculation.

The main consequence is that the polaritons are now generally superpositions of all of the exciton states.
As before, we would utilize the LP, however its cavity fraction $|C|^2$ will now generally be diminished in comparison to the single-exciton case.
This has the effect of reducing the output coupling and thus the transduction efficiency.
More subtly, the constituent exciton states also have their own microwave couplings and generally have inter-exciton microwave dipole moments that drive population between them.
This provides multiple pathways for the transduction to occur resulting in interference that can both enhance or diminish the effective microwave coupling.
While non-trivial, these effects can generally be taken as first-order perturbations provided that the other exciton states are sufficiently far detuned.

\clearpage
\section{Transducer design}

\subsection{Finite-element method simulations}
Finite-element method simulations (COMSOL) were performed to compute the optical and electrostatic field profiles of the ridge waveguide transducer design.
The optical simulations show a well-confined fundamental transverse-electric (TE) mode with effective index $n_\mathrm{eff} = 3.314$ and a group index $n_g = 4.75$ (primarily due to material dispersion, geometric group index is only about $3.72$).
Propagation loss in the waveguide corresponds to a total quality factor $Q = 2.20 \times 10^6$, with substrate-limited $Q$ of $6.35\times10^6$ and electrode-absorption-limited $Q$ of $3.35\times10^6$.
For the device simulations in the main text, we assume a more conservative intrinsic $Q_i = 5\times10^5$ which includes realistic fabrication loss.
The electrostatic simulation gives a capacitance-per-unit-length of $C^\prime = 0.0496$\,fF/\textmu m.
The capacitor is well within the lumped-element regime.
The waveguide width of 500\,nm is chosen to enable high confinement of the optical mode while limiting the in-plane field component of the electrostatic/microwave field to no more than 5\% of the out-of-plane component.
This is important as the in-plane field can drive additional broadening through ionization and the in-plane Stark shift.

The ridge waveguide is chosen to be 25\,\textmu m in length.
This gives a free spectral range of the optical Fabry-Perot modes of about 1\,THz at 810\,nm which is substantially larger than the expected optical coupling rate $g/2\pi < 200$\,GHz near the operation point such that only one mode needs to be considered on resonance.
The Bragg coupling and reflector mirrors are designed with transmission of $7.5$\% and $0.01$\% to maintain a loaded $Q^\mathrm{opt}\approx 30{,}000$.
Our design has a nominal penetration depth of about 3.3\,\textmu m per mirror which slightly increases the length of the cavity mode.
The total capacitance also follows as $C=1.24$\,fF.
For operation at 30\,GHz, this requires a series inductor of about 23\,nH which is achievable with high-kinetic-inductance nanowires with low parasitic capacitance.

\subsection{Quantum well simulations}
The QW design is simulated using a two-step variational solver.
The first step directly computes the electron and hole QW subbands by solving the finite-difference Schr\"odinger equation in 1-d.
The second step feeds the lowest two electron and hole subbands into a variational solver based on the algorithm of Dignam and Sipe~\cite{dignam1991exciton}.
This method treats the exciton as a superposition of $1s$-like states formed between each of the four pairs of subbands with the pairwise Bohr radii as the variational parameter.
The algorithm enables the computation of all four exciton states simultaneously by leveraging their orthogonality.
We minimize the total energy of all four states simultaneously which achieves reasonable accuracy on all states.
Minimizing only the lowest energy state can increase the expected optical coupling of that state by up to 40\% at larger biases but does not produce usable data for the other states.
This process is repeated for both the heavy and light hole masses.

The optical and microwave couplings are computed from the resulting wavefunctions using the formula given in Sec.~\ref{sec:si:hamiltonian:optical-interaction} and~\ref{sec:si:hamiltonian:microwave-interaction} and the device parameters determined from the finite-element method simulations.
The optical mode is assumed to have sinusoidal spatial profile along the propagation axis with an effective length extended by the mirror penetration ($6.6$\,\textmu m for both mirrors).
The mode area is roughly $0.0523$\,\textmu m$^2$ and the zero-point field is $E_\mathrm{zpf} = \sqrt{\hbar\omega / 2\epsilon_0 n_\mathrm{eff}^2 A L_\mathrm{eff}}$ where the group dispersion due to the GaAs excitonic response is neglected (as to not count it twice).
The inter-band dipole element for the heavy hole is assumed to be $d_{vc}/e=0.581$\,nm~\cite{gu2013relation} while the light-hole is reduced by the $k\cdot p$ selection rule factor $1/\sqrt{3}$~\cite{chuang2009physics}.
The field at the QW layer is calculated to be about $0.56$\,MV/m per volt applied at the electrodes such that the target bias field at $1$\,MV/m corresponds to an applied dc bias of about $1.8$\,V.
The inter-exciton microwave coupling rates are also computed while the mixing between the heavy and light hole states is assumed to be zero.

\clearpage
\section{Quantum well reflectance spectroscopy}

\subsection{Sample and measurements}
The GaAs/AlGaAs heterostructure was grown via molecular beam epitaxy on a semi-insulating GaAs substrate with GaAs buffer layers.
The measured QW consists of a 30-nm GaAs layer confined by Al$_{0.27}$Ga$_{0.73}$As barriers.
The center of the QW lies approximately at 280\,nm beneath the top surface.
We examine the lowest heavy-hole (HH) and light-hole (LH) excitons of the QW.

Reflectance spectra were obtained in a Montana Instruments cryostat at 6.7\,K.
A Thorlabs L820P100 laser diode operated at a low injection current of 25\,mA provided a continuous emission spectrum spanning approximately 810--830\,nm.
The optical power incident on the sample was below 10\,nW.
The weak probe was used to minimize excitation-induced broadening, including broadening associated with exciton-exciton interactions.
The representative spectra were recorded at zero applied bias with an integration time of one second.

The reflectance spectrum is measured in a Princeton Instruments Acton SpectraPro SP-2750 spectrometer with a 1600 lines/mm diffraction grating.
A narrow continuous-wave (CW) laser near 820\,nm measured with the same spectrometer slit setting mapped to approximately two pixels along the dispersion direction.
The wavelength resolution of the spectrometer near the HH resonance is approximately $0.011$\,nm/pixel corresponding to an instrument limited resolution of about $0.023$\,nm ($10$\,GHz, 42\,\textmu eV).

\subsection{Data processing and fitting}
The detector dark count offset of 650 counts is removed prior to the remaining analysis.
The resulting spectrum $S(\lambda)$ contained the reflection spectrum, scaled by the spectral envelope of the source, and with a multiplicative interference fringe pattern.
The fringe frequencies are identified by a Fourier transform and the constituent sinusoidal components were fitted along with a slow phase correction and slow baseline.
This oscillation could be removed and the resulting clean spectrum could be normalized by the background envelope (obtained by fitting the regions excluding the exciton resonances).
The resulting dimensionless reflection spectrum $\tilde R$ has an approximately unity off-resonance background, but does not establish an absolute scale.

The absolute non-resonant reflectance $R_\mathrm{bg}$ could not be independently measured and so we treat it as an assumed parameter with the normalized reflectance given by $R(E) = R_\mathrm{bg}\tilde R(E)$.
The analysis was repeated for $R_\mathrm{bg}=0.25$ and $0.30$ to assess sensitivity to the normalization.
The reflectance is modeled using a local field correction which accounts for interference between the surface reflection and the QW's excitonic response~\cite{andreani1991radiative,ivchenko2005optical}
\begin{subequations}
    \begin{align}
        R_\mathrm{model}(E) &= \abs{\frac{\sqrt{R_\mathrm{bg}} + r_X(E)}{1 + \sqrt{R_\mathrm{bg}} r_X(E)}}^2,\\
        r_X(E) &= \sum_{k} \frac{i\hbar \Gamma_{0,k} \exp(i\phi_k) }{E - E_k - i\hbar(\Gamma_{0,k} + \Gamma_{i,k})},
    \end{align}
\end{subequations}
where $\Gamma_{0,k}$ and $\Gamma_{i,k}$ denote the radiative and non-radiative half-width parameters, respectively, and $\phi_k$ is the relative phase.
The full spectrum is fit directly to this model with two resonances $k=\{\mathrm{HH},\mathrm{LH}\}$.
The pole-width is defined $W_k = 2\hbar(\Gamma_{0,k} + \Gamma_{i,k})$ which corresponds to the FWHM of the total exciton response, including both homogeneous and inhomogeneous broadening.

\subsection{Background sensitivity and interpretation}
Both assumed backgrounds yield the same goodness of fit, $R^2 \approx 0.96$.
However, the HH radiative half-width changes from $29.0$ to $34.1$\,\textmu eV, while the nonradiative half-width changes from $22.7$ to $16.0$\,\textmu eV.
The non-radiative component can be ascribed to a number of different sources including genuine non-radiative decay (homogeneous broadening), inhomogeneous broadening, and finite instrument resolution. 
The total pole width remains approximately constant at $103.5$ to $100.2$\,\textmu eV, respectively.
Similar changes are observed in the LH line widths.
Consequently, the radiative/non-radiative decomposition of the line width is primarily sensitive to the off-resonant reflectance.
These ranges describe the sensitivity to the assumed background and are not confidence intervals or independently established bounds on $R_\mathrm{bg}$.
The total and the decomposed radiative/nonradiative line widths are close in scale to the expected spectrometer resolution and so will require higher-resolution techniques (such as resonant excitation) to fully resolve.

\begin{figure}
    \centering
    \includegraphics{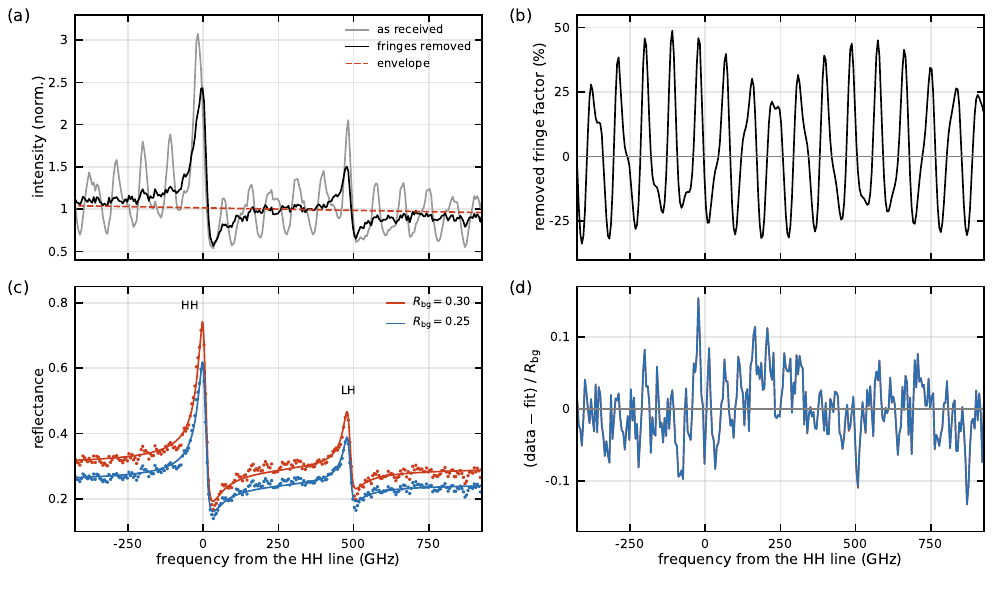}
    \caption{
        \textbf{Processing of reflectance data.}
        (a) Background subtracted and fringe-corrected signals with fitted smooth envelope.
        (b) The removed multiplicative fringe contribution.
        (c) Reflectance for the two assumed backgrounds with fits.
        (d) Fit residuals divided by the background $R_\mathrm{bg}$.
        The two curves are nearly identical.
        All panels use the same reference energy, chosen as the fitted HH energy for $R_\mathrm{bg} = 0.30$.
    }
    \label{fig:si-reflectance}
\end{figure}

\begin{table}
    \centering
    \caption{
        HH and LH line widths in \textmu eV, all full widths at half maximum.
        $2\hbar\Gamma_{0,k}$ is the radiative width in a uniform medium, $2\hbar\Gamma_{i,k}$ the non-radiative width,
        and $W_k = 2\hbar(\Gamma_{0,k} + \Gamma_{i,k})$ the total width in a uniform medium.
        Errors are conditional $1\sigma$ fit errors; none is given for $W_k$ since the fit does not provide the covariance of $\Gamma_{0,k}$ and $\Gamma_{i,k}$.
    }
    \begin{tabular*}{0.7\textwidth}{@{\extracolsep{\fill}}lllll}
        \hline
        $R_\mathrm{bg}$ & Exciton & $2\hbar\Gamma_{0,k}$ (\textmu eV) & $2\hbar\Gamma_{i,k}$ (\textmu eV) & $W_k$ (\textmu eV) \\
        \hline
        0.30 & HH & $68.3 \pm 1.4$ & $31.9 \pm 2.1$ & $100.2$ \\
             & LH & $30.3 \pm 1.5$ & $65.7 \pm 5.5$ & $96.0$ \\
        0.25 & HH & $58.1 \pm 1.2$ & $45.5 \pm 2.2$ & $103.5$ \\
             & LH & $25.9 \pm 1.2$ & $71.5 \pm 5.7$ & $97.5$ \\
        \hline
    \end{tabular*}
\end{table}

\clearpage
\section{Disorder and cavity protection} \label{sec:si:disorder}
\subsection{Modeling disorder}
Inhomogeneous broadening of the exciton arises from disorder in the QW dimensions/interfaces and the local strain/electromagnetic environment.
A standard model of the disorder~\cite{savona2007disorder} assumes an effective center-of-mass (CM) potential $V(\vb R)$ which is a Gaussian-distributed stochastic field with zero mean and correlation
\begin{equation}
    \ev{V(\vb R)V(\vb R^\prime)} = \sigma_V^2 f(|\vb R- \vb R^\prime|),
\end{equation}
where $\sigma_V^2$ is the variance of the potential and $f(R)$ is a normalized correlation function with characteristic decay length $\Lambda_c$.
In principle, the model can be generalized to any stochastic field with probability distribution function $P(V)$ and corresponding correlation function.

In the presence of disorder, the CM states $\Phi_{\vb K}(\vb R)$ are now admixtures of the non-disordered eigenstates and are generally localized~\cite{savona2007disorder,hegarty1984localized}.
Consequently, each of the disordered system's eigenstates at frequencies $\omega_n$ may generally have finite projection onto the momentum-matched ``bright'' state(s) of the non-disordered system and thus couple optically to the cavity.
By completeness, if a single state was momentum-matched in the non-disordered system at rate $G$, then this gives the oscillator sum rule
\begin{equation}
    G^2 = \sum_n |g_n|^2,
\end{equation}
where $g_n$ is the coupling rate of the $n$-th eigenstate of the disordered system.
It is difficult to predict \textit{a priori} the correlations between the optical coupling $g_n$ and the frequency $\omega_n$ without a detailed model of the underlying disorder potential $V(\vb R)$, which is itself difficult to measure.
Thus, we do not presume any specific underlying disorder and treat the problem heuristically.

We consider a generic QCSE Hamiltonian extended to a disordered system with many exciton states $\hat e_n$ 
\begin{equation}
    \hat H = \omega_a \hat a^\dagger \hat a
    + \omega_b \hat b^\dagger \hat b
    + \sum_n \omega_n \hat e_n^\dagger \hat e_n
    - \sum_n \Big(g_n \hat a^\dagger \hat e_n + g_n^* \hat a \hat e_n^\dagger\Big)
    - \sum_n \xi_n \hat e_n^\dagger\hat e_n \Big(\hat b+\hat b^\dagger\Big),
\end{equation}
where $\hat a$ and $\hat b$ are the optical/microwave field operators.
We define the bright state as the specific superposition that couples directly to the optical mode
\begin{equation}
    \ket{B} = \frac{1}{G} \sum_n g_n^* \ket{n} \qquad\longleftrightarrow\qquad \hat B = \frac{1}{G}\sum_n g_n \hat e_n.
\end{equation}
The bright state is not an eigenstate of the system, which has important implications for transduction that will be discussed later.
The local density of states projected onto the bright state $\rho(\omega)$ is defined as
\begin{equation}
    \rho(\omega) = \sum_n p_n \delta(\omega-\omega_n)
\end{equation}
where $p_n = |g_n|^2/G^2$ is the weight of the exciton state $\ket{n}$ in the bright state.
Completeness requires $\int\dd{\omega}\rho(\omega) = 1$.
Assuming that the disorder only shifts the frequencies and does not modify the intrinsic homogeneous line width of each exciton, i.e.\ $\gamma_n = \gamma$, the exciton response is encapsulated by the Green's function
\begin{equation}
    K(\omega) = \int \dd{x} \frac{\rho(x)}{\omega - x + i\gamma/2}.
\end{equation}
A discrete ensemble has identically $K(\omega) = \sum_n p_n / (\omega - \omega_n + i\gamma/2)$, and so, for a sparsely populated ensemble, the response function appears as several peaks with randomly distributed positions and amplitudes.
If the frequency spacing between the coupled excitons ($p_n>0$) is small compared to the homogeneous line width $\gamma$, the exciton response function $K(\omega)$ smooths the overlapping peaks, blurring the sub-linewidth structure of the density of states $\rho(\omega)$.
In this sense, there are generally a large number of distinct $\rho(\omega)$ which reproduce a given response $K(\omega)$ to within some arbitrary error $\epsilon$.
As will be shown below, all relevant exciton dynamics enters only through the response function $K(\omega)$.
Thus, provided $K(\omega)$ is known to within $\epsilon$, we do not need to know the specific density of states $\rho(\omega)$ --- nor the underlying disorder which determines it.
Thus going forward, we \textit{assume} a density of states $\rho(\omega)$ that gives rise to a response $K(\omega)$ that approximates a physically relevant model.

\subsection{Cavity protection}
First we review the standard cavity protection result \cite{diniz2011protecting,houdre1996rabi,whittaker1998inhomogeneous,zhong2017cavityprotected} in the absence of the microwave field and its interactions with the exciton.

\subsubsection{Setup}
The Heisenberg-Langevin equations are
\begin{subequations}
    \begin{align}
        \dv{\hat a}{t} &= - \bigg(i\omega_a + \frac{\kappa}{2}\bigg) \hat a + i\sum_n g_n \hat e_n - \sqrt{\kappa_\mathrm{ext}}\hat a_\mathrm{in,ext} - \sqrt{\kappa_\mathrm{int}} \hat a_\mathrm{in,int}, \\
        \dv{\hat e_n}{t} &= - \bigg(i\omega_n + \frac{\gamma}{2}\bigg) \hat e_n + i g_n^* \hat a - \sqrt{\gamma}\hat e_{n,\mathrm{in}}.
    \end{align}
\end{subequations}
Fourier transform the equations and define the denominators $D_a(\omega) = \omega_a - \omega - i\kappa/2$ and $D_n(\omega) = \omega_n - \omega - i\gamma/2$.
Solving for each of the exciton equations yields
\begin{equation}
    \hat e_n = \frac{g_n^*\hat a + i\sqrt{\gamma}\hat e_{n,\mathrm{in}}}{D_n(\omega)}.
\end{equation}
Inserting these back into the cavity equation produces two new terms: a self-energy term
\begin{equation}
    \sum_n \frac{|g_n|^2\hat a}{D_n(\omega)} = - G^2 K(\omega)\hat a,
\end{equation}
and a noise term
\begin{equation}
    \hat F_a(\omega) = \sum_n \frac{ig_n\sqrt{\gamma}\hat e_{n,\mathrm{in}}}{D_n(\omega)}.
\end{equation}
The self-energy term renormalizes the optical cavity frequency and loss; define the generalized denominator
\begin{equation}
    \label{eq:si:disorder:generalized-denominator}
    \mathcal D_a(\omega) \equiv D_a(\omega) + G^2 K(\omega) 
    = \Big(\omega_a + G^2\Re K(\omega) - \omega\Big) - \frac{i}{2}\Big(\kappa - 2G^2\Im K(\omega)\Big).
\end{equation}
The cavity equation can then be solved and inserted into the input-output equation to get the scattering matrix row
\begin{align}
    \hat a_\mathrm{out,ext} &= \hat a_\mathrm{in,ext} + \sqrt{\kappa_\mathrm{ext}} \hat a , \nonumber\\
    \label{eq:si:disorder:unpumped-cavity-reflection}
    &= \bigg(1 + \frac{i\kappa_\mathrm{ext}}{\mathcal D_a(\omega)} \bigg) \hat a_\mathrm{in,ext}
    + \frac{i\sqrt{\kappa_\mathrm{ext}\kappa_\mathrm{int}}}{\mathcal D_a(\omega)} \hat a_\mathrm{in,int}
    + \frac{\sqrt{\kappa_\mathrm{ext}}}{\mathcal D_a(\omega)} \hat F_a(\omega),
\end{align}
where the first term's coefficient is identified as the complex reflection amplitude of the cavity.
Finally, it is straightforward to confirm that the new exciton noise operator has the commutation relation
\begin{equation}
    [\hat F_a(\omega),\hat F_a^\dagger(\omega^\prime)] = -2G^2\Im K(\omega) \delta(\omega-\omega^\prime)
\end{equation}
as required by the fluctuation-dissipation theorem and so the system is completely characterized in terms of the exciton response $K(\omega)$.

\subsubsection{Linearization}
The solutions given in the last section are exact for any general exciton response $K(\omega)$, however they are consequently somewhat opaque to interpretation.
Here we make some approximations to write the generalized denominator $\mathcal D_a(\omega)$ in a simple-pole form that describes the polariton.
To do this we will linearize the denominator about the polariton resonance itself.
Since the original denominator $D_a(\omega) = \omega_a - \omega - i\kappa/2$ is already linear, we only need to linearize the exciton response $K(\omega)$.

We will seek to expand the response in both frequency and loss at a generic frequency point $\omega_*$ in the limit $\gamma\to0^+$.
First we can split the response into its real and imaginary parts
\begin{equation}
    K(\omega;\gamma) = \int\dd{x}\rho(x) \bigg[
        \frac{\omega-x}{(\omega-x)^2 + \gamma^2/4} - i \frac{\gamma/2}{(\omega-x)^2 + \gamma^2/4}
    \bigg].
\end{equation}
The response depends on $\omega$ and $\gamma$ only through $\omega+i\gamma/2$, so $\partial_\gamma K = (i/2)\partial_\omega K$.
In the limit as $\gamma\to0$ from above, the real part of the integrand collapses to $\rho(x)/(\omega-x)$ in the tails $|\omega-x|\gg \gamma$ while symmetrically excluding the region $|x-\omega|\lesssim \gamma$, which yields the principal value (PV) component.
The imaginary part of the integrand is a Lorentzian at $\omega$ with normalized area $\pi$ multiplying $\rho(x)$.
In the limit of $\gamma\to0$ the Lorentzian tends to a delta function at $\omega$.
Thus evaluating at the expansion point $\omega_*$
\begin{equation}
    \label{eq:si:disorder:k-expansion-zeroth-order}
    K(\omega_*;0^+) = \mathcal P \int\dd{x}\frac{\rho(x)}{\omega_*-x} - i\pi\rho(\omega_*),
\end{equation}
where $\mathcal P$ denotes the principal value component.
The response is expanded as
\begin{equation}
    K(\omega;\gamma) = K(\omega_*;0^+) + \pdv{K(\omega_*;0^+)}{\omega}\Big(\omega - \omega_* + i\frac{\gamma}{2}\Big),
\end{equation}
where we used $\partial_\gamma K = (i/2)\partial_\omega K$ to combine the $\omega$ and $\gamma$ terms.
The derivative is evaluated as
\begin{align*}
    \pdv{K(\omega;\gamma)}{\omega} &= - \int\dd{x} \frac{\rho(x)}{(\omega - x + i\gamma/2)^2} \\
    &= - \left.\frac{\rho(x)}{\omega-x + i\gamma/2}\right|_{-\infty}^{+\infty} + \int\dd{x}\frac{\rho^\prime(x)}{\omega - x + i\gamma/2}, \\
    &= \int\dd{x}\frac{\rho^\prime(x)}{\omega - x + i\gamma/2},
\end{align*}
where the first line directly applies the $\omega$ derivative, the second integrates by parts, and the third enforces the boundary conditions.
Note that this step assumes that the density of states is itself differentiable.
The limit as $(\omega,\gamma)\to(\omega_*,0^+)$ follows as before
\begin{equation}
    \pdv{K(\omega_*;0^+)}{\omega} = \mathcal P \int\dd{x}\frac{\rho^\prime(x)}{\omega_* - x} - i\pi \rho^\prime(\omega_*).
\end{equation}
Defining the shorthand notation $\mathrm{PV}_f \equiv \mathcal P \int\dd{x} f(x)/(\omega_*-x)$ we can write the expanded form of $K(\omega)$ as
\begin{equation}
    \label{eq:si:disorder:linearized-exciton-response}
    K(\omega) \approx \Big[\mathrm{PV}_\rho - i\pi\rho(\omega_*)\Big] 
        + \Big[\mathrm{PV}_{\rho^\prime} - i\pi\rho^\prime(\omega_*)\Big] \Big(\omega-\omega_*+i\frac{\gamma}{2}\Big)
\end{equation}
and thus the generalized denominator Eq.~\eqref{eq:si:disorder:generalized-denominator} takes the linearized form $\mathcal D_a(\omega) = A + B(\omega- \omega_*)$ where
\begin{align*}
    A &= \Big(\omega_a - \omega_* + G^2\mathrm{PV}_\rho + G^2\frac{\gamma}{2}\pi\rho^\prime\Big)
        - \frac{i}{2}\Big( \kappa + 2\pi G^2 \rho - G^2\gamma\mathrm{PV}_{\rho^\prime}\Big), \\
    B &= \Big(-1 + G^2\mathrm{PV}_{\rho^\prime}\Big) - iG^2\pi \rho^\prime.
\end{align*}
In this form the denominator has a single zero at the complex frequency $\omega_* - A/B$.
This then maps to a Lorentzian resonance in the optical mode at real frequency $\omega_* -\Re(A/B)$ with line width $2\Im(A/B)$.

\subsubsection{Resonance condition and cavity protection}
So far $\omega_*$ is an arbitrary expansion point.
To expand at the polariton, we require that $\omega_*$ is itself the resonance.
This requires $\Re(A) = 0$, which gives the resonance condition
\begin{equation}
    \label{eq:si:disorder:resonance-condition-general}
    \omega_* = \omega_a + G^2 \mathrm{PV}_\rho + G^2 \frac{\gamma}{2}\pi \rho^\prime.
\end{equation}
There are generally multiple $\omega_*$ which satisfy the resonance condition.
To specify the polariton we assume that $G$ is sufficiently large compared to the characteristic width of the density of states $\rho(\omega)$ such that the polariton resonance occurs out in the tails of the distribution.
Here both the density of states $\rho(\omega)$ and its derivative can be assumed to asymptote to zero and so the derivative term $\rho^\prime$ can be dropped.
Similarly, the resonance $\omega_*$ occurs far away from the bulk of $\rho(\omega)$ and so the PV integral is dominated by the overlap in far tails.
This allows us to expand the PV integral in terms of the resonance separation from the density's mean $\bar\omega$ as
\begin{equation}
    \label{eq:si:disorder:pv-rho-approximate}
    \mathrm{PV}_\rho = \mathcal P \int\dd{x} \frac{\rho(x)}{\omega_* - \bar\omega} \bigg[
        1 + \frac{x - \bar\omega}{\omega_* - \bar\omega} + \frac{(x - \bar\omega)^2}{(\omega_* - \bar\omega)^2} + \cdots
    \bigg]
    \approx \frac{1}{\omega_* - \bar\omega},
\end{equation}
where the last step truncates to the lowest order and uses the fact that $\int\dd{x}\rho(x) = 1$.
This is essentially equivalent to assuming all of the density, from the perspective of the polariton, exists at a $\bar\omega$, and so the resonance condition takes the form
\begin{equation}
    (\omega_* - \omega_a) (\omega_* - \bar\omega) = G^2,
\end{equation}
matching the homogeneous system with the exciton frequency $\omega_e=\bar\omega$.
The additional terms in this expansion, plus the dropped derivative term give rise to additional small shifts to the polariton frequency.
Thus for our analytical model we take $\omega_* = \omega_-$ as the frequency of the LP in the non-disordered system.

Dropping the other derivative terms and picking the LP resonance as the expansion point, the denominator becomes
\begin{equation}
    \mathcal D_a(\omega) = \Big[-1 + G^2\mathrm{PV}_{\rho^\prime} \Big] (\omega - \omega_-) - \frac{i}{2}\Big[\kappa + 2\pi G^2\rho - G^2 \gamma \mathrm{PV}_{\rho^\prime} \Big].
\end{equation}
Note that the principal value of the derivative $\mathrm{PV}_{\rho^\prime}$ does not drop out because it is a convolution.
To simplify further we can leverage the Hellmann-Feynman theorem which states that, for an eigenstate $\ket{\lambda}$ with energy $E_\lambda$, one has $\dv*{E_\lambda}{\lambda} = \ev{\dd{\hat H}\!/\!\dd{\lambda}}{\lambda}$.
For the lower polariton with frequency $\omega_-$ and optical cavity at $\omega_a$ we thus have
\begin{equation*}
    \dv{\omega_-}{\omega_a} = \ev{\hat a^\dagger \hat a}{a_-} = |C|^2,
\end{equation*}
where $|C|^2$ is the Hopfield cavity fraction for the LP.
Differentiating the resonance condition Eq.~\eqref{eq:si:disorder:resonance-condition-general} we have
\begin{equation}
    \dv{\omega_-}{\omega_a} = 1 + G^2 \mathrm{PV}_{\rho^\prime} \dv{\omega_-}{\omega_a}
    \qquad\implies\qquad
    \dv{\omega_-}{\omega_a} = |C|^2 = \frac{1}{1 - G^2\mathrm{PV}_{\rho^\prime}},
\end{equation}
where the derivative falls through the PV via integration by parts.
Consequently we can define the generalized Hopfield exciton fraction via
\begin{equation}
    |X|^2 = 1 - |C|^2 = \frac{-G^2\mathrm{PV}_{\rho^\prime}}{1 - G^2\mathrm{PV}_{\rho^\prime}},
\end{equation}
which implies that $G^2\mathrm{PV}_{\rho^\prime} = - |X|^2/|C|^2$.
Note that this behaves like the exciton fraction but for the entire ensemble, rather than any individual exciton state.
The linearized denominator becomes
\begin{equation}
    \mathcal D_a(\omega) = \frac{1}{|C|^2}(\omega_- -\omega) - \frac{i}{2}\bigg[\kappa + 2\pi G^2\rho + \frac{|X|^2}{|C|^2}\gamma\bigg].
\end{equation}

Now the cavity reflection coefficient can be read off from Eq.~\eqref{eq:si:disorder:unpumped-cavity-reflection} as
\begin{equation}
    r(\omega) = 1 + \frac{i|C|^2\kappa_\mathrm{ext}}{(\omega_- - \omega) - i\tilde\kappa_-/2}
\end{equation}
which matches the expected single-pole polariton reflection spectrum with a renormalized line width
\begin{equation}
    \tilde\kappa_- = |C|^2 \kappa + |X|^2\gamma + 2\pi |C|^2 G^2\rho(\omega_-).
\end{equation}
This is the standard cavity protection \cite{diniz2011protecting,houdre1996rabi,whittaker1998inhomogeneous,zhong2017cavityprotected} result which shows that the polariton line width is broadened by $2\pi |C|^2 G^2 \rho(\omega_-)$ due to the exciton ensemble.
This additional contribution can be suppressed provided $\rho(\omega_-)$ can be suppressed.
This is robustly achieved by increasing the collective coupling $G$ at fixed detuning, provided the distribution $\rho(\omega)$ falls off faster than $\omega^{-2}$.
Alternatively, at fixed $G$, one can modify the cavity-exciton detuning to further move the LP away from the exciton ensemble, however the line width may not necessarily decrease if the cavity line width $\kappa$ exceeds the exciton homogeneous line width $\gamma$.

\subsection{Cavity protection in transduction} \label{sec:si:disorder:transduction}
We now consider the effect of the inhomogeneity on transduction.
There are two different mechanisms:
\begin{enumerate}
    \item \textbf{Bright state dephasing:} A photon in the cavity creates the bright state superposition which, because it is not an eigenstate, dephases over time due to the inhomogeneously broadened exciton ensemble.
    This rotates the bright state into the orthogonal dark state ensemble which can no longer couple back to the cavity.
    The dark state excitons then emit outside of the cavity before the ensemble can rephase back into the bright state.
    Thus, the dark state ensemble acts as a non-Markovian bath which creates loss and can inject noise back into the system.
    This is the same mechanism that the standard cavity protection addresses and assumed to be the dominant mechanism in the QCSE transducer.
    
    \item \textbf{Misaligned rotations:} At first order in the linearized regime, the EO interaction manifests as a unitary rotation between the microwave fluctuations $\delta\hat b$ and a specific superposition of the exciton fluctuations $\delta\hat e_n$
    \begin{equation}
        \delta\hat C = \frac{1}{\mathcal N}\sum_n \xi_n \bar e_n^* \delta\hat e_n
    \end{equation}
    where $\mathcal N = \sqrt{\sum_n \xi_n^2|\bar e_n|^2}$ is the normalization.
    This is analogous to the bright mode $\hat B$, but for the linearized exciton-microwave interaction.
    In general, $\xi_n$ is stochastic (inherited from the underlying disorder) may also have its own distribution and correlations with $\omega_n$ and $g_n$.
    As a result, the microwave bright state $\delta\hat C$ will not in general be aligned to the optical bright state $\delta\hat B$ (here as the linearized fluctuation).
    Since the polariton (in the cavity-protected limit) is a superposition of the bright state and optical mode (e.g.\ $\delta\hat a_- = X^*\delta\hat B - C\delta\hat a$), the microwave interaction rotating $\delta\hat C \leftrightarrow \delta\hat b$ will generically rotate some component of the polariton $\delta\hat a_-$ into the dark state ensemble which manifests as a loss in the transduction efficiency.

    Fortunately, this effect can be largely suppressed by engineering the system properly.
    Because the coupling $\xi_n$ scales only with the static dipole times the local microwave field and \textit{not} the area or optical mode overlap (see Sec.~\ref{sec:si:hamiltonian:microwave-interaction}), the EO coupling rate should be largely uniform provided the microwave/bias field is itself uniform.
    Thus we can approximate $\xi_n \approx \xi$ for all states $n$.
    Furthermore, the parametric enhancement scales as the steady-state exciton field $\bar e_n^*$ which arises from the coherent optical pumping.
    The optical pumping specifically drives the bright state itself such that $\bar e_n \propto g_n^*$.
    Thus
    \begin{equation}
        \delta\hat C \approx \frac{\xi}{\mathcal N}\sum_n g_n \delta\hat e_n \sim \delta\hat B,
    \end{equation}
    and the rotations are realigned.
    To reiterate, this requires only that the microwave and bias field are uniform over the exciton states, which is generally straightforward to achieve.
    Thus we neglect this effect and take $\xi_n = \xi$ going forward.
\end{enumerate}

\subsubsection{Mean-field equations}
Assuming that the microwave couplings are uniform we can compute the Heisenberg-Langevin equations as Eq.~\eqref{eq:si:efficiency:heisenberg-langevin}.
As before we can linearize the equations assuming a drive at the pump frequency $\omega_\mathrm{pump}$.
Let $\hat a = (\bar a + \delta\hat a)\exp(-i\omega_\mathrm{pump}t)$, $\hat a_\mathrm{in,ext} = (\bar s + \delta\hat a_\mathrm{in,ext})\exp(-i\omega_\mathrm{pump}t)$, and so on.
The microwave mode does not rotate at the pump frequency.
In the frequency domain the zeroth-order mean-field equations are
\begin{subequations}
    \begin{align}
        D_a(\omega_\mathrm{pump}) \bar a &= \sum_n g_n\bar e_n + i\sqrt{\kappa_\mathrm{ext}}\bar s, \\
        D_n(\omega_\mathrm{pump}) \bar e_n &= g_n^*\bar a,
    \end{align}
\end{subequations}
where we renormalize the microwave field at its self-consistent shifted frequency $\bar b = 0$.
This is exactly the mean-field version of the standard cavity protection model discussed in the prior section.
The cavity field can be linearized about the polariton
\begin{equation}
    \bar a = \frac{i\sqrt{\kappa_\mathrm{ext}} \bar s}{\mathcal D_a(\omega_\mathrm{pump})}
    = \frac{i|C|^2\sqrt{\kappa_\mathrm{ext}} \bar s}{\omega_- - \omega_\mathrm{pump} - i\tilde\kappa_-/2},
\end{equation}
and the exciton field follows straightforwardly
\begin{equation}
    \bar e_n = \frac{ig_n^* |C|^2\sqrt{\kappa_\mathrm{ext}}\bar s}{\big(\omega_n - \omega_\mathrm{pump} - i\gamma/2\big)\big(\omega_- - \omega_\mathrm{pump} - i\tilde\kappa_-/2\big)},
\end{equation}
which, out in the tails where $\omega_n - \omega_\mathrm{pump} \approx \bar\omega - \omega_\mathrm{pump}$, scales exactly in proportion to $g_n^*$ only.
This assumes that the linearization is valid to within the detuning $|\omega_- - \omega_\mathrm{pump}| \approx \omega_b$.

\subsubsection{First-order equations}
The first order equations follow directly.
Fourier transforming to the new slow frequency $\delta$ in the rotated frame gives
\begin{subequations}
    \begin{align}
        D_a(\omega_\mathrm{pump} + \delta) \delta\hat a &= 
            \sum_n g_n\delta\hat e_n
            + i\sqrt{\kappa_\mathrm{ext}}\delta\hat a_\mathrm{in,ext}
            + i\sqrt{\kappa_\mathrm{int}}\delta\hat a_\mathrm{in,int}, \\
        D_n(\omega_\mathrm{pump} + \delta) \delta\hat e_n &= 
            g_n^*\delta\hat a
            + \xi\bar e_n \Big(\delta\hat b + \delta\hat b^\dagger\Big)
            + i\sqrt{\gamma}\delta\hat e_{n,\mathrm{in}}, \\    
        D_b(\delta) \delta \hat b &= 
            \xi\sum_n\Big(\bar e_n^* \delta\hat e_n + \bar e_n\delta\hat e_n^\dagger\Big)
            + i\sqrt{\Gamma_\mathrm{ext}}\delta\hat b_\mathrm{in,ext}
            + i\sqrt{\Gamma_\mathrm{int}}\delta\hat b_\mathrm{in,int},
    \end{align}
    with the counter-rotating terms
    \begin{align}
        D_a^*(\omega_\mathrm{pump} - \delta) \delta\hat a^\dagger &= 
            \sum_n g_n^* \delta\hat e_n^\dagger
            - i\sqrt{\kappa_\mathrm{ext}}\delta\hat a_\mathrm{in,ext}^\dagger
            - i\sqrt{\kappa_\mathrm{int}}\delta\hat a_\mathrm{in,int}^\dagger, \\
        D_n^*(\omega_\mathrm{pump} - \delta) \delta\hat e_n^\dagger &= 
            g_n\delta\hat a^\dagger
            + \xi\bar e_n^* \Big(\delta\hat b + \delta\hat b^\dagger\Big)
            - i\sqrt{\gamma}\delta\hat e_{n,\mathrm{in}}^\dagger, \\    
        D_b^*(-\delta) \delta \hat b^\dagger &= 
            \xi\sum_n\Big(\bar e_n^* \delta\hat e_n + \bar e_n\delta\hat e_n^\dagger\Big)
            - i\sqrt{\Gamma_\mathrm{ext}}\delta\hat b_\mathrm{in,ext}^\dagger
            - i\sqrt{\Gamma_\mathrm{int}}\delta\hat b_\mathrm{in,int}^\dagger.
    \end{align}
\end{subequations}
Note that the notation is $\hat a^\dagger \equiv [\hat a(-\delta)]^\dagger = \hat a^\dagger(\delta)$ and so on, which follows from the definition of the Fourier transform on the operators.
Care must be taken with the commutation relations $[\hat a,\hat a^\dagger] = [\hat a(\omega),(\hat a(-\omega^\prime))^\dagger] = \delta(\omega+\omega^\prime)$.
The denominators are the same as before
\begin{align*}
    D_a(\omega) &= \omega_a - \omega - i\kappa/2, \\
    D_n(\omega) &= \omega_n - \omega - i\gamma/2, \\
    D_b(\omega) &= \omega_b - \omega - i\Gamma/2. 
\end{align*}

At this point we eliminate the exciton equations in the same manner as before.
Solving directly
\begin{subequations}
    \begin{align}
        \delta\hat e_n &= \frac{g_n^*\delta\hat a + \xi\bar e_n (\delta\hat b + \delta\hat b^\dagger) + i\sqrt{\gamma}\delta\hat e_{n,\mathrm{in}} }{D_n(\omega_\mathrm{pump}+\delta)}, \\
        \delta\hat e_n^\dagger &= \frac{g_n\delta\hat a^\dagger + \xi\bar e_n^* (\delta\hat b + \delta\hat b^\dagger) - i\sqrt{\gamma}\delta\hat e_{n,\mathrm{in}}^\dagger }{D_n^*(\omega_\mathrm{pump}-\delta)},
    \end{align}
\end{subequations}
which can be re-inserted into the optical and microwave equations.
As before, this generates a new self-energy term which renormalizes the frequency and line width and has a matching noise term, now in both the microwave and optical equations.
However, we now also get new terms which correspond to the direct couplings between the optical and microwave operators via the traced-out exciton modes.
We examine each equation separately below.

\vspace{1em}
\noindent\textbf{Co-rotating optical mode (signal)} 
Straightforward substitution yields
\begin{align*}
    D_a(\omega_\mathrm{pump} + \delta) \delta\hat a &= 
            \underbrace{\sum_n \frac{|g_n|^2}{D_n(\omega_\mathrm{pump}+\delta)}}_{-G^2K(\omega_\mathrm{pump}+\delta)} \delta\hat a
            + \underbrace{\sum_n \frac{\xi g_n \bar e_n}{D_n(\omega_\mathrm{pump}+\delta)}}_{\mathrm{Define:\ } \xi J(\delta)} \Big(\delta\hat b + \delta\hat b^\dagger\Big)
            + \underbrace{\sum_n \frac{ig_n\sqrt{\gamma}\delta\hat e_{n,\mathrm{in}}}{D_n(\omega_\mathrm{pump}+\delta)}}_{\hat F_a(\omega_\mathrm{pump}+\delta)}
            \\
            &\qquad\qquad
            + i\sqrt{\kappa_\mathrm{ext}}\delta\hat a_\mathrm{in,ext}
            + i\sqrt{\kappa_\mathrm{int}}\delta\hat a_\mathrm{in,int}.
\end{align*}
This has the same form as before with the addition of the new coupling to the microwave quadrature with rate
\begin{equation}
    J(\delta) = \sum_n \frac{g_n \bar e_n}{D_n(\omega_\mathrm{pump}+\delta)}.
\end{equation}
It is possible to relate this coupling rate back to the exciton response $K(\omega)$ as
\begin{align*}
    J(\delta) &= \bar a\sum_n \frac{|g_n|^2}{D_n(\omega_\mathrm{pump})D_n(\omega_\mathrm{pump}+\delta)}, \\
    &= \frac{\bar a}{\delta} \sum_n \bigg[ \frac{|g_n|^2}{D_n(\omega_\mathrm{pump}+\delta)} - \frac{|g_n|^2}{D_n(\omega_\mathrm{pump})} \bigg], \\
    &= -\frac{G^2\bar a}{\delta}\Big[K(\omega_\mathrm{pump}+\delta) - K(\omega_\mathrm{pump})\Big].
\end{align*}
The first line inserts $\bar e_n = g_n^*\bar a/D_n(\omega_\mathrm{pump})$, the second uses partial fractions, and the third identifies the two terms with the exciton response at different frequencies.

\vspace{1em}
\noindent\textbf{Counter-rotating optical mode (idler)}
Similarly
\begin{align*}
    D_a^*(\omega_\mathrm{pump} - \delta) \delta\hat a^\dagger &= 
            \underbrace{\sum_n \frac{|g_n|^2}{D_n^*(\omega_\mathrm{pump}-\delta)}}_{-G^2K^*(\omega_\mathrm{pump}-\delta)} \delta\hat a^\dagger
            + \underbrace{\sum_n \frac{\xi g_n^* \bar e_n^*}{D_n^*(\omega_\mathrm{pump}-\delta)}}_{\mathrm{Define:\ } \xi J^*(-\delta)} \Big(\delta\hat b + \delta\hat b^\dagger\Big)
            + \underbrace{\sum_n \frac{-ig_n^*\sqrt{\gamma}\delta\hat e_{n,\mathrm{in}}^\dagger}{D_n^*(\omega_\mathrm{pump}-\delta)}}_{[\hat F_a(\omega_\mathrm{pump}-\delta)]^\dagger}
            \\
            &\qquad\qquad
            - i\sqrt{\kappa_\mathrm{ext}}\delta\hat a_\mathrm{in,ext}^\dagger
            - i\sqrt{\kappa_\mathrm{int}}\delta\hat a_\mathrm{in,int}^\dagger.
\end{align*}

\vspace{1em}
\noindent\textbf{Co-rotating microwave mode}
This takes on a structurally similar form but with the addition of the extra conjugate from the idler part,
\begin{align*}
    D_b(\delta) \delta \hat b &= 
            \underbrace{\sum_n \frac{\xi\bar e_n^*g_n^*}{D_n(\omega_\mathrm{pump}+\delta)} }_{\mathrm{Define:\ } \xi\tilde J(\delta)}
                \delta\hat a
            + \underbrace{\sum_n \frac{\xi\bar e_ng_n}{D_n^*(\omega_\mathrm{pump}-\delta)} }_{\xi \tilde J^*(-\delta)}
                \delta\hat a^\dagger
            + \underbrace{\sum_n \bigg(
                    \frac{\xi^2|\bar e_n|^2}{D_n(\omega_\mathrm{pump}+\delta)}
                    +
                    \frac{\xi^2|\bar e_n|^2}{D_n^*(\omega_\mathrm{pump}-\delta)}
                \bigg)}_{\mathrm{Define:\ }\xi^2[\Pi(\delta) + \Pi^*(-\delta)]}
                \Big(\delta\hat b + \delta\hat b^\dagger\Big)
            \\
            &\qquad\qquad
            + \underbrace{\sum_n\frac{i\xi\bar e_n^*\sqrt{\gamma}\delta\hat e_{n,\mathrm{in}}}{D_n(\omega_\mathrm{pump}+\delta)}}_{\mathrm{Define:\ }\hat F_b(\omega_\mathrm{pump}+\delta)}
            + \underbrace{\sum_n\frac{-i\xi\bar e_n\sqrt{\gamma}\delta\hat e_{n,\mathrm{in}}^\dagger}{D_n^*(\omega_\mathrm{pump}-\delta)}}_{[\hat F_b(\omega_\mathrm{pump}-\delta)]^\dagger}
            + i\sqrt{\Gamma_\mathrm{ext}}\delta\hat b_\mathrm{in,ext}
            + i\sqrt{\Gamma_\mathrm{int}}\delta\hat b_\mathrm{in,int}.
\end{align*}
The microwave self-energy now has the form $\xi^2[\Pi(\delta) + \Pi^*(-\delta)]$ which also serves as an exciton-mediated coupling rate with the counter-rotating $\hat b^\dagger$.
Similar to the optical self-energy, this renormalizes the microwave frequency via the real part and the line width via the imaginary part.
It is straightforward to confirm that the commutator of the noise operators is
\begin{equation}
    [\hat F_b(\omega_\mathrm{pump}+\delta), (\hat F_b(\omega_\mathrm{pump}+\delta^\prime))^\dagger] = 2\xi^2\Im\Pi(\delta)\delta(\delta-\delta^\prime),
\end{equation}
as required by the fluctuation-dissipation theorem.
Using a similar partial fraction expansion, it is also possible to write the microwave self-energy in terms of the exciton response function as
\begin{equation*}
    \Pi(\delta) = - \frac{1}{\delta}\sum_n|\bar e_n|^2 - G^2|\bar a|^2\bigg[ \frac{K(\omega_\mathrm{pump}+\delta)  - K^*(\omega_\mathrm{pump}) }{\delta(\delta+i\gamma)} \bigg],
\end{equation*}
such that
\begin{equation}
    \xi^2\Big(\Pi(\delta) + \Pi^*(-\delta)\Big) = - \frac{\xi^2G^2|\bar a|^2}{\delta(\delta+i\gamma)}
    \Big[
        K(\omega_\mathrm{pump}+\delta) + K^*(\omega_\mathrm{pump}-\delta) - K(\omega_\mathrm{pump}) - K^*(\omega_\mathrm{pump})
    \Big].
\end{equation}
Likewise the optical coupling terms have rates
\begin{equation}
    \tilde J(\delta) = \sum_n \frac{g_n^*\bar e_n^*}{D_n(\omega_\mathrm{pump}+\delta)}
    = - \frac{G^2\bar a^*}{\delta + i\gamma}\Big[K(\omega_\mathrm{pump}+\delta) - K^*(\omega_\mathrm{pump})\Big].
\end{equation}
It is instructive to observe that the signal-microwave coupling is asymmetric with the signal equation coupling via $J(\delta)$ while the microwave equation couples back via $\tilde J(\delta) \neq J^*(\delta)$.
It is this asymmetry that leads to the nonreciprocal transduction response.
The physical origin of this effect is that the optical and microwave modes couple both dispersively via direct coupling through the coherent exciton population and dissipatively via their shared coupling to the exciton bath.
The dissipative coupling breaks the time-reversal symmetry and interferes with the coherent coupling asymmetrically to generate the nonreciprocity.
It should be noted that this holds in general for any response $K(\omega)$ and thus occurs even in the single-exciton system.
However, in the single-exciton case, a polariton with sufficient splitting has $D_n(\omega_-) \approx \omega_e - \omega_-$, where the $-i\gamma$ dissipative part of the coupling is negligible compared to the detuning such that $\tilde J(\omega) \approx J^*(\omega)$ and reciprocity is recovered.
In contrast, the disordered system will generally have excitons at frequencies $\omega_n$ such that $(\omega_n - \omega_-) \lesssim \gamma$ and thus the dissipative coupling becomes significant.

\vspace{1em}
\noindent\textbf{Counter-rotating microwave mode}
Similarly
\begin{align*}
    D_b^*(-\delta) \delta \hat b^\dagger &= 
            \underbrace{\sum_n \frac{\xi\bar e_n^*g_n^*}{D_n(\omega_\mathrm{pump}+\delta)} }_{\xi\tilde J(\delta)}
                \delta\hat a
            + \underbrace{\sum_n \frac{\xi\bar e_ng_n}{D_n^*(\omega_\mathrm{pump}-\delta)} }_{\xi \tilde J^*(-\delta)}
                \delta\hat a^\dagger
            + \underbrace{\sum_n \bigg(
                    \frac{\xi^2|\bar e_n|^2}{D_n(\omega_\mathrm{pump}+\delta)}
                    +
                    \frac{\xi^2|\bar e_n|^2}{D_n^*(\omega_\mathrm{pump}-\delta)}
                \bigg)}_{\xi^2[\Pi(\delta) + \Pi^*(-\delta)]}
                \Big(\delta\hat b + \delta\hat b^\dagger\Big)
            \\
            &\qquad\qquad
            + \underbrace{\sum_n\frac{i\xi\bar e_n^*\sqrt{\gamma}\delta\hat e_{n,\mathrm{in}}}{D_n(\omega_\mathrm{pump}+\delta)}}_{\hat F_b(\omega_\mathrm{pump}+\delta)}
            + \underbrace{\sum_n\frac{-i\xi\bar e_n\sqrt{\gamma}\delta\hat e_{n,\mathrm{in}}^\dagger}{D_n^*(\omega_\mathrm{pump}-\delta)}}_{[\hat F_b(\omega_\mathrm{pump}-\delta)]^\dagger}
            - i\sqrt{\Gamma_\mathrm{ext}}\delta\hat b_\mathrm{in,ext}^\dagger
            - i\sqrt{\Gamma_\mathrm{int}}\delta\hat b_\mathrm{in,int}^\dagger,
\end{align*}
which is identical in all but the leading denominator and the preexisting noise terms because the excitons entered as a quadrature.

\vspace{1em}
\noindent\textbf{Summary}
The full set of equations with the excitons eliminated from the dynamics are
\begin{subequations}
    \label{eq:si:disorder:dynamical-equations-exact}
    \begin{align}
        \mathcal D_a(\omega_\mathrm{pump}+\delta) \delta\hat a &=
            \xi J(\delta) \Big(\delta\hat b + \delta\hat b^\dagger\Big)
            + \hat F_a(\omega_\mathrm{pump} + \delta) 
            + i\sqrt{\kappa_\mathrm{ext}}\delta\hat a_\mathrm{in,ext}
            + i\sqrt{\kappa_\mathrm{int}}\delta\hat a_\mathrm{in,int}, \\
        \mathcal D_a^*(\omega_\mathrm{pump}-\delta) \delta\hat a^\dagger &=
            \xi J^*(-\delta) \Big(\delta\hat b + \delta\hat b^\dagger\Big)
            + [\hat F_a(\omega_\mathrm{pump} - \delta)]^\dagger 
            - i\sqrt{\kappa_\mathrm{ext}}\delta\hat a_\mathrm{in,ext}^\dagger
            - i\sqrt{\kappa_\mathrm{int}}\delta\hat a_\mathrm{in,int}^\dagger, \\
        \mathcal D_b(\delta) \delta\hat b &=
            \xi^2\Big[\Pi(\delta) + \Pi^*(-\delta)\Big] \delta\hat b^\dagger
            + \xi \tilde J(\delta) \delta\hat a
            + \xi \tilde J^*(-\delta) \delta\hat a^\dagger \nonumber\\
            &\qquad\qquad
            + \hat F_b(\omega_\mathrm{pump} + \delta) 
            + [\hat F_b(\omega_\mathrm{pump} - \delta)]^\dagger 
            + i\sqrt{\Gamma_\mathrm{ext}}\delta\hat b_\mathrm{in,ext}
            + i\sqrt{\Gamma_\mathrm{int}}\delta\hat b_\mathrm{in,int}, \\
        \mathcal D_b^*(-\delta) \delta\hat b^\dagger &=
            \xi^2\Big[\Pi(\delta) + \Pi^*(-\delta)\Big] \delta\hat b
            + \xi \tilde J(\delta) \delta\hat a
            + \xi \tilde J^*(-\delta) \delta\hat a^\dagger \nonumber\\
            &\qquad\qquad
            + \hat F_b(\omega_\mathrm{pump} + \delta) 
            + [\hat F_b(\omega_\mathrm{pump} - \delta)]^\dagger 
            - i\sqrt{\Gamma_\mathrm{ext}}\delta\hat b_\mathrm{in,ext}^\dagger
            - i\sqrt{\Gamma_\mathrm{int}}\delta\hat b_\mathrm{in,int}^\dagger ,
    \end{align}
\end{subequations}
where the generalized denominators are
\begin{subequations}
    \begin{align}
        \mathcal D_a(\omega) &= D_a(\omega) + G^2 K(\omega), \\
        \mathcal D_b(\delta) &= D_b(\delta) - \xi^2\Big[\Pi(\delta) + \Pi^*(-\delta)\Big]
    \end{align}
\end{subequations}
and the couplings are
\begin{subequations}
    \begin{align}
        J(\delta) &= 
            -\frac{G^2\bar a}{\delta} \Big[K(\omega_\mathrm{pump} + \delta) - K(\omega_\mathrm{pump})\Big], \\
        \tilde J(\delta) &= 
            -\frac{G^2\bar a^*}{\delta+i\gamma} \Big[K(\omega_\mathrm{pump} + \delta) - K^*(\omega_\mathrm{pump})\Big], \\
            \xi^2\Big[\Pi(\delta) + \Pi^*(-\delta)\Big] &= 
            -\frac{\xi^2G^2|\bar a|^2}{\delta(\delta+i\gamma)}
            \Big[
                K(\omega_\mathrm{pump}+\delta) + K^*(\omega_\mathrm{pump}-\delta) - K(\omega_\mathrm{pump}) - K^*(\omega_\mathrm{pump})
            \Big].
    \end{align}
\end{subequations}
The new noise operators have commutators
\begin{subequations}
    \label{eq:si:disorder:commutators-exact}
    \begin{align}
        [\hat F_a(\omega), \hat F_a(\omega^\prime)] = [\hat F_b(\omega), \hat F_b(\omega^\prime)] = [\hat F_a(\omega), \hat F_b(\omega^\prime)] &= 0, \\
        [\hat F_a(\omega), (\hat F_a(\omega^\prime))^\dagger] &= 
            -2 G^2\Im K(\omega)\delta(\omega-\omega^\prime), \\
        [\hat F_b(\omega_\mathrm{pump}+\delta), (\hat F_b(\omega_\mathrm{pump}+\delta^\prime))^\dagger] &= 
            2 \xi^2\Im \Pi(\delta)\delta(\delta-\delta^\prime), \\
        [\hat F_a(\omega_\mathrm{pump}+\delta), (\hat F_b(\omega_\mathrm{pump}+\delta^\prime))^\dagger] &= 
            -i\xi\Big[J(\delta) - \tilde J^*(\delta)\Big] \delta(\delta-\delta^\prime),
    \end{align}
\end{subequations}
where the last commutator is straightforwardly obtained by a partial fraction expansion.
This specifies the full system and enables straightforward numeric solutions given a known exciton response $K(\omega)$.
The ``exact'' values given in the main text Fig.~5(c,d) evaluate these equations using an exciton response $K(\omega)$ derived from a continuum density of states $\rho(\omega)$ that takes the assumed form of a Student's $t$-distribution.

\subsubsection{Linearized regime}
As before we can make approximations to make the interpretation more straightforward.
The first step is to linearize the optical denominators at the polariton frequency as we did before
\begin{equation}
    \mathcal D_a(\omega) \approx \frac{1}{|C|^2}\Big(\omega_- - \omega -i\frac{\tilde\kappa_-}{2}\Big),
\end{equation}
where $\omega_-$ is the LP frequency and $\tilde\kappa_-$ is the renormalized LP line width including the exciton loss.
As before, this assumes validity only within the vicinity of the polariton $\omega_-$.
Specifically, this assumes that the single-pole form holds out to as far as the idler at $\omega_i \approx \omega_- - 2\omega_b$.
For red-detuned pumping of the LP this is generally valid as both $K(\omega)$ and its derivatives are expected to asymptotically approach zero in the tails.
However, blue-detuned pumping of the LP (for amplification) or red-detuned pumping of the UP can be problematic as it requires the single-pole approximation to hold in a frequency band where $K(\omega)$ may change substantially.

The remaining coupling coefficients and noise terms depend on only the slow frequency $\delta$ via the bare exciton denominators as $1/D_n(\omega_\mathrm{pump}\pm\delta)$.
The remaining terms thus only vary over scales on the order of the homogeneous line width $\gamma$.
Transduction is generally limited to the microwave bandwidth $\Gamma\ll\gamma$ and so for all relevant probe frequencies $\delta$, the variation of the coefficients is of order $|\delta - \omega_b|/\gamma \ll 1$.
Consequently we can approximate the coefficients by expanding to zeroth order with $\delta = \omega_b$.

As before we take the expansion with $K(\omega;\gamma) \to K(\omega_*;0^+)$ as given by Eq.~\eqref{eq:si:disorder:k-expansion-zeroth-order} where $\omega_*$ is an arbitrary expansion point.
Here we will expand with $\delta = \omega_b$ at the pump $\omega_\mathrm{pump}$, signal $\omega_\mathrm{pump} + \omega_b = \omega_-$, and idler $\omega_\mathrm{pump} - \omega_b = \omega_i$ frequencies.
We find
\begin{align}
    \label{eq:si:disorder:linearized-coupling-1}
    J(\delta) &\approx 
        - G^2\bar a \bigg[\frac{\Re K(\omega_\mathrm{pump} + \omega_b) - \Re K(\omega_\mathrm{pump})}{\omega_b} \bigg] 
        + i \frac{\pi G^2\bar a}{\omega_b} \Big[\rho(\omega_\mathrm{pump} + \omega_b) - \rho(\omega_\mathrm{pump})\Big], \nonumber\\
    &\approx 
        - G^2\bar a \pdv{\Re K(\omega_-)}{\omega}
        + i \frac{\pi G^2\bar a}{\omega_b} \Big[\rho(\omega_-) - \rho(\omega_\mathrm{pump})\Big], \nonumber\\
    &\approx
        \bar a\frac{|X|^2}{|C|^2} + i \frac{\pi G^2\bar a}{\omega_b} \Big[\rho(\omega_-) - \rho(\omega_\mathrm{pump})\Big],
\end{align}
where the first line separates the real and imaginary parts via Eq.~\eqref{eq:si:disorder:k-expansion-zeroth-order}, the second identifies the real part finite difference with the derivative evaluated at the polariton, and the third uses the fact that $G^2\Re(\pdv*{K(\omega_-)}{\omega}) = G^2\mathrm{PV}_{\rho^\prime} = -|X|^2/|C|^2$.
The approximation of the finite difference as the derivative carries an error of order $\omega_b/|\omega_\mathrm{pump} - \bar\omega|\ll 1$ which can be seen from the leading order approximation of $\Re K = \mathrm{PV}_\rho$ in Eq.~\eqref{eq:si:disorder:pv-rho-approximate}.
Similarly one finds
\begin{equation}
    \label{eq:si:disorder:linearized-coupling-2}
    \tilde J(\delta) \approx \bar a^*\frac{|X|^2}{|C|^2} + i\frac{\pi G^2\bar a^*}{\omega_b}\Big[\rho(\omega_-) + \rho(\omega_\mathrm{pump})\Big],
\end{equation}
where we take the additional approximation of $\omega_b + i\gamma \approx \omega_b$ to drop the $i\gamma$ in the leading coefficient.

The microwave self-energy term can be approximated
\begin{align}
    \xi^2\Big[\Pi(\delta) + \Pi^*(-\delta)\Big] &\approx 
         -\frac{\xi^2G^2|\bar a|^2}{\omega_b^2}
        \Big[
            \Re K(\omega_\mathrm{pump}+\omega_b) + \Re K^*(\omega_\mathrm{pump}-\omega_b) - 2\Re K(\omega_\mathrm{pump})
        \Big] \nonumber\\
        &\qquad\qquad + i\frac{\pi \xi^2 G^2|\bar a|^2}{\omega_b^2}\Big[\rho(\omega_\mathrm{pump}+\omega_b) - \rho(\omega_\mathrm{pump}-\omega_b)\Big]
    \nonumber\\
    &\approx 
        -\xi^2G^2|\bar a|^2 \pdv[2]{\Re K(\omega_\mathrm{pump})}{\omega}
        + i\frac{\pi \xi^2 G^2|\bar a|^2}{\omega_b^2}\Big[\rho(\omega_-) - \rho(\omega_i)\Big],
    \nonumber\\
    &\approx 
        -\frac{2\xi^2G^2|\bar a|^2}{(\omega_\mathrm{pump} - \bar\omega)^3}
        + i\frac{\pi \xi^2 G^2|\bar a|^2}{\omega_b^2}\Big[\rho(\omega_-) - \rho(\omega_i)\Big],        
\end{align}
where the first line splits the real and imaginary parts, the second approximates real part's centered second difference with the second derivative of $\Re K(\omega)$, and the third uses Eq.~\eqref{eq:si:disorder:pv-rho-approximate} to approximate the derivative as
\begin{equation*}
    \pdv[2]{\Re K(\omega_\mathrm{pump})}{\omega} \approx \left.\pdv[2]{\omega}\bigg(\frac{1}{\omega - \bar\omega}\bigg)\right|_{\omega=\omega_\mathrm{pump}}
    = \frac{2}{(\omega_\mathrm{pump} - \bar\omega)^3}.
\end{equation*}
From this we can identify the renormalized microwave frequency $\tilde\omega_b = \omega_b + \delta\omega_b$ and line width $\tilde\Gamma = \Gamma + \delta\Gamma$ where
\begin{subequations}
    \begin{align}
        \delta\omega_b &\approx \frac{2\xi^2G^2|\bar a|^2}{(\omega_\mathrm{pump} - \bar\omega)^3} , \\
        \delta\Gamma &\approx \frac{2\pi \xi^2 G^2|\bar a|^2}{\omega_b^2}\Big[\rho(\omega_-) - \rho(\omega_i)\Big].
    \end{align}
\end{subequations}
Given $G\approx|\omega_\mathrm{pump} - \bar\omega|$ the frequency shift is roughly $\delta\omega_b\approx -2\xi^2|\bar a|^2 / G$.
With this the microwave denominator can be generalized as
\begin{equation}
    \mathcal D_b(\delta) = \tilde\omega_b - \delta - i \frac{\tilde\Gamma}{2}.
\end{equation}
With all of these approximations, the system of equations reduces to a linear system with constant coefficients and generalized denominators in the form of simple poles with renormalized frequencies and line widths.

\subsubsection{Transduction efficiency and noise}
We now analytically compute the transduction efficiency and noise in the linearized regime.
Note that these solutions are strictly valid in the limit where the density of states $\rho(\omega)$ is continuous and small near the frequencies of interest.
The discrete case fails for reasons that will be discussed in the next sections.
To facilitate the analytical solution, we first take the RWA to drop the counter-rotating microwave mode $\hat b^\dagger$ from the dynamics.
This is valid in the limit where $\tilde\Gamma/2 \ll 2\omega_b$ which manifestly holds for high-$Q$ microwave modes.
The consolidated dynamical equations follow from the linearization simplifications of Eqs.~\eqref{eq:si:disorder:dynamical-equations-exact} as
\begin{subequations}
    \begin{align}
        \mathcal D_a(\omega_\mathrm{pump}+\delta) \delta\hat a &=
            \xi J(\omega_b) \delta\hat b
            + \hat F_a(\omega_-) 
            + i\sqrt{\kappa_\mathrm{ext}}\delta\hat a_\mathrm{in,ext}
            + i\sqrt{\kappa_\mathrm{int}}\delta\hat a_\mathrm{in,int}, \\
        \mathcal D_a^*(\omega_\mathrm{pump}-\delta) \delta\hat a^\dagger &=
            \xi J^*(-\omega_b) \delta\hat b
            + [\hat F_a(\omega_i)]^\dagger 
            - i\sqrt{\kappa_\mathrm{ext}}\delta\hat a_\mathrm{in,ext}^\dagger
            - i\sqrt{\kappa_\mathrm{int}}\delta\hat a_\mathrm{in,int}^\dagger, \\
        \mathcal D_b(\delta) \delta\hat b &=
            \xi \tilde J(\omega_b) \delta\hat a
            + \xi \tilde J^*(-\omega_b) \delta\hat a^\dagger
            + \hat F_b(\omega_-) 
            + [\hat F_b(\omega_i)]^\dagger
            + i\sqrt{\Gamma_\mathrm{ext}}\delta\hat b_\mathrm{in,ext}
            + i\sqrt{\Gamma_\mathrm{int}}\delta\hat b_\mathrm{in,int}.
    \end{align}
\end{subequations}
The first step is to eliminate the idler
\begin{equation}
    \delta\hat a^\dagger = 
        \frac{\xi J^*(-\omega_b)}{\mathcal D_a^*(\omega_\mathrm{pump}-\delta)} \delta\hat b
        + \frac{[\hat F_a(\omega_i)]^\dagger}{\mathcal D_a^*(\omega_\mathrm{pump}-\delta)}
        - i\frac{\sqrt{\kappa_\mathrm{ext}}}{\mathcal D_a^*(\omega_\mathrm{pump}-\delta)}\delta\hat a_\mathrm{in,ext}^\dagger
        - i\frac{\sqrt{\kappa_\mathrm{int}}}{\mathcal D_a^*(\omega_\mathrm{pump}-\delta)}\delta\hat a_\mathrm{in,int}^\dagger
\end{equation}
which enters the microwave equation as a second self-energy term and additional noise terms (dropping frequency arguments where unambiguous)
\begin{align}
    \bigg[\mathcal D_b(\delta) - \frac{\xi^2 J^*(-\omega_b) \tilde J^*(-\omega_b)}{\mathcal D_a^*}\bigg] \delta\hat b &=
        \xi \tilde J(\omega_b) \delta\hat a
        - i\frac{\xi \tilde J^*\sqrt{\kappa_\mathrm{int}}}{\mathcal D_a^*} \delta\hat a_\mathrm{in,int}^\dagger
        - i\frac{\xi \tilde J^*\sqrt{\kappa_\mathrm{ext}}}{\mathcal D_a^*} \delta\hat a_\mathrm{in,ext}^\dagger
        + \frac{\xi \tilde J^*[\hat F_a(\omega_i)]^\dagger}{\mathcal D_a^*}
        \nonumber\\
        &\qquad\qquad
        + \hat F_b(\omega_-) 
        + [\hat F_b(\omega_i)]^\dagger
        + i\sqrt{\Gamma_\mathrm{ext}}\delta\hat b_\mathrm{in,ext}
        + i\sqrt{\Gamma_\mathrm{int}}\delta\hat b_\mathrm{in,int}.
\end{align}
The self-energy term $\Sigma_\mathrm{idler} \approx \xi^2 J^*\tilde J^*/\mathcal D_a^*(\omega_i)$ is suppressed as $\tilde\kappa_-/4\omega_b \ll 1$ with weak dispersion in the sideband-resolved regime and so we can absorb it as a small second renormalization to the microwave pole
\begin{equation}
    \tilde{\mathcal D}_b(\delta) = \mathcal D_b(\delta) - \Sigma_\mathrm{idler} = 
        (\omega_b + \delta\omega_b - \Re \Sigma_\mathrm{idler}) - \delta - i\frac{(\Gamma + \delta\Gamma + 2\Im\Sigma_\mathrm{idler})}{2}.
\end{equation}

From here the analysis proceeds as usual: the microwave equation can be solved to eliminate $\hat b$ from the signal equation $\delta\hat a$ to determine the output extrinsic field.
Now, since both $\mathcal D_a(\omega_\mathrm{pump} + \delta)$ and $\tilde{\mathcal D}_b(\delta)$ are resonant when $\delta \approx \omega_b$ the self-energy terms are genuine couplings.
The resulting output optical field is
\begin{align}
    \delta\hat a_\mathrm{out,ext} =&{}
        -\frac{i\xi^2 J\tilde J^*\sqrt{\kappa_\mathrm{ext}\kappa_\mathrm{int}}}{\mathcal D_a^*(\mathcal D_a\tilde{\mathcal D}_b - \xi^2 J \tilde J)} \delta\hat a_\mathrm{in,int}^\dagger
        -\frac{i\xi^2 J\tilde J^*\kappa_\mathrm{ext}}{\mathcal D_a^*(\mathcal D_a\tilde{\mathcal D}_b - \xi^2 J \tilde J)} \delta\hat a_\mathrm{in,ext}^\dagger
        +\frac{\xi^2 J\tilde J^*\sqrt{\kappa_\mathrm{ext}}[\hat F_a(\omega_i)]^\dagger}{\mathcal D_a^*(\mathcal D_a\tilde{\mathcal D}_b - \xi^2 J \tilde J)}
        \nonumber\\
        &
        +\frac{i\xi J\sqrt{\kappa_\mathrm{ext}\Gamma_\mathrm{int}}}{(\mathcal D_a\tilde{\mathcal D}_b - \xi^2 J \tilde J)} \delta\hat b_\mathrm{in,int}
        +\frac{i\xi J\sqrt{\kappa_\mathrm{ext}\Gamma_\mathrm{ext}}}{(\mathcal D_a\tilde{\mathcal D}_b - \xi^2 J \tilde J)} \delta\hat b_\mathrm{in,ext}
        \nonumber\\
        &
        +\frac{\xi J\sqrt{\kappa_\mathrm{ext}} \hat F_b(\omega_-)}{(\mathcal D_a\tilde{\mathcal D}_b - \xi^2 J \tilde J)}
        +\frac{\xi J\sqrt{\kappa_\mathrm{ext}}[\hat F_b(\omega_i)]^\dagger}{(\mathcal D_a\tilde{\mathcal D}_b - \xi^2 J \tilde J)}
        \nonumber\\
        &
        + \bigg[1 + \frac{i\kappa_\mathrm{ext}\tilde{\mathcal D}_b}{(\mathcal D_a\tilde{\mathcal D}_b - \xi^2 J \tilde J)}\bigg] \delta\hat a_\mathrm{in,ext}
        + \frac{i\sqrt{\kappa_\mathrm{ext}\kappa_\mathrm{int}}\tilde{\mathcal D}_b }{(\mathcal D_a\tilde{\mathcal D}_b - \xi^2 J \tilde J)} \delta\hat a_\mathrm{in,int}
        + \frac{\sqrt{\kappa_\mathrm{ext}}\tilde{\mathcal D}_b \hat F_a(\omega_-)}{(\mathcal D_a\tilde{\mathcal D}_b - \xi^2 J \tilde J)}.
\end{align}
from which the microwave-to-optical scattering matrix element can be read off as
\begin{equation}
    S_{ab}(\delta) = \frac{i\xi J(\omega_b) \sqrt{\kappa_\mathrm{ext}\Gamma_\mathrm{ext}}}{\mathcal D_a(\omega_\mathrm{pump}+\delta)\tilde{\mathcal D}_b(\delta) - \xi^2 J(\omega_b) \tilde J(\omega_b)},
\end{equation}
which gives the transduction efficiency as $\eta_{ab} = |S_{ab}|^2$.
The output noise flux follows as $n_{a,\mathrm{out}} = \ev*{\delta\hat a_\mathrm{out,ext}^\dagger \delta\hat a_\mathrm{out,ext}}$.
Subtracting the transduction path's noise, dividing by the efficiency $\eta_{ab}$, and simplifying gives the added noise
\begin{align}
    n_{a,\mathrm{add}} &= 
        \underbrace{\frac{\Gamma_\mathrm{int}}{\Gamma_\mathrm{ext}} n_\mathrm{th}}_{\mathrm{Thermal}}
        + \underbrace{\frac{\xi^2|\tilde J(-\omega_b)|^2 (\kappa_\mathrm{int} + \kappa_\mathrm{ext})}{\Gamma_\mathrm{ext}|\mathcal D_a(\omega_\mathrm{pump}-\delta)|^2}}_{\mathrm{Optical\ amplification}}
        + \underbrace{\frac{2\pi \xi^2 G^2 |\tilde J(-\omega_b)|^2}{\Gamma_\mathrm{ext}|\mathcal D_a(\omega_\mathrm{pump}-\delta)|^2} \rho(\omega_i)}_{\mathrm{Exciton\ idler\ amplification}}
        \nonumber\\
        &\qquad
        + \underbrace{\frac{2\xi^2}{\Gamma_\mathrm{ext}}\Im\bigg[\frac{\tilde J(-\omega_b)}{\mathcal D_a(\omega_\mathrm{pump}-\delta)} \Big(J(-\omega_b) - \tilde J^*(-\omega_b) \Big) \bigg]}_{\mathrm{Idler\ amplification\ interference}}
        + \underbrace{\frac{2\xi^2}{\Gamma_\mathrm{ext}} \Im\Pi(-\omega_b)}_{\mathrm{Ex.\ pump/idler\ amp.}},
\end{align}
where we dropped the normal-ordered optical terms since their occupation is zero $\ev{\hat a^\dagger \hat a}=0$, utilized the commutators Eqs.~\eqref{eq:si:disorder:commutators-exact} as, e.g., $\ev{\hat a \hat a^\dagger} = [\hat a,\hat a^\dagger]$, and kept the microwave thermal occupation $\ev*{(\hat b(\omega))^\dagger\hat b(\omega^\prime)} = n_\mathrm{th}\delta(\omega-\omega^\prime)$.
The first two terms correspond to the typical thermal and idler amplification noise sources in the non-disordered system.
The remaining terms correspond to noise injected by the exciton at either the pump/idler, and the interference of that noise with the optical pathway at the idler frequency.
In principle, similar exciton noise terms exist also in the non-disordered case, arising from the far-detuned homogeneous tail of the exciton (they were effectively dropped in the earlier analysis when the UP was removed from the dynamics).
In the disordered system, such terms can be more significant as exciton states may exist near the idler/pump and inject noise directly.

The analysis can also be done in the opposite direction for optical-to-microwave transduction with some subtle changes.
First the transduction efficiency becomes
\begin{equation}
    S_{ba}(\delta) = \frac{i\xi \tilde J(\omega_b) \sqrt{\kappa_\mathrm{ext}\Gamma_\mathrm{ext}}}{\mathcal D_a(\omega_\mathrm{pump}+\delta)\tilde{\mathcal D}_b(\delta) - \xi^2 J(\omega_b) \tilde J(\omega_b)},
\end{equation}
which is identical up to the swap of $J(\omega_b)\to\tilde J(\omega_b)$ in the numerator.
This is the origin of the nonreciprocal transduction efficiencies since $\tilde J \neq J^*$.
The noise is also similarly calculated but takes on a slightly different form
\begin{align}
    n_{b,\mathrm{add}} &= 
        \underbrace{\frac{1}{\eta_{ba}(\delta)} \bigg|1 + i\frac{\Gamma_\mathrm{ext}\mathcal D_a(\omega_\mathrm{pump}+\delta)}{\mathcal D_a(\omega_\mathrm{pump}+\delta)\tilde{\mathcal D}_b(\delta) - \xi^2 J(\omega_b) \tilde J(\omega_b)}\bigg|^2 n_\mathrm{th}}_{\mathrm{Extrinsic\ thermal\ (reflection)}}
        + \underbrace{\frac{1}{\eta_{ba}(\delta)}\frac{\Gamma_\mathrm{ext}\Gamma_\mathrm{int}|\mathcal D_a(\omega_\mathrm{pump}+\delta)|^2}{|\mathcal D_a(\omega_\mathrm{pump}+\delta)\tilde{\mathcal D}_b(\delta) - \xi^2 J(\omega_b) \tilde J(\omega_b)|^2} n_\mathrm{th}}_{\mathrm{Intrinsic\ thermal}}
        \nonumber\\
        &\qquad
        + \underbrace{R\frac{\xi^2|\tilde J(-\omega_b)|^2 (\kappa_\mathrm{int} + \kappa_\mathrm{ext})}{\Gamma_\mathrm{ext}|\mathcal D_a(\omega_\mathrm{pump}-\delta)|^2}}_{\mathrm{Optical\ amplification}}
        + \underbrace{R\frac{2\pi \xi^2 G^2 |\tilde J(-\omega_b)|^2}{\Gamma_\mathrm{ext}|\mathcal D_a(\omega_\mathrm{pump}-\delta)|^2} \rho(\omega_i)}_{\mathrm{Exciton\ idler\ amplification}}
        \nonumber\\
        &\qquad
        + \underbrace{R\frac{2\xi^2}{\Gamma_\mathrm{ext}}\Im\bigg[\frac{\tilde J(-\omega_b)}{\mathcal D_a(\omega_\mathrm{pump}-\delta)} \Big(J(-\omega_b) - \tilde J^*(-\omega_b) \Big) \bigg]}_{\mathrm{Idler\ amplification\ interference}}
        + \underbrace{R\frac{2\xi^2}{\Gamma_\mathrm{ext}} \Im\Pi(-\omega_b)}_{\mathrm{Ex.\ pump/idler\ amp.}}.
\end{align}
The first two terms are the thermal noise injected from the extrinsic and intrinsic microwave reservoirs coupling out via the extrinsic path.
They take the form of the pure microwave reflection and transmission scaled by $1/\eta_{ba}(\delta)$.
The second and third lines are the exact same amplification terms from the optical noise $n_{a,\mathrm{add}}$ but are scaled by
\begin{equation}
    R = \frac{\Gamma_\mathrm{ext} |\mathcal D_a(\omega_\mathrm{pump}+\delta)|^2}{\kappa_\mathrm{ext} \xi^2 |J(\omega_b)|^2} \frac{\eta_{ab}}{\eta_{ba}}
    = \frac{\Gamma_\mathrm{ext} |\mathcal D_a(\omega_\mathrm{pump}+\delta)|^2}{\kappa_\mathrm{ext} \xi^2 |\tilde J(\omega_b)|^2}.
\end{equation}
This scaling factor exists because all of the amplification noise derives from the generation of an idler and microwave photon pair.
For this to become optical noise it must be transduced into the optical frequency and emitted into the extrinsic channel.
However, the microwave noise can leak directly out of the extrinsic channel and so its amplification noise terms need to be rescaled accordingly.
Starting with the optical output, undoing the emission divides by $\kappa_\mathrm{ext}$.
Further dividing by $|\xi J(\omega_b)/\mathcal D_a|^2$ undoes the up-conversion of the microwave noise quanta into the optical signal frequency.
Multiplying by $\Gamma_\mathrm{ext}$ corresponds to the emission of the microwave noise into the microwave extrinsic channel.
Lastly, rescaling the normalization by the ratio of the efficiencies gives the final added noise.

\subsubsection{Interpretation}
We will focus on the microwave-to-optical transduction process going forward.
Defining the susceptibilities
\begin{subequations}
    \begin{align}
        \chi_-(\delta) &= \frac{-i\tilde \kappa_-}{2|C|^2\mathcal D_a(\omega_\mathrm{pump}+\delta)}, \\
        \chi_b(\delta) &= \frac{-i\tilde \Gamma}{2\tilde{\mathcal D}_b(\delta)},
    \end{align}
\end{subequations}
the transduction efficiency can be written
\begin{equation}
    \eta_{ab}(\delta) = 4 \tilde\eta_- \tilde\eta_b \frac{|\big(J(\omega_b)/\tilde J(\omega_b)\big)\tilde{\mathcal C}| \, |\chi_-(\delta)|^2|\chi_b(\delta)|^2}{|1 + \tilde{\mathcal C} \chi_-(\delta)\chi_b(\delta)|^2},
\end{equation}
where $\tilde\eta_- = |C|^2\kappa_\mathrm{ext}/\tilde\kappa_-$ and $\tilde\eta_b = \Gamma_\mathrm{ext}/\tilde\Gamma$ are the output coupling efficiencies for the polariton and microwave mode using the re-normalized line widths, respectively.
The effective cooperativity is
\begin{equation}
    \tilde{\mathcal C} \equiv \frac{4\xi^2|C|^2J(\omega_b)\tilde J(\omega_b)}{\tilde\kappa_-\tilde\Gamma},
\end{equation}
which is now generally complex valued which slightly skews the response in frequency.
Nevertheless, the transduction response is essentially the same as the non-disordered system except with the line widths broadened to the renormalized values.

In the limit that the density of states is effectively flat over the frequency band of interest, i.e. $\rho(\omega_-) \approx \rho(\omega_\mathrm{pump}) \approx \rho(\omega_i)$, the coupling rates Eqs.~\eqref{eq:si:disorder:linearized-coupling-1} and~\eqref{eq:si:disorder:linearized-coupling-2} are approximated as
\begin{align*}
    J(\omega_b)
        &\approx \bar a \frac{|X|^2}{|C|^2}, \\
    \tilde J(\omega_b)
        &\approx \bar a^* \frac{|X|^2}{|C|^2} + i \frac{2 \pi G^2 \bar a^*}{\omega_b} \rho(\omega_-).
\end{align*}
The transduction efficiency reduces to
\begin{equation}
    \eta_{ab}(\delta) = 4 \tilde\eta_- \tilde\eta_b \frac{\mathcal C |\chi_-(\delta)|^2|\chi_b(\delta)|^2}{|1 + \mathcal C ( 1 + i\theta) \chi_-(\delta)\chi_b(\delta)|^2},
\end{equation}
where $\mathcal C = 4\xi^2|X|^4 |\bar a_-|^2 / \tilde \kappa_- \tilde\Gamma$ is the clean system cooperativity with renormalized line widths and polariton occupation $|\bar a_-|^2 = |\bar a|^2 / |C|^2$.
The imaginary part is
\begin{equation}
    \theta = \frac{2\pi G^2 |C|^2 }{|X|^2 \omega_b} \rho(\omega_-),
\end{equation}
reproducing the result from Eq.~(8) in the main text.
The noise follows from these simplifications
\begin{align}
    n_\mathrm{add} &= 
        \underbrace{
            \frac{\Gamma_\mathrm{int}}{\Gamma_\mathrm{ext}}n_\mathrm{th}
        }_{\mathrm{Thermal}}
        +
        \underbrace{
            \frac{4\xi^2|\bar a|^2 |X|^4}{|C|^2\tilde\kappa_-\Gamma_\mathrm{ext}}|\chi_-(-\omega_b)|^2
        }_{\mathrm{Optical\ idler\ amp.}}
        -
        \underbrace{
            \frac{32\pi G^2 \xi^2|\bar a|^2|X|^2}{\tilde\kappa_-^2\Gamma_\mathrm{ext}}\rho(\omega_i) |\chi_-(-\omega_b)|^2
        }_{\mathrm{Idler\ interference}}
        +
        \underbrace{
            \frac{2\pi G^2\xi^2}{\Gamma_\mathrm{ext}} \frac{|\bar a|^2}{\omega_b^2} \Big[ \rho(\omega_i) + \rho(\omega_\mathrm{pump}) \Big]
        }_{\mathrm{Exciton\ idler\ and\ pump\ amp.}},
\end{align}
which reproduces the result from Eq.~(9) in the main text.
In the limit that $G^2\rho(\omega_-)\to 0$, the line widths reduce to their homogeneous values, the couplings become symmetric $J(\omega_b) = \tilde J^*(\omega_b)$, and the exciton loss terms vanish, reducing the efficiency and noise to the non-disordered system values.
Thus the transduction process is cavity protected in the same manner as the polariton itself.

\clearpage
\subsection{Discrete limit}
As discussed previously, the exciton dynamics enters only through the exciton response $K(\omega)$ which can generally be approximated to within some finite error by a large number of distinct density of states $\rho(\omega)$.
Thus, we argued that $K(\omega)$ is best treated as a measured quantity in practice, rather than something to derive from a microscopic model of the disorder center-of-mass (CM) potential $V(\vb R)$.
Our prior analysis, including the linearized model presented in the main text, did so by assuming that the exciton response $K(\omega)$ and underlying density of states $\rho(\omega)$ possessed smooth asymptotic behavior.
However, such assumptions may not hold in practice, as will be discussed in this section.

\begin{figure}[b]
    \centering
    \includegraphics{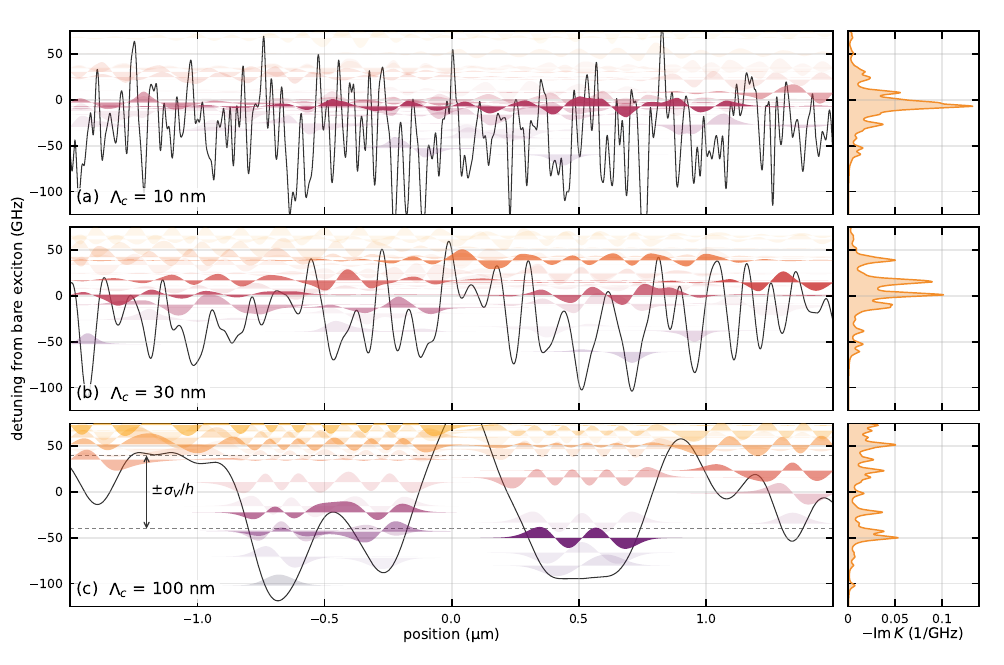}
    \caption{
        \textbf{Center-of-mass exciton states with disorder.}
        The eigenstates (left) and representative dissipative exciton response $-\Im K(\omega)$ (right) for a single instance of a Gaussian noise CM potential profile $V(\vb R)/h$ with Gaussian correlation length $\Lambda_c$ of (a) 10\,nm, (b) 30\,nm, and (c) 100\,nm.
        The noise has standard deviation $\sigma_V/h = 40$\,GHz and the overlap is taken against an optical mode with an effective wavelength of 250\,nm (roughly corresponding to the actual wavelength).
        All $K(\omega)$ assume a homogeneous broadening of $\gamma=5$\,GHz.
    }
    \label{fig:si-disorder-states}
\end{figure}

\subsubsection{Structure of disorder in the low-energy tail}
To estimate the exciton response directly, we utilize the standard model~\cite{savona2007disorder} that the CM potential $V(\vb R)$ takes on the form of a Gaussian stochastic field with correlation $\ev{V(\vb R)V(\vb R^\prime)} = \sigma_V^2 f(|\vb R - \vb R^\prime|)$ where $\sigma_V^2$ is the Gaussian variance and $f(R)$ is a unit-normalized correlation function with characteristic length scale $\Lambda_c$.
For our purposes we will assume $f(R)=\exp(-R^2/(2\Lambda_c^2))$ such that the noise is Gaussian correlated.

Figure~\ref{fig:si-disorder-states} shows the calculated CM energy eigenstates and the corresponding exciton response function $K(\omega)$ for three different 1-d CM potential realizations $V(\vb R)$ (more comment on the 2-d case below).
The potentials are Gaussian random fields with zero mean, standard deviation $\sigma_V/h = 40$\,GHz, and varied correlation length $\Lambda_c$.
The wavefunctions (plotted at their respective energies) have opacity that corresponds to their optical coupling weight $p_n = |g_n|^2/G^2$.

\begin{figure}
    \centering
    \includegraphics{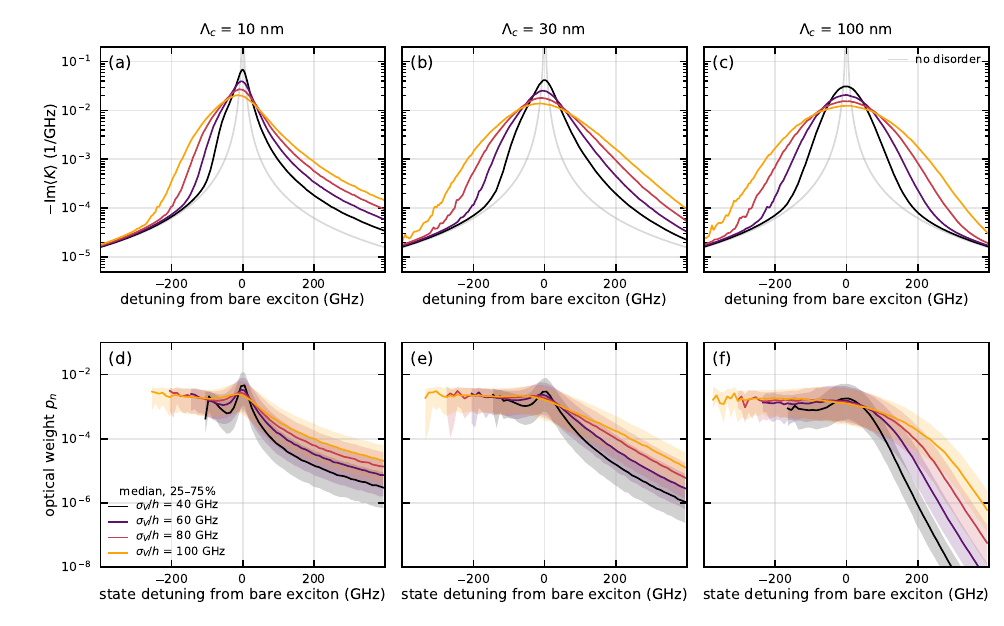}
    \caption{
        \textbf{Inhomogeneous broadening in 1-d.}
        Ensemble-averaged dissipative exciton response $-\Im\ev{K(\omega)}$ of the Gaussian potential profile over 1000 realizations at each potential width $\sigma_V/h = \{40,60,80,100\}$\,GHz and with Gaussian correlation lengths $\Lambda_c$ of (a) 10\,nm, (b) 30\,nm, and (c) 100\,nm.
        (d--f) The median and quartiles of the optical weight distribution as a function of the state's shift for each correlation length $\Lambda_c$ and amplitude $\sigma_V$.
    }
    \label{fig:si-disorder-ensemble}
\end{figure}

The short-correlation-length case $\Lambda_c = 10$\,nm (Fig.~\ref{fig:si-disorder-states}a) can be considered the ``quantum limit'' in which the exciton cannot be readily confined within any local minima as the kinetic energy scale $\hbar^2/2M\Lambda_c^2 \gg \sigma_V$.
In this limit, the bound states are spread over multiple nearby local minima which averages the effective potential preventing the formation of deep bound states and shortening the low-energy tail.
In contrast, the high-spatial-frequency components of the potential enables mixing of the high-momentum states of the exciton dispersion which extends the high-frequency tail of the inhomogeneous distribution.
This is shown in Fig.~\ref{fig:si-disorder-ensemble}a in which the ensemble averaged exciton response $-\Im\ev{K(\omega)}$ has a strongly asymmetric profile.
Nevertheless, the fast oscillations of the potential are averaged out which appears as a motional narrowing of the exciton peak~\cite{savona2007disorder,whittaker1996motional}.

The long-correlation-length case $\Lambda_c = 100$\,nm (Fig.~\ref{fig:si-disorder-states}c) corresponds to the situation where the potential variations are slow enough that the local minima support a large number of locally bound states, i.e.\ $\hbar^2/2M\Lambda_c^2 \ll \sigma_V$.
In this limit the states effectively follow the Gaussian distribution of the potential which reproduces a symmetric, Gaussian distribution in the inhomogeneous broadening Fig.~\ref{fig:si-disorder-ensemble}c.
The intermediate case $\Lambda_c = 30$\,nm (Fig.~\ref{fig:si-disorder-states}b) is a combination of both wherein some local minima can host individual deeply bound states.

Figure~\ref{fig:si-disorder-ensemble}d--f show the distribution of the optical weight against the frequency detuning of each exciton state.
The weight of the median state tends to plateau on the low-frequency end of the distribution irrespective of the specific parameters that generated the distribution.
In the classical limit $\Lambda_c \gtrsim 30$\,nm, this can be understood by the fact that these states arise as the low-energy states of the local minima of the potential.
Since the spatial correlation forces most local minima to take a similar shape (with different absolute energies), the states within each local minima tend to have similar spatial profile and thus similar overlap with the optical mode.
Thus, the low-energy tail of the exciton response does not generally smoothly asymptote but is instead formed from a handful of mostly isolated peaks with similar coupling strengths and a frequency distribution that roughly samples the underlying distribution.

An important caveat to the modeling thus far is that we have only considered 1-d motion.
The treatment of the disorder is straightforwardly extended into 2-d, however there are a few notable differences.
First is that the extra dimension allows for more local minima to form per unit length along the waveguide axis.
This manifests as an increase in the number of states, which now sample the distribution more densely with weaker individual couplings.
The other significant difference comes from the well-known fact that the bound states of the potential are more weakly bound in 2-d compared to 1-d.
As a result, states only weakly bound within local minima in 1-d can escape causing the low-energy tails to diminish in comparison to otherwise identical disorder in 1-d.
This effect is most prominent in the short-correlation-length potentials where all deeply bound states are only weakly bound.

\begin{figure}
    \centering
    \includegraphics{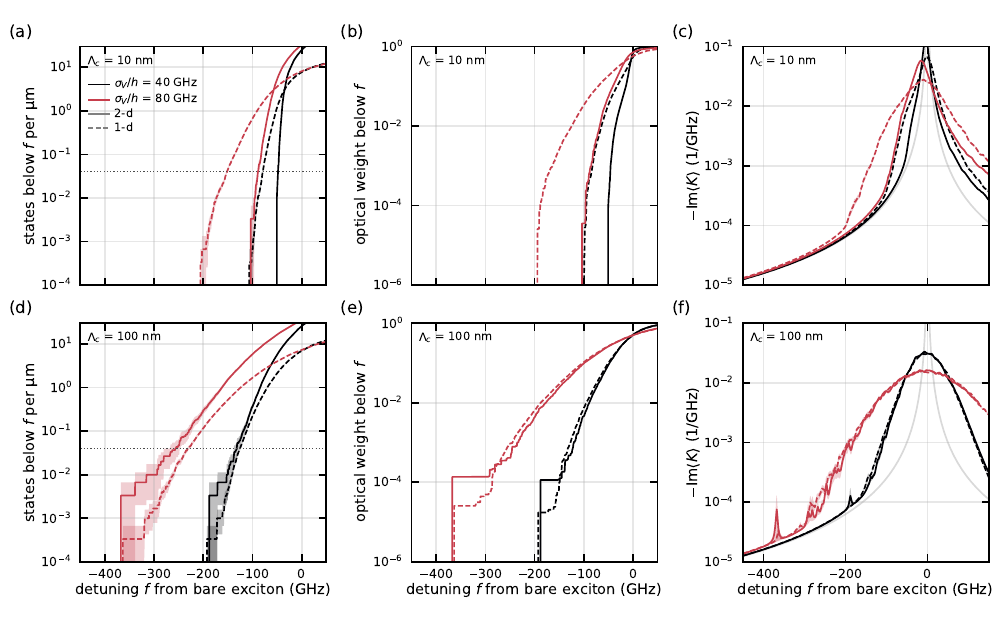}
    \caption{
        \textbf{Disorder in 1-d versus 2-d.}
        All plots directly compare the ensemble averaged statistics of the 1-d (dashed) and 2-d (solid) disorder problem.
        (a) The number of states per micrometer below a frequency $f$ relative to the bare exciton, (b) the optical weight $\sum_n |g_n|^2/G^2$ below a frequency $f$ relative to the bare exciton, (c) the average dissipative exciton response $-\Im\ev{K(\omega)}$, for correlation $\Lambda_c = 10$\,nm.
        (d--f) The same but for $\Lambda_c = 100$\,nm.
        Shorter correlation lengths give significantly different average behavior due to the weaker binding in 2-d.
    }
    \label{fig:si-disorder-2d}
\end{figure}

These effects are illustrated in Fig.~\ref{fig:si-disorder-2d}, which plots the statistics of a 1-d and 2-d system against each other in the quantum and classical regimes.
As can be seen in Fig.~\ref{fig:si-disorder-2d}d, the spatial density of states below a given frequency is generally substantially higher in the classical limit due to the increased number of potential minima per unit length.
On the other hand, in the quantum limit (Fig.~\ref{fig:si-disorder-2d}a), although there are more states near zero detuning, the tail collapses much faster in 2-d due to the weaker binding enhancing the effect of motional narrowing.
This drives a similar collapse in the optical weight below a given frequency (Fig.~\ref{fig:si-disorder-2d}b) in the quantum limit, whereas the weight distribution remains effectively identical in the classical limit (Fig.~\ref{fig:si-disorder-2d}e).
Similarly, the total exciton response in the quantum limit is strongly modified (Fig.~\ref{fig:si-disorder-2d}c) while the classical limit is effectively identical (Fig.~\ref{fig:si-disorder-2d}f).

\subsubsection{Transduction in the discrete limit}

\begin{figure}
    \centering
    \includegraphics{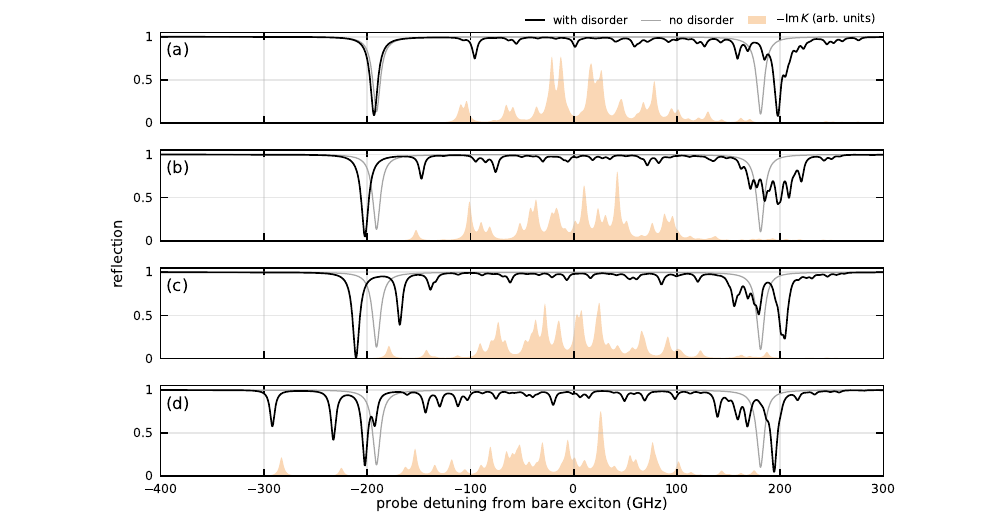}
    \caption{
        \textbf{Polaritons with discrete disorder.}
        Calculated polariton spectra for four different realizations of a potential with $\sigma_V/h=80$\,GHz and $\Lambda_c = 30$\,nm.
        The cavity parameters (including coupling) match the typical values of the device from the main text (specifically Fig.~5).
        (a) An instance where the states are clustered tightly and the LP lands approximately where it would in the non-disordered system.
        (b) A single weakly coupled exciton state lands near the LP causing a dispersive shift.
        (c) A strongly coupled exciton state lands almost on the LP, hybridizing with it to form a pair of split peaks.
        (d) Many exciton states land near the LP and the polariton cannot be cleanly distinguished.
    }
    \label{fig:si-disorder-polariton}
\end{figure}

\begin{figure}
    \centering
    \includegraphics{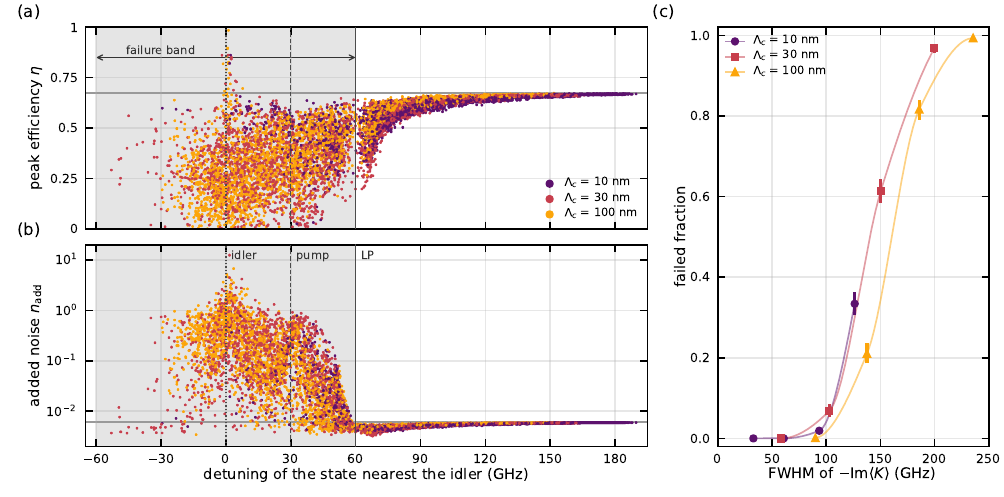}
    \caption{
        \textbf{Transduction with discrete disorder.}
        (a) The peak transduction efficiency and (b) added noise at the impedance matched point as a function of the distance between the idler line and the nearest exciton state.
        The pump is placed relative to the brightest optical pole to account for dispersive shifts.
        The data includes all $12{,}000$ computed realizations of a 25-\textmu m-long 1-d waveguide with varying $\sigma_V$ and $\Lambda_c$.
        The noise increases substantially when an exciton state lands at the idler or the pump lines.
        We define a $4\omega_b$-wide band centered at the idler in which the devices are classified as ``failures''.
        (c) Plots the fraction of failed devices against the FWHM of the exciton response $-\Im\ev{K(\omega)}$ for different $\Lambda_c$.
        The same set of disorder amplitudes $\sigma_V$ are used throughout, but the motional narrowing modifies the measured line width.
    }
    \label{fig:si-disorder-failure}
\end{figure}

Provided these models reasonably approximate the physical system, the exciton response function $K(\omega)$ will not be the smooth, asymptotically decaying version of the main text, but will rather consist of a cluster of discrete and overlapping peaks with varied heights.
Near the LP, the exciton response can have substantial value and dispersion which affects the polariton much more strongly than the linearized model would otherwise suggest.
Figure~\ref{fig:si-disorder-polariton} plots the polariton spectra of four different realizations of the exciton ensemble as computed from the disorder model above.
They illustrate the general effects that the discrete disorder can have on the polariton.
In Fig.~\ref{fig:si-disorder-polariton}a, the ensemble is tightly grouped around the mean bare-exciton value and the LP almost exactly matches the clean system.
In Fig.~\ref{fig:si-disorder-polariton}b, a single exciton state in the low-energy tail lands close to the LP causing a large dispersive shift on the order of a line width.
If an exciton state lands within a line width of the LP as in Fig.~\ref{fig:si-disorder-polariton}c, then the LP splits, hybridizing with that exciton branch.
Finally, once many states start to form near the LP, the structure becomes much more complicated and it becomes difficult to unambiguously identify the LP.
Thus, in the discrete limit, the effect of the excitons is not to smoothly broaden the LP, but rather induce a combination of hybridization and dispersive shifts.

With respect to transduction, small dispersive shifts are largely negligible as the pump frequency can always be tuned to match the LP.
In comparison to the continuum case, since the LP is not generally broadened by the dispersive shifts, the performance can essentially match that of the non-disordered system even in the presence of finite disorder.
However, hybridization can reduce the efficiency by reducing the effective photon fraction $|C|^2$ of the LP.
This is shown in Fig.~\ref{fig:si-disorder-failure}a as a steady decrease in the peak transduction efficiency as the lowest energy exciton state approaches the LP.
However, beyond a reduction in efficiency, the discrete case has a far more pernicious failure mode arising from the increased added noise when the exciton states form below the LP.
This is seen in Fig.~\ref{fig:si-disorder-failure}b as a several order-of-magnitude increase in the added noise when an exciton state lands below the LP in a $4\omega_b$ band around the idler.
This noise injection from the excitons arises from the parametric gain/amplification process being driven on the excitons in the tail and was a familiar issue in the continuous case as well.
However, unlike the continuous case, the discrete exciton states are generally more strongly coupled and so their contribution to the noise is substantially larger.
This is why the added noise on average peaks when an exciton lands on the idler or pump lines wherein the amplification noise is resonantly enhanced rather than being sideband suppressed.
Thus, disorder in the discrete case arises not as a smooth reduction in the performance but a discrete failure mode.
Since this loosely corresponds to the probability that an exciton state lands in the failure band the fraction of failed devices roughly follows the underlying distribution of the exciton states themselves as shown in Fig.~\ref{fig:si-disorder-failure}c.

\subsubsection{Transition to the continuum limit}
Finally we show how the discrete limit transforms into the continuum limit as the number of exciton states is increased.
To do so we consider a heuristic model where the exciton frequencies $\omega_n$ are Gaussian distributed and the weights are uniform, i.e.\ $p_n = 1/N$, where $N$ is the number of exciton states in the device.
In the limit of $N\to\infty$, the exciton response $K(\omega)$ maps to a Voigt line shape from which the continuum limit can be calculated exactly via the methods described above.

\begin{figure}
    \centering
    \includegraphics{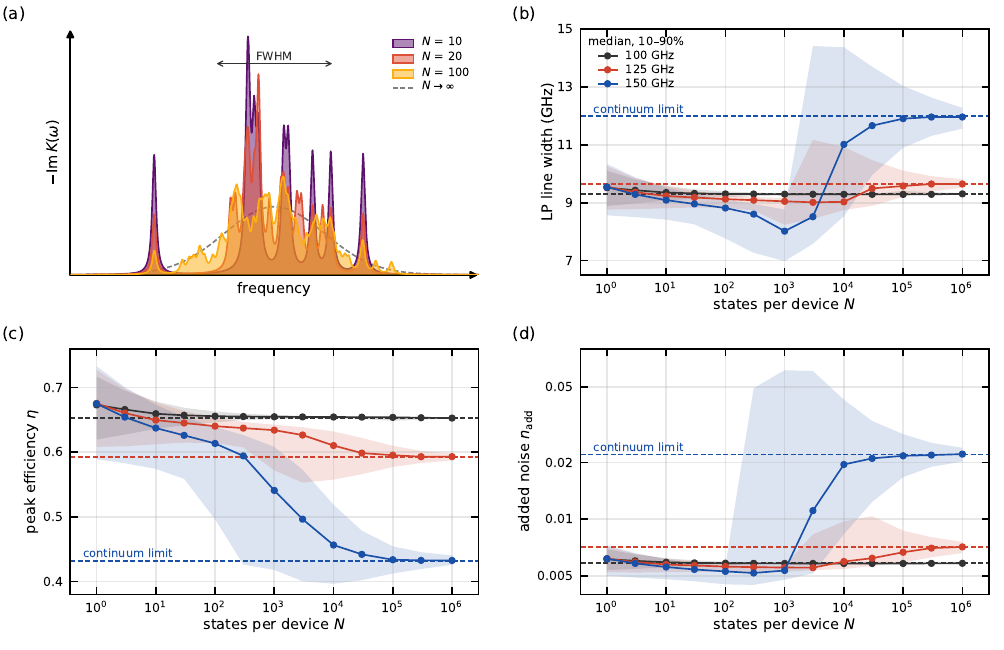}
    \caption{
        \textbf{Discrete-to-continuum limit.}
        (a) An illustration of the dispersive exciton response function as the number of states $N$ is increased.
        Here states are not modeled from the disorder itself but are given equal optical weight $g_n = G/\sqrt{N}$ and frequency distributed according to a Gaussian.
        States are continually added such that all states present in $N=10$ are also in $N=20$ and $N=100$, however their individual weights proportionally decrease.
        As $N\to\infty$ the response converges to the Voigt shape implied by the Gaussian distribution.
        (b) The median and 10/90-percentiles of the LP line width as the number of states is increased via Monte Carlo simulation.
        (c) The median and 10/90-percentiles of the peak transduction efficiency and (d) added noise at the impedance matched point as the number of states increases.
    }
    \label{fig:si-disorder-nsweep}
\end{figure}

Figure~\ref{fig:si-disorder-nsweep}a shows an example exciton response as the number of samples $N$ is increased.
For small $N$, the response appears as a series of discrete Lorentzian responses.
As more states are added the weight of any single peak goes down and more states begin to overlap in the center until the response converges to the continuum limit.
We track the polariton line width, peak transduction efficiency, and added noise at the impedance-matched point as the number of states is increased for different disorder line widths of $\{100,125,150\}$\,GHz in Fig.~\ref{fig:si-disorder-nsweep}b--d.

At $N=1$, the model maps to the non-disordered system with a random exciton-cavity detuning.
The variance between devices is large at this point because the cavity fraction of the exciton varies widely.
As $N$ increases, the median system performance begins to drift due to the random dispersive shifts becoming more common, however the variance actually narrows slightly as increased sampling better averages the exciton response.
However, the probability that an exciton state lands in the failure band eventually begins to increase substantially and the variance grows correspondingly.
The variance then peaks roughly when about half the devices contain a single exciton in the failure band.
This is where the discrete failure modes become most prominent in the statistics and the device performance begins to change rapidly.
For our particular operation point, this corresponds to $N\sim10^3$ for the simulated distributions which explains the change in the curvature near this point in Fig.~\ref{fig:si-disorder-nsweep}b--d.
With further increasing $N$, the failure band becomes increasingly populated and system response smooths out.
Once the failure band is populated so densely that individual states cannot be resolved, the median response converges on the continuum limit and the variance shrinks.

\clearpage
\bibliography{main}